\documentclass[twocolumn]{aastex631}

\definecolor{webgreen}{rgb}{0,.5,0}
\definecolor{webbrown}{rgb}{.6,0,0}

\newcommand{\marknotes}[1]{\noindent {\textsc{\textcolor{magenta}{#1}}}}

\graphicspath{{./}{figures/}}

\begin{document}

\title[SPGvsMPG]{Atmospheric Circulation in Simulations of the AGN-CGM Connection at Halo Masses $\sim 10^{13.5} \, M_\odot$}

\correspondingauthor{Deovrat Prasad, G. Mark Voit}
\email{deovratd@msu.edu,voit@msu.edu}

\author[0000-0003-1255-6375]{Deovrat Prasad}
\affiliation{Department of Physics and Astronomy, Michigan State University, MI, US}

\author[0000-0002-3514-0383]{G. Mark Voit}
\affiliation{Department of Physics and Astronomy, Michigan State University, MI, US}

\author[0000-0002-2786-0348]{Brian W. O'Shea}
\affiliation{Department of Physics and Astronomy, Michigan State University, MI, US}
\affiliation{Department of Computational Mathematics, Science, and Engineering, Michigan State University, MI, US}

\begin{abstract}
Coupling between active galactic nuclei (AGNs) and the circumgalactic medium (CGM) is critical to the interplay between radiative cooling and feedback heating in the atmospheres of the universe's most massive galaxies. This paper presents a detailed analysis of numerical simulations showing how kinetic AGN feedback with a strong momentum flux interacts with the CGM. Our analysis shows that 
large scale CGM circulation driven by that momentum flux plays an important role in reconfiguring the galactic atmosphere and regulating the atmosphere's central entropy level. We find that most of the AGN’s energy output goes into lifting of circumgalactic gas rather than heating of atmospheric gas within the galaxy, consequently reconfiguring the circumgalactic medium (CGM) by replacing low entropy gas originally in the core with higher entropy gas from larger radii. Circulation of the CGM on $\sim 10–100$ kpc scales therefore plays a critical role in preventing over-cooling of gas in these simulated galaxies but leads to elevated entropy profiles $\sim 1–10$ kpc compared to the observed entropy profiles of massive elliptical galaxies in the same mass range. The simulations also show that our choices of accretion efficiency and jet opening angle significantly affect the AGN-CGM coupling. Reducing the jet opening angle to 1/4 of the fiducial opening angle increases the jet momentum flux, enabling it to drill through to larger radii without effectively coupling with the CGM at the center ($r < 5$ kpc). Outflows with a lower momentum flux decelerate and thermalize the bulk of their energy at smaller radii ($r \lesssim 10$ kpc). 
\end{abstract}
\keywords{Galaxies, Cooling Flow, Super Massive Black Hole, AGN Feedback, Stellar Feedback, Circumgalactic Medium }

\section{Introduction}
\label{sec:introduction}

Active galactic nuclei (AGN) feedback appears necessary for balancing radiative cooling of the circumgalactic medium (CGM) surrounding massive galaxies. Without AGN feedback, the CGM of a massive elliptical galaxy inevitably develops a strong central cooling flow that feeds low-entropy gas into the galaxy, where it fuels star formation \citep{fabian1994,lewis2000}.
AGN feedback can limit or prevent star formation by doing two things: (1) it can provide at least enough energy to the CGM to compensate for radiative cooling \citep{McNamara2007}, and (2) 
it can operate without exciting thermal instabilities that lead to overproduction of cold gas clouds capable of forming stars \citep{pizz2005,mccourt12, sharma12}.
Point (2) implies that the heat input into the CGM cannot be overly centralized, because overheating of the central gas stimulates convection and therefore drives thermal instabilities in the CGM \citep{Field1965,Balbus1986ApJ...303L..79B,Balbus1988ApJ...328..395B,voit2017}. X-ray observations show that AGN feedback in massive elliptical galaxies can accomplish both (1) and (2) without greatly raising the entropy and cooling time of gas at the centers of $10^{13} M_\odot$--$10^{14} \, M_\odot$ halos \citep{birzan2004,rafferty2006, voit15N, voit2015}.

Within the central few kiloparsecs (kpc) of many nearby massive ellipticals, entropy levels in the ambient medium are observed to go below 5 keV cm$^2$ (\citealt{werner2012,werner2014, babyk2018}), corresponding to a central cooling time $t_{\rm cool} \lesssim 100$~Myr. In order to remain in such a state, the feedback loop that limits cooling must be able to tune itself. Otherwise, AGN feedback would overheat the ambient galactic atmosphere in its vicinity. Self-regulation of AGN feedback therefore requires a closed feedback loop that tunes itself on a timescale of less than 100 Myr (\citealt{Gaspari2012_ellipticals,wang2019,prasad2020}). 

\citealt{prasad2020} (Paper I hereafter) introduced a suite of simulations designed to explore the dependence of AGN feedback on halo mass and galactic environment. It implemented an identical feedback mechanism in halos ranging in mass from $10^{12} \, M_\odot$ to $10^{15} \, M_\odot$ to test how the same feedback mechanism would respond in different galactic environments. The AGN feedback mechanism was bipolar kinetic outflows fueled by accretion of cold gas with a fraction $10^{-4}$ of the rest mass transformed into feedback energy. We found that the results of AGN feedback depended on (1) halo mass, (2) CGM pressure, (3) the stellar velocity dispersion of the central galaxy, and (4) the balance between radiative cooling and SNIa heating. 

This paper focuses on a subset of those simulations centering on halos of mass $\sim 10^{13.5} \, M_\odot$, in which AGN feedback couples interestingly with CGM pressure and SNIa heating. The simulations of those halos displayed two different modes of self-regulation. In the Single Phase Galaxy (SPG) simulation, self-regulation was tight, leading to a nearly steady state while producing very little extended multiphase gas. On the other hand, self-regulation in the Multi-phase Galaxy (MPG) simulation was weaker, with AGN feedback switching back and forth between high-power and low-power modes. 

According to the “black hole feedback valve” hypothesis of \citealt{voit2020}, the fine-tuning mechanism observed in the Single Phase Galaxy simulation of Paper I is enabled by two things:
\begin{enumerate}
    \item A deep central gravitational potential, corresponding to a stellar velocity dispersion $\sigma_v > 240 \, {\rm km \, s^{-1}}$.
    \item Narrow bipolar outflows from the AGN that enable kinetic power to drill through the ambient medium at 1--10~kpc and thermalize the outflow energy in the CGM at $> 10$~kpc.
\end{enumerate}
Requirement (1) allows SNIa heating to maintain a steady-state atmosphere consisting of a cooling flow at small radii and a heated outflow that sits atop the cooling flow.  Requirement (2) allows AGN feedback to regulate the inner cooling flow by coupling AGN feedback directly to the CGM, thereby regulating CGM pressure without disrupting the ambient medium within the galaxy. 

\begin{figure*}[!t] 
 \includegraphics[width=3.0in,height=2.8in]{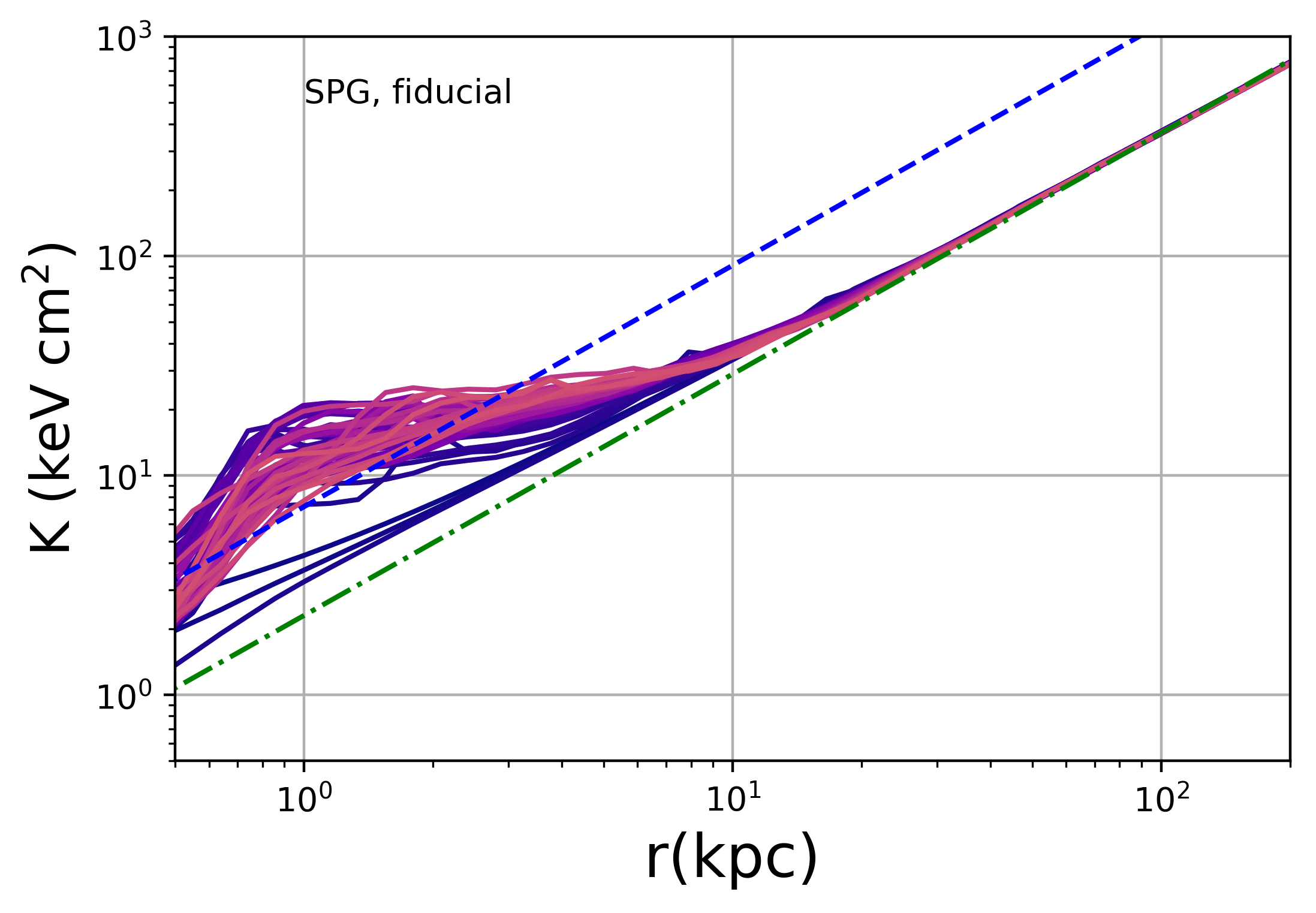}
 \includegraphics[width=3.6in,height=2.8in]{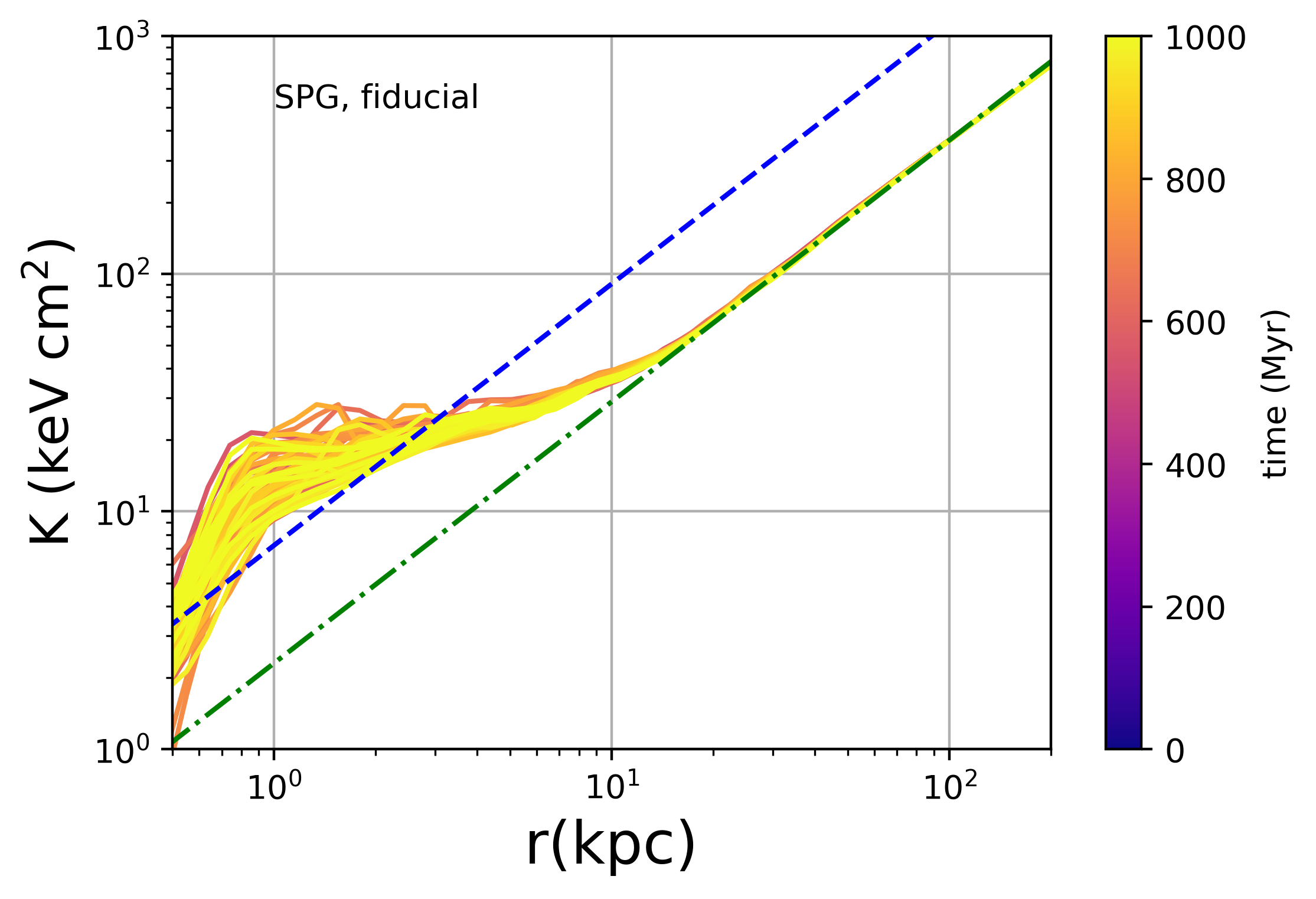}
 \includegraphics[width=3.3in,height=2.8in]{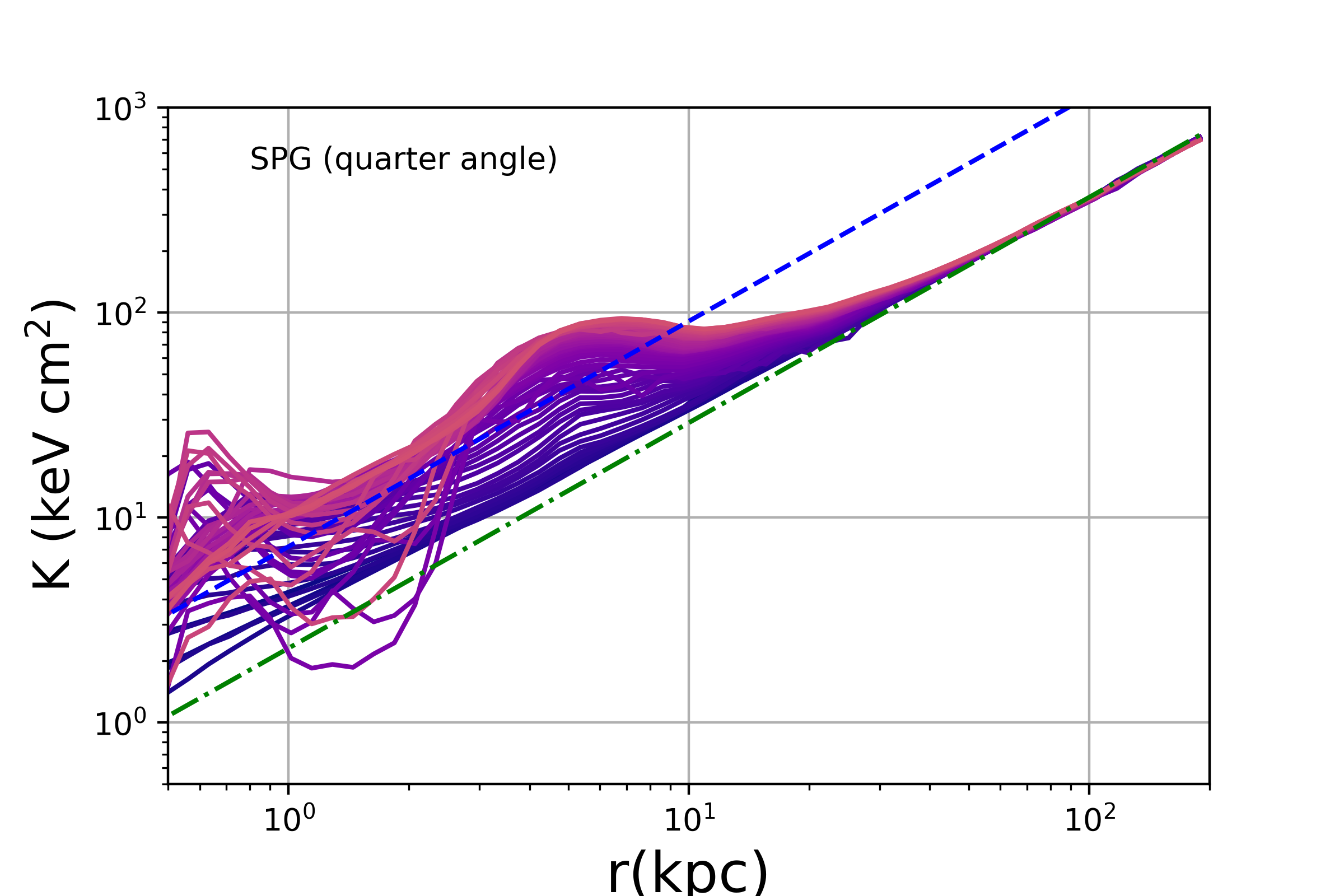}
 \includegraphics[width=4.1in,height=2.8in]{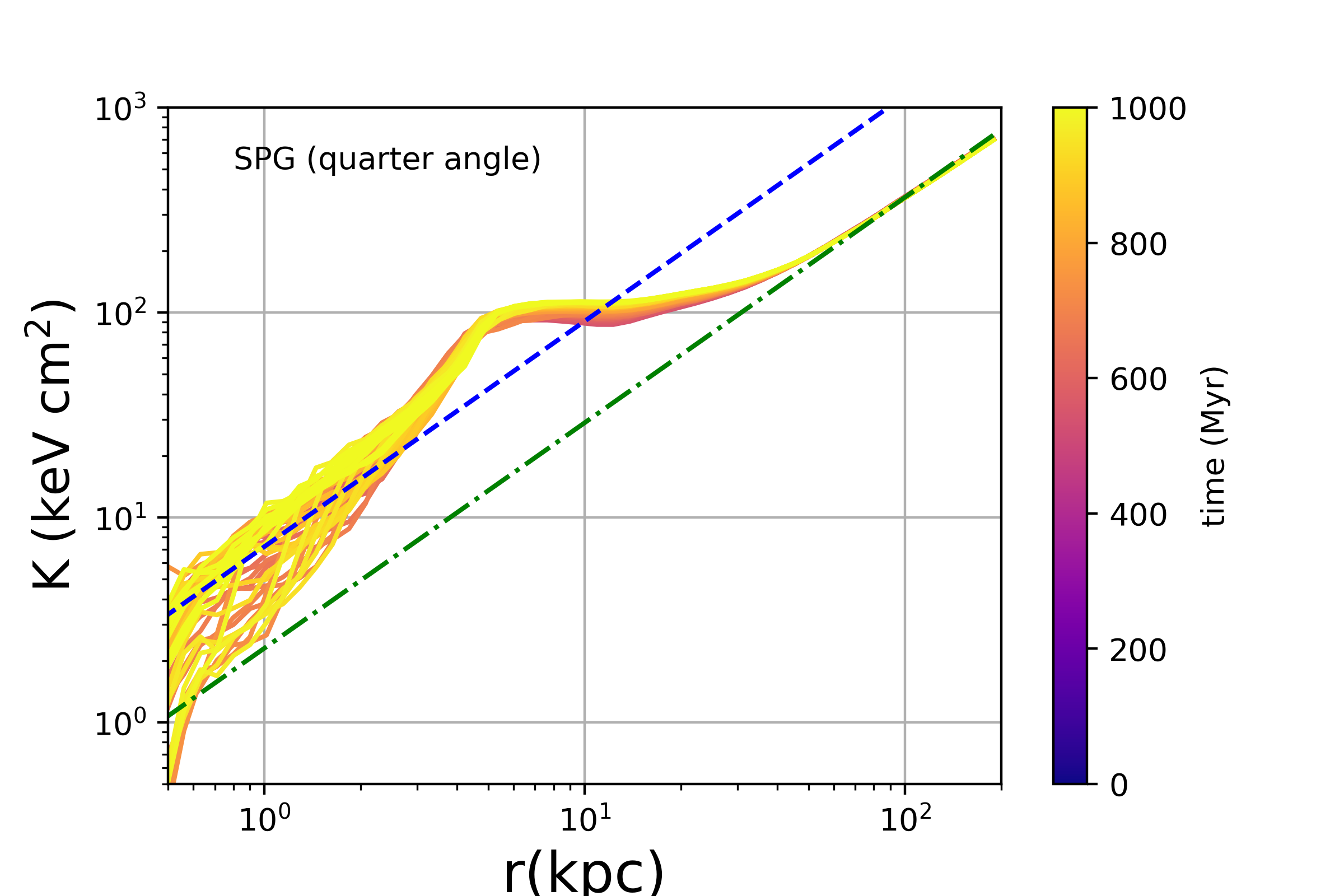}
 \caption{Evolution of mass-weighted mean entropy profiles in the single phase galaxy (SPG) simulations from Paper I, with the fiducial simulation on the top and a simulation with narrower jets, one-quarter the width of the fiducial jets, on the bottom.  The dot-dashed green lines are similar to the observed entropy profiles of single-phase massive elliptical galaxies (such as NGC~4472 and NGC~4261), in that they have $K \propto r^{1.1}$ and a similar normalization. Those lines are also similar to the initial entropy profiles in our SPG simulations.
 Colored solid lines representing 10~Myr intervals show how the entropy profiles evolve during the first 0.5~Gyr (left panels) and from 0.5-1 Gyr (right panels).  The dashed blue lines show entropy profiles with the same slope as the green dashed lines but an entropy normalization 3 times greater. In the fiducial simulation, the entropy excess relative to the initial state becomes greatest at 1--3~kpc.  In the simulation with narrower jets, an entropy excess develops farther from the center, peaking near 7~kpc. 
} 
 \label{fig:entropy}
\end{figure*}

The qualitative behavior of the simulations in Paper I aligns with the hypothesis presented in \citet{voit2020}.  However, the simulated atmospheres do not agree in detail with observations because simulated entropy levels in the central few kiloparsecs are elevated relative to observations (Figure \ref{fig:entropy}).  This result suggests that the simulated bipolar outflows are coupling too strongly to the ambient medium within the galaxy. 

The Appendix of Paper I presented a preliminary test of the strong-coupling hypothesis that explored the consequences of changing the jet opening angle $\theta_{\rm jet}$. Reducing the jet width in the SPG simulations to 1/2 and 1/4 of the original opening angle showed that those changes did indeed affect the entropy profile at 1-10 kpc. However, they did not improve overall agreement of the simulated entropy profiles with observations. Instead, the narrower jets produced excess entropy relative to observations at $\sim 10$~kpc (see right panel of Figure~\ref{fig:entropy}).

This paper more deeply explores how AGN feedback in those simulations self-regulates and why they develop entropy excesses develop at $r\sim 1-10$~kpc. Our broad goal is to clarify the features that AGN feedback needs to have in order to produce galactic atmospheres in better agreement with X-ray observations. A more specific goal is to test the black hole feedback valve hypothesis, which proposes that bipolar jets fueled by a cooling flow at $\lesssim 1$~kpc closely couple CGM pressure with AGN feedback in galaxies having $\sigma_v > 240 \, {\rm km \,  s^{-1}}$.  

The analysis of simulated AGN feedback presented in this paper shows that the AGN's power output is not tuning itself to balance radiative cooling, as seems to happen in the cores of galaxy clusters. Instead, the AGN suppresses central cooling by reconfiguring the galaxy's atmosphere via large scale circulation patterns that raise entropy levels at $< 10$~kpc. Each simulation begins with a powerful bipolar outburst that lifts low-entropy gas out of the central region along the polar axis, allowing higher-entropy gas to sink inward along directions more than $\sim 30^\circ$ from the polar axis. That circulation pattern gradually boosts the ambient atmospheric entropy at $r\sim 1$--10~kpc until SNIa heating exceeds radiative cooling within the galaxy. This entropy boost results in AGN feedback settling into a low-power state maintained by SNIa heating that continually sweeps ejected stellar gas out of the galaxy's center. During the low-power state, the swept-out gas accumulates $r\sim 30$--100~kpc from the galaxy's center. 
Gas accumulation in the CGM slowly raises the CGM pressure and eventually leads to another strong AGN outburst that once again reconfigures the galaxy's atmosphere through large-scale circulation rather than direct heating.

The paper proceeds as follows.  Section \ref{sec:analysis} introduces the analysis methods to be applied in subsequent sections.  Section \ref{sec:AGN} examines how AGN feedback self-regulates in the simulations from Paper I, paying particular attention to the role of stellar heat input and what ultimately happens to AGN feedback energy.  Section \ref{sec:circ} looks more closely at how AGN feedback drives large-scale circulation in these simulated galactic atmospheres and how circulation affects the radial distribution of entropy. Section \ref{sec:disc} discusses whether our results can be considered realistic, explores their dependence on assumptions about feedback efficiency and jet width, and compares them to some prior simulations of bipolar AGN feedback in massive galaxies. Section \ref{sec:conc} summarizes the paper's main findings.

\section{Simulations and Analysis Methods}
\label{sec:analysis}
This section briefly describes the AGN feedback simulations presented in Paper I and introduces the methods this paper adopts for analyzing them.

\subsection{Simulations}
\label{sec:sim}

The simulations we analyze here were carried out with the Enzo code (\citealt{Bryan2014,Enzo_2019}). The gravitational potential confining each simulated galactic atmosphere consists of three components: a dark-matter potential that follows an NFW profile, a stellar potential that follows a Hernquist profile, and a supermassive black hole (SMBH) potential approximated with a Paczyinski-Witta profile.  In the SPG simulations, a central galaxy of a total stellar mass $M_* = 2 \times 10^{11} \, M_\odot$ and central velocity dispersion $\sigma_v \approx 280 \, {\rm km \, s^{-1}}$ is embedded in a halo of mass $M_{\rm halo} = 10^{13.6} \, M_\odot$ ($c_{200}=7.5$) and has a central black hole mass $M_{\rm BH} = 2.6 \times 10^9 \, M_\odot$. In the MPG simulations, a central galaxy of a total stellar mass $M_* = 1.2 \times 10^{11} \, M_\odot$ and central velocity dispersion $\sigma_v \approx 230 \, {\rm km \, s^{-1}}$ is embedded in a halo of mass $M_{\rm halo} = 10^{13.6} \, M_\odot$ ($c_{200}=7.5$) and has a central black hole mass $M_{\rm BH} = 4.6 \times 10^8 \, M_\odot$. 

Initially, the galactic atmospheres are spherically symmetric, homogeneous at each radius, and close to hydrostatic equilibrium.  Atmospheric structure at $t = 0$ depends on atmospheric entropy profiles that were chosen to mimic two particular galaxies.  The SPG simulations are meant to be similar to NGC~4472 and have an initial entropy profile $K(r) = 1.5 + 400 (r / 100 \, {\rm kpc})^{1.05}$, where $K \equiv kT n_e^{-2/3}$ is in units of keV~cm$^2$. The MPG simulations are meant to be similar to NGC 5044 and have an initial entropy profile $K(r) = 1.3 + 150 (r / 100 \, {\rm kpc})^{1.05}$ in units of keV~cm$^2$.  The lower circumgalactic entropy in the MPG simulations results in greater circumgalactic pressure, density, and gas mass. Consequently, the MPG's CGM more strongly confines the central galaxy's atmosphere and suffers greater radiative losses.

Two separate energy sources can mitigate radiative cooling of those atmospheres: stellar heating and AGN feedback.  Stellar heating consists of feedback due to stars formed during the course of the simulation as well as heating from the preexisting old stellar population. Feedback from stars formed during the simulation follows the prescription from \citet{Bryan2014}. Feedback due to the old stellar population is modelled with steady, spherically symmetric injection of thermal energy into the simulation domain at a rate proportional to the stellar mass density. The old stellar population also injects kinetic energy as it sheds gas mass in the form of stellar winds and SN Ia explosions. Our simulations assume that this kinetic energy immediately thermalizes.
Steady stellar heating (SNIa heating plus thermalization of the kinetic energy of ejected stellar gas from the old stellar population) injects thermal energy at a specific rate $\sim 10^{30} \, {\rm erg \, s^{-1}} \, M_\odot^{-1}$, producing $10^{41.4} \, {\rm erg \, s^{-1}}$ in the SPG simulations and $10^{41.2} \, {\rm erg \, s^{-1}}$ in the MPG simulations.
AGN feedback power depends on the amount of cold gas within the central 0.5~kpc and pumps energy into the galactic atmosphere via bipolar jets.  The power during a particular time step is
\begin{equation}
    \dot{E_{\rm AGN}} = \epsilon_{\rm AGN} \dot{M}_{\rm acc} c^2
\end{equation}
where $\epsilon_{\rm AGN} = 10^{-4}$ is the efficiency factor applied in all simulations presented in Paper~I.\footnote{Section \ref{sec:comparison} discusses the rationale for this choice of $\epsilon_{\rm AGN}$ and its consequences for jet momentum flux and AGN-CGM coupling. }
The AGN accretion rate $\dot{M}_{\rm acc}$ is taken to be the cold gas mass at $r < 0.5$~kpc divided by 1~Myr, and $c$ is the speed of light.  Feedback energy output emanates from the ends of two cylinders extending 1~kpc from the galaxy's center and is 90\% kinetic and 10\% thermal.  Gas mass is conserved, because the outward mass flow equals the gas mass subtracted from cold clouds within 0.5~kpc during each time step.  In the fiducial simulations, the jet opening angle is $\theta_{\rm jet} = 27^\circ$ from the polar axis.  In the quarter-angle jet simulations, all parameters are the same, except that $\theta_{\rm jet} = (27/4)^\circ \approx 7^\circ$.

\subsection{Energy Flow}
\label{sec:ber_met}

In the simulations of Paper~I, the impact of AGN feedback on the surrounding atmosphere turns out to depend critically on the flow pattern of atmospheric gas.  This paper therefore examines propagation of the Bernoulli specific energy
\begin{equation}
    \epsilon_{\rm B} = \frac{u+P}{\rho}+ \frac{v^2}{2} + \phi 
    \; \; ,
\end{equation}
where $\rho$ is gas mass density, $u$ is thermal energy density, $P$ is pressure, $v^2/2$ is the specific kinetic energy of bulk flow, and $\phi$ is the gravitational potential.  The Bernoulli energy flux of gas moving with velocity \textbf{v} is
\begin{equation}
    {\bf F_B} = \epsilon_{\rm B}\rho{\bf v}
    \; \; .
\end{equation}
To account for the energy flow through a spherical shell at radius $r$, we define
\begin{equation}
    \dot{E}_{\rm B} = \int_{4\pi} \epsilon_{\rm B} \rho v_r r^2 d\Omega 
\end{equation}
to be the radial Bernoulli energy flow. However, unlike the thermal and kinetic components of Bernoulli energy, the gravitational energy flowing outward through the shell is not able to offset radiative cooling at larger radii.  We therefore define
\begin{equation}
    \dot{E}_{\rm flow} = \dot{E}_{\rm B} - \dot{M} \phi  
    \; \; ,
\end{equation}
in which $\dot{M}$ is the net outflow rate of gas mass through the shell at $r$, to be the non-gravitational component of Bernoulli energy flow.

The total energy $E(r,t)$ of gas within a sphere of radius $r$ changes at the rate
\begin{equation}
    \frac{\partial E}{\partial t} = -\dot{E}_{\rm B} + \dot{E}_{\rm AGN} + \dot{E}_* + \dot{E}_{\phi *} - \dot{E}_{\rm rad} 
\end{equation}
where $\dot{E}_{\rm AGN}$ is the rate of AGN energy input, $\dot{E}_*$ is the rate of non-gravitational stellar energy input, $\dot{E}_{\phi *}$ represents potential energy gains owing to stellar mass ejection that adds gas to the galaxy's atmosphere, and $\dot{E}_{\rm rad}$ represents radiative energy losses. Stellar energy input has both supernova and kinetic components, so that 
\begin{equation}
    E_\star(r) = \int_0^r \left( \epsilon_{\rm SN} + \frac{3}{2}\sigma_v^2 \right) \dot{\rho}_\star dV
    \; \; ,
\end{equation}
where $\dot{\rho}_*$ is the stellar mass loss rate per unit volume and $\epsilon_{\rm SN}$ is the specific SN energy per unit mass loss from the entire stellar population. The total energy ejected from SNIa explosions assumes $10^{51}$ erg/SNIa at a specific rate of $3\times10^{-14}$ SNIa yr$^{-1}$ M$_\odot^{-1}$, following \citet{voit2015}. 
Radiative losses are determined by
\begin{equation}
    \dot{E}_{\rm rad} (r,t) = \int_0^r n_e n_i \Lambda (T) \, dV
    \; \; ,
\end{equation}
where $n_e$ and $n_i$ are electron and ion number densities, respectively, and $\Lambda (T)$ is the usual temperature-dependent radiative cooling function, assuming solar metallicity. In our analysis, we generally smooth the AGN power according to
\begin{equation}
    \dot{E}_{\rm AGN} (t,\tau) = \frac {1} {\tau} \int_0^t \dot{E}_{\rm AGN}(t^\prime) 
    \, e^{-(t - t^\prime)/\tau} \, dt^\prime
\end{equation}
with $\tau = 100$~Myr because $\dot{E}_{\rm AGN}(t)$ varies greatly on timescales of $\ll 100$~Myr. 

\subsection{Radial Distributions}
\label{sec:distributions}

The analysis of \S \ref{sec:AGN} examines how the local contributions to the energy budget in the vicinity of radius $r$ compare with the total energy flow through the system.  We therefore express the radial distributions of radiative losses and stellar heating via
\begin{eqnarray}
\label{eq:tild}
   \tilde{L}_{\rm rad}(r) & \equiv & \frac{\partial \dot{E}_{\rm rad}}{\partial \ln r} \\
   \tilde{P}_*(r) & \equiv & \frac{\partial \dot{E}_*}{\partial \ln r} 
    \; \; .
\end{eqnarray}
Dividing each of these quantities by $r$ gives the instantaneous rate of energy gain or loss $(\partial \dot{E} / \partial r) \Delta r$ within a radial shell of thickness $\Delta r$. Comparing the two quantities shows whether radiative losses dominate stellar heating, or vice versa, within a particular radial shell.

\section{AGN Feedback}
\label{sec:AGN}

This section describes the key results from our single phase galaxy and multi-phase galaxy simulations. We show the initial stages of evolution of the fiducial SPG and MPG runs in Section \ref{sec:init}. Section \ref{sec:regulation} presents the results during the self-regulation phase of the fiducial SPG and MPG simulations. This is followed by Section \ref{sec:config}, where we show that the strong AGN feedback leads to atmospheric reconfiguration in the fiducial runs. In Section \ref{sec:accumulation} we look at  the evolution of the stellar ejecta in the fiducial SPG and MPG runs. Finally, we discuss the probable cause for the excess entropy in our simulations in Section \ref{sec:ent}. \\

\subsection{Initial Outburst}
\label{sec:init}

\begin{figure*}[!t]
 \includegraphics[width=1.72in,height=1.5in]{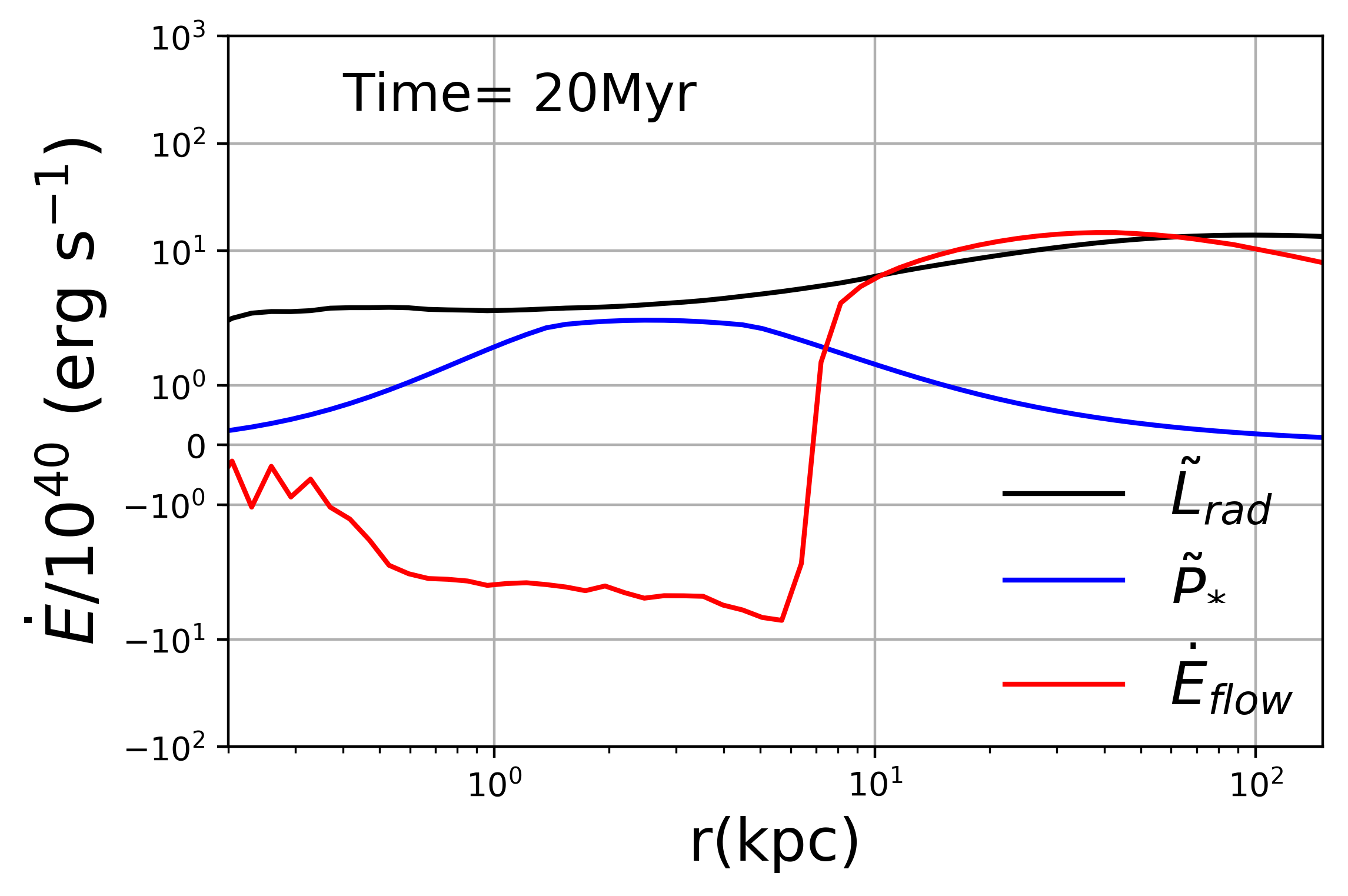}
 \includegraphics[width=0.75\textwidth]{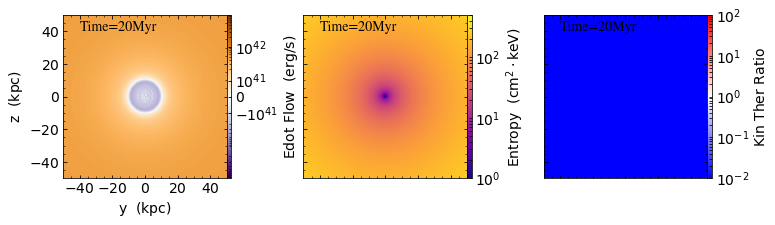}
 \includegraphics[width=1.72in,height=1.5in]{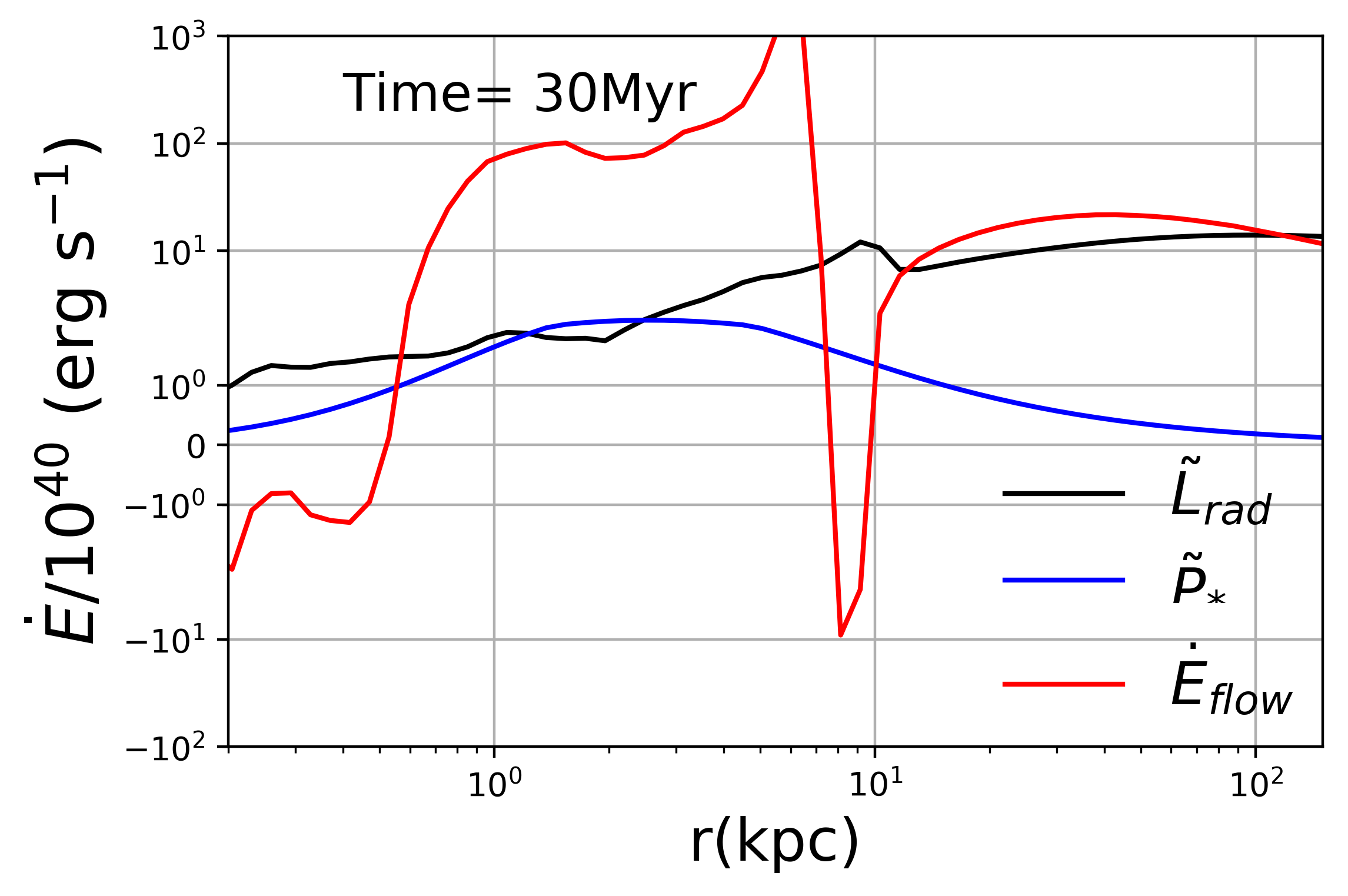}
 \includegraphics[width=0.75\textwidth]{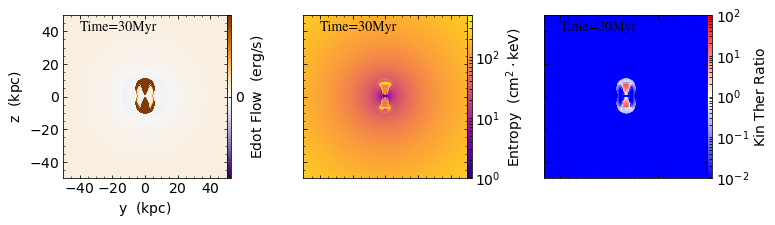}
 \includegraphics[width=1.75in,height=1.5in]{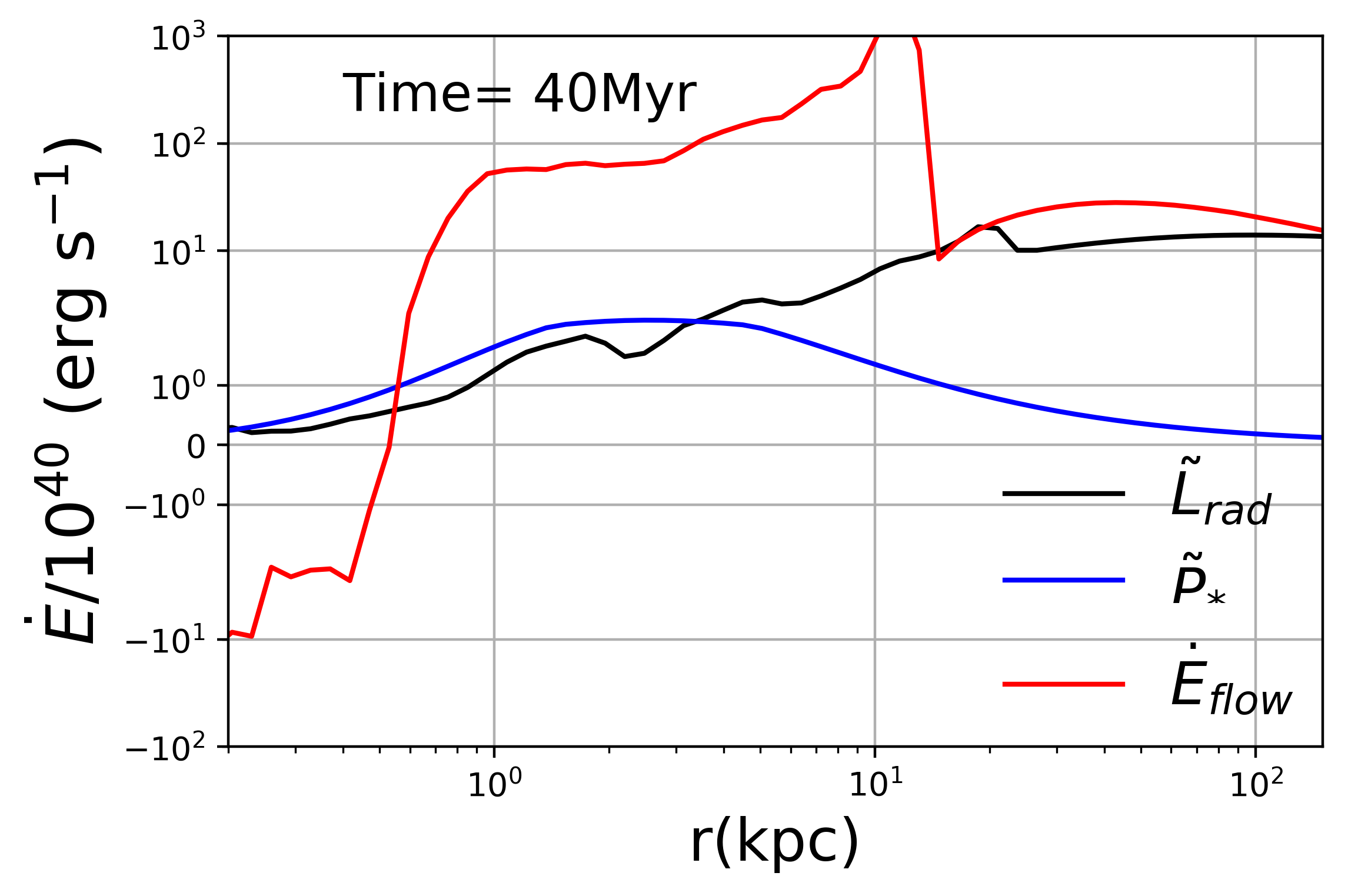}
 \includegraphics[width=0.75\textwidth]{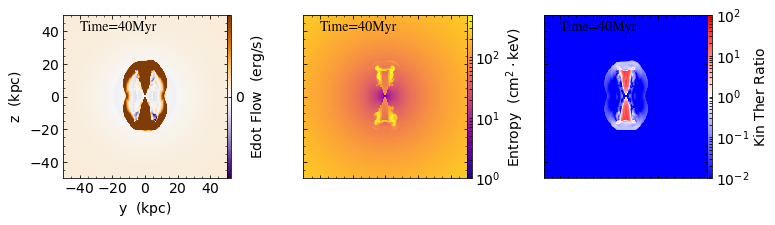}
 \caption{
 Evolution of the initial outburst in the fiducial SPG simulation, starting just before the outburst ($t = 20$~Myr) and proceeding through the time when the initial shock front is propagating through the galaxy ($t = 30$~Myr) to the time when stellar heating starts to exceed radiative cooling within the galaxy ($t = 40$~Myr). Panels in the left column show the radial distributions of various energy sources and sinks.  Black lines in those panels show the radial distribution of radiative losses ($\tilde{L}_{\rm rad}$). Blue lines show the radial distribution of power input from the galaxy's aging stellar population ($\tilde{P}_*$).  Red lines show the non-gravitational component of Bernoulli power ($\dot{E}_{\rm flow}$) flowing through radius $r$. Panels in the second column illustrate the polar distribution of $\dot{E}_{\rm flow}$ by showing $4 \pi r^2 (\epsilon_{\rm B} - \phi) \rho v_r$ over a 100 kpc $\times$ 100 kpc box.  
 Panels in the third column show the polar distribution of gas entropy ($K$) over the same area.
 Panels in the fourth column show the flow's ratio of kinetic to thermal energy in the galaxy's rest frame.  
}  
 \label{fig:init_spg}
\end{figure*}

After initialising the galaxy in near-hydrostatic equilibrium as described in Paper I, we let the galaxies evolve  for $t\sim1.5$ Gyr with stellar and AGN feedback switched on from the start. As described in \S \ref{sec:sim} the stellar feedback includes the star formation and feedback due to young stars as well as SNIa type feedback due to old stars.

\subsubsection{Fiducial SPG}

Figure \ref{fig:init_spg} shows the evolution of the single phase galaxy during $t\lesssim 40$ Myr. Its left column shows the radial distributions of radiative cooling (black), SNIa heating (blue) and non-gravitational Bernoulli power (red) in the initial stages of the fiducial SPG run. Other columns show snapshots of the non-gravitational Bernoulli energy flux (second column from left) and mass flux (third column from left), along with maps of kinetic energy to thermal energy ratio (right column).

The figure shows the evolution of the single phase galaxy's atmosphere at the following stages:
\begin{itemize}

    \item \textbf{20 Myr:} Gas within $r \approx 7$~kpc is cooling and flowing inward, because its initial cooling time is $\sim 20$~Myr.  Gas outside of that radius is moving outward because the initial state is not perfectly hydrostatic.  It is slightly over-pressured, relative to gravity, and so the atmosphere outside of the cooling region initially expands.  During this epoch, radiative losses locally exceed stellar power at all radii and AGN feedback has not yet begun because cold gas clouds have not yet formed.
    
     \item \textbf{30 Myr:} Formation of cold gas clouds within the accretion zone ($r < 0.5$~kpc) has initiated AGN feedback.  The initial feedback impulse has propagated to $\sim 7$~kpc, almost reaching the outer edge of the initial cooling flow, which is now at $r \approx 10$~kpc.  Gas propelled along the jet axis by the AGN is strongly shocked as it encounters resistance from the CGM, while the shock propagating into the CGM is relatively weak along the jet axis and has become subsonic in the equatorial regions.  Consqeuently, much of the gas at $< 10$~kpc but far from the jet axis still has $K < 10 \, {\rm keV \, cm^2}$. 
     
     \item \textbf{40 Myr:} The initial feedback impulse has propagated beyond $\sim 15$~kpc, and high-entropy bubbles are starting to form at that radius. Low-entropy gas still persists in the equatorial regions.  However, the initial AGN impulse has lowered the gas density within $r < 3$~kpc enough for stellar heating to exceed radiative cooling in that region.
          
\end{itemize}

During this crucial period, the initial AGN outburst has flipped the inner atmosphere of the galaxy from a cooling-dominated state to a heating-dominated state without strongly disrupting the inner 10~kpc because most of the AGN energy is drilling to larger radii through bipolar channels.

\subsubsection{Fiducial MPG}
 
\begin{figure*}[!t]
\includegraphics[width=1.7in,height=1.6in]{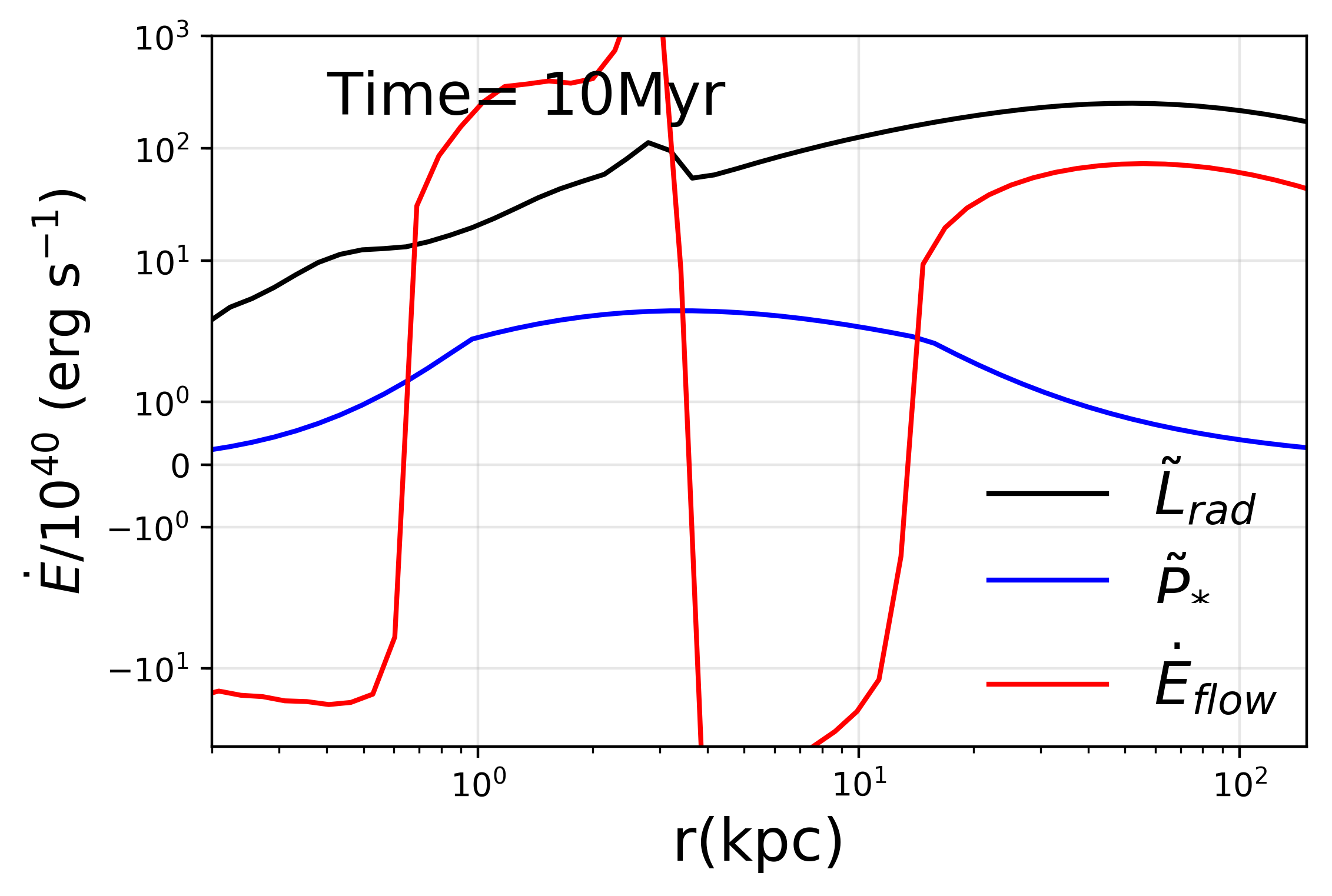}
\includegraphics[width=0.75\textwidth]{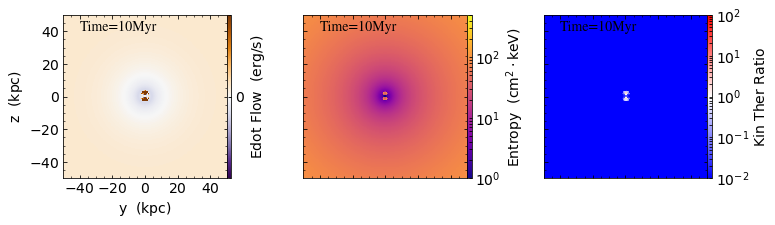}
\includegraphics[width=1.7in,height=1.6in]{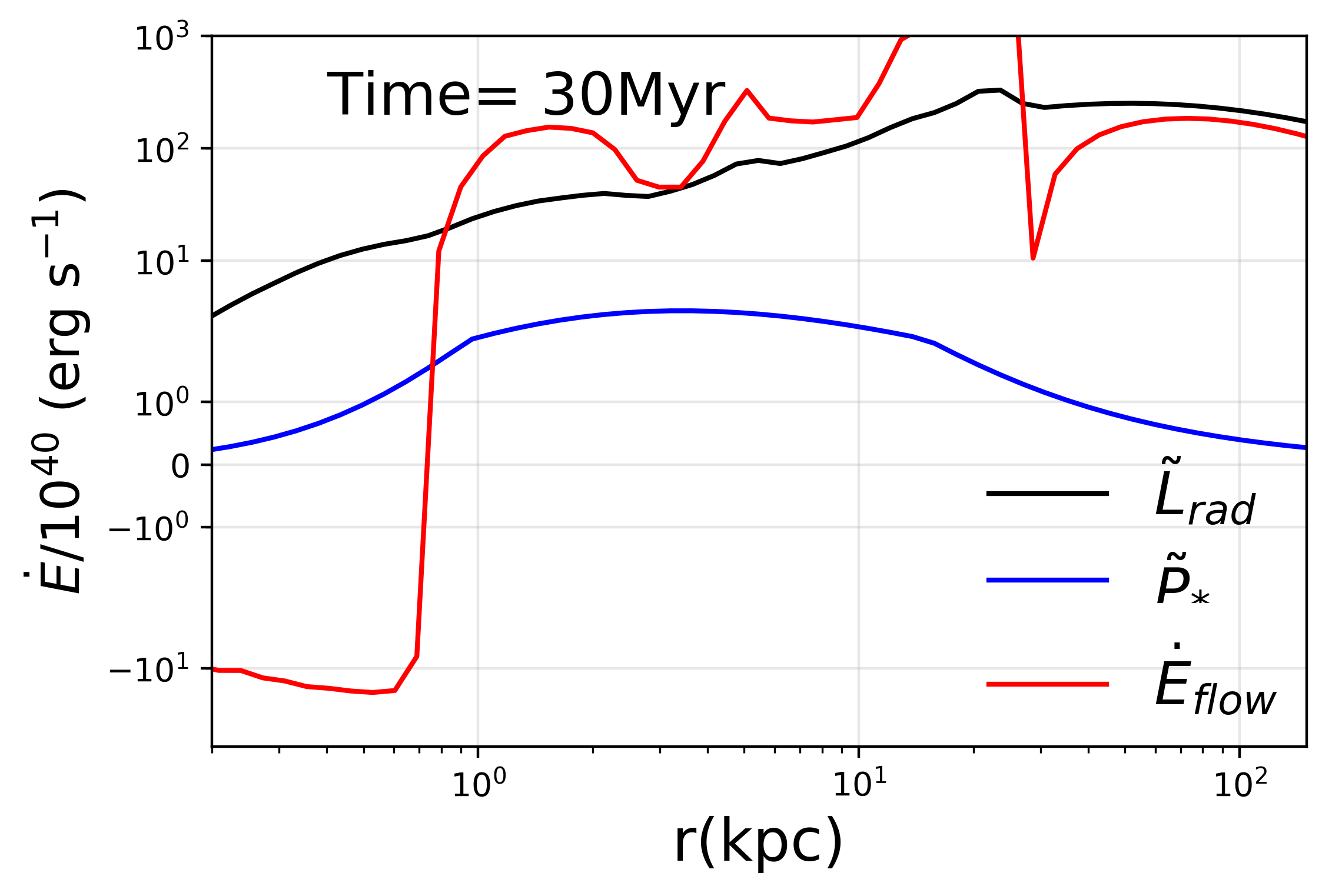}
\includegraphics[width=0.75\textwidth]{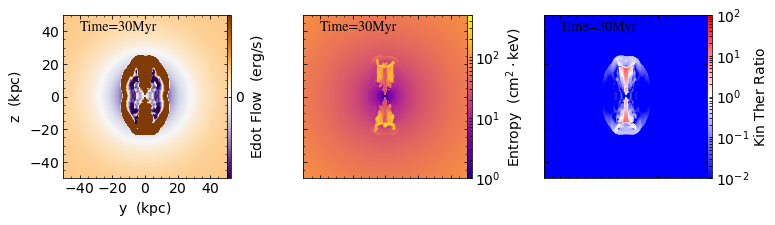}
\includegraphics[width=1.7in,height=1.6in]{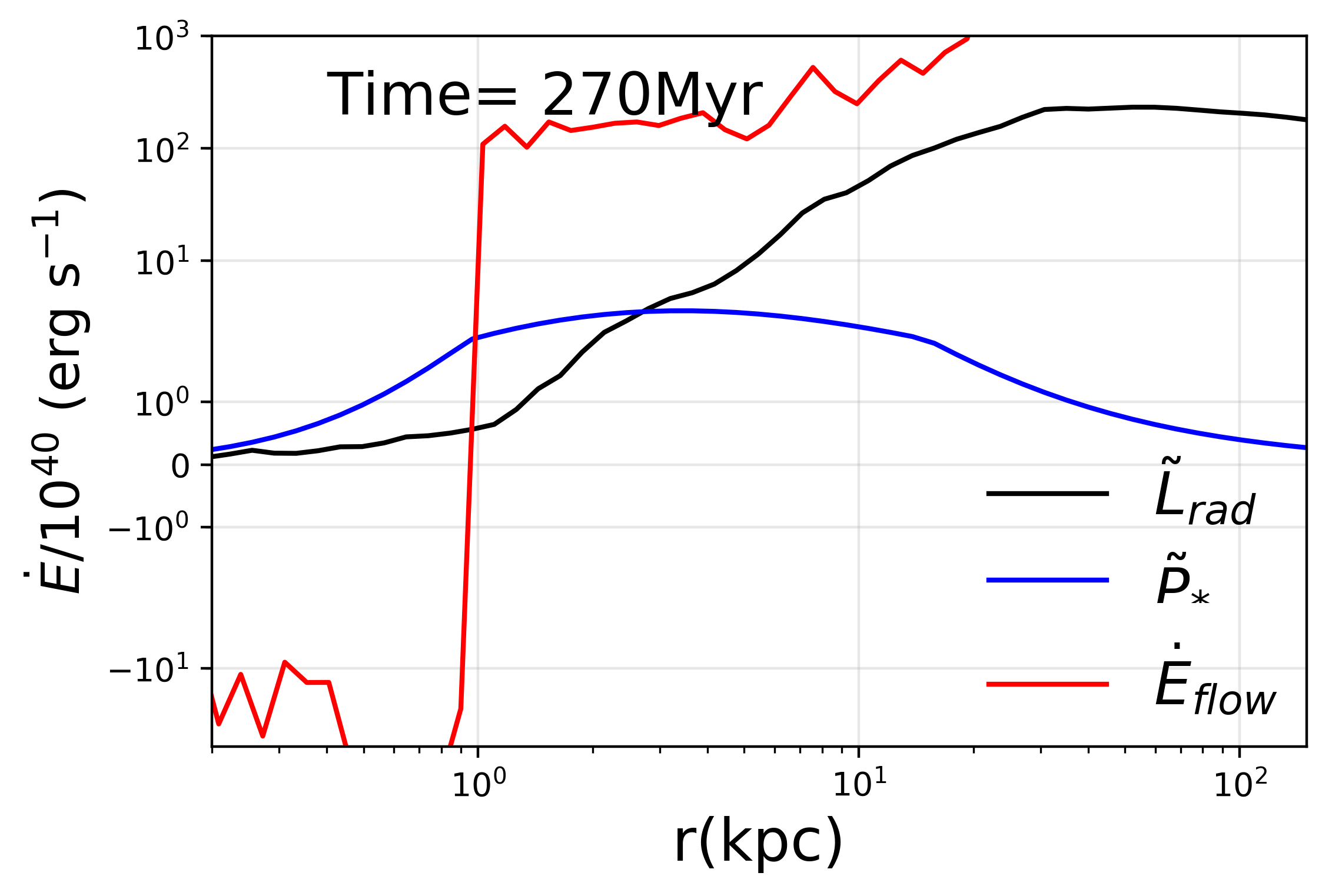}
\includegraphics[width=0.75\textwidth]{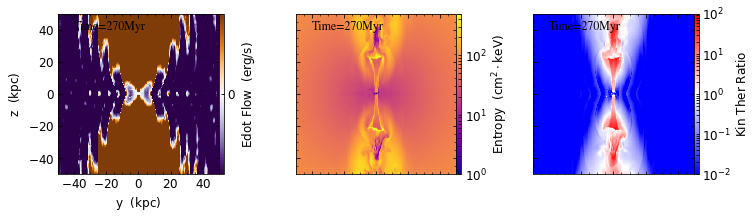}
 \caption{
 Evolution of the initial outburst in the fiducial MPG simulation, starting at $t = 10$ Myr, shortly after the AGN outburst begins and proceeding through the time when the initial shock front is propagating through the galaxy ($t = 30$~Myr) to the time when stellar heating starts to exceed radiative cooling within the galaxy ($t = 270$~Myr). All figure elements are identical to those in Figure~\ref{fig:init_spg}. 
 }
 \label{fig:init_mpg}
\end{figure*}
 
Figure~\ref{fig:init_mpg} shows the evolution of the multi-phase galaxy during $t\lesssim 270$ Myr. Its panels show the same quantities as in Figure~\ref{fig:init_spg}. Unlike the fiducial SPG run, AGN feedback in the MPG run begins at $t < 10$ Myr. That is because the cooling time within $r<0.5$ kpc is $t_{\rm cool} \lesssim 10$ Myr. The initial jet outburst results in a non-gravitational energy flow of $\sim 10^{44}$ erg s$^{-1}$ during the first AGN outburst. 

The figure shows the evolution of the multiphase galaxy at the following stages: 
\begin{itemize}

    \item \textbf{10 Myr:} 
    Cold gas accumulation near the central black hole
    has already initiated the first AGN outburst. The jet has propagated through the central $\sim3$~kpc into the initial cooling flow, which extends to $r \approx 10$~kpc at this moment, with a non-gravitational energy flow exceeding $10^{43}$ erg s$^{-1}$. The jet outburst faces resistance from the cooling CGM with much of the gas at $< 10$~kpc still having $K < 10 \, {\rm keV \, cm^2}$.
    
     \item \textbf{30 Myr:} Formation of cold gas clouds within $r < 10$~kpc continues to power a strong AGN outburst. (Note $t_{\rm cool} \lesssim 30$ Myr at $r<10$ kpc.) The initial feedback impulse has propagated to $\sim 20$~kpc.  Gas propelled along the jet axis by the AGN is strongly shocked as it encounters resistance from the CGM, while the shock propagating into the CGM is relatively weak along the jet axis and has become subsonic in the equatorial regions. Consequently, much of the gas at $< 10$~kpc but away from the jet axis still has $K < 10 \, {\rm keV \, cm^2}$. 
     
     \item \textbf{270 Myr:} The powerful AGN jet ($P_{\rm jet} > 10^{44}$ erg s$^{-1}$) has propagated to $r>200$ kpc with jet plasma kinetically dominated along the jet axis. 
     The AGN jets have also entrained significant amounts of low entropy gas, as is seen from the entropy plot. 
     The AGN activity by this stage has allowed $\tilde{P}_{\rm SNIa}$ to become comparable to the $\tilde{L}_{\rm rad}$ in the central $r<5$ kpc. 
     Stellar heating can therefore start sweeping atmospheric gas out of the galaxy and into the CGM.
\end{itemize}

\subsection{Self-Regulation}
\label{sec:regulation}

In both of the fiducial simulations, stellar heating plays an important role in self-regulation of AGN feedback, even though the power supplied by the AGN ($P_{\rm jet}$) is considerably greater than the power supplied by stellar heating ($P_{\rm SNIa}$). This section looks in more detail at the evolution of the critical interplay between radiative cooling and the two energy sources---stellar heating and AGN feedback.

\subsubsection{Fiducial SPG}
\label{sec:spg_evol}
\begin{figure*}[!t]
 \includegraphics[width=2.3in,height=1.6in]{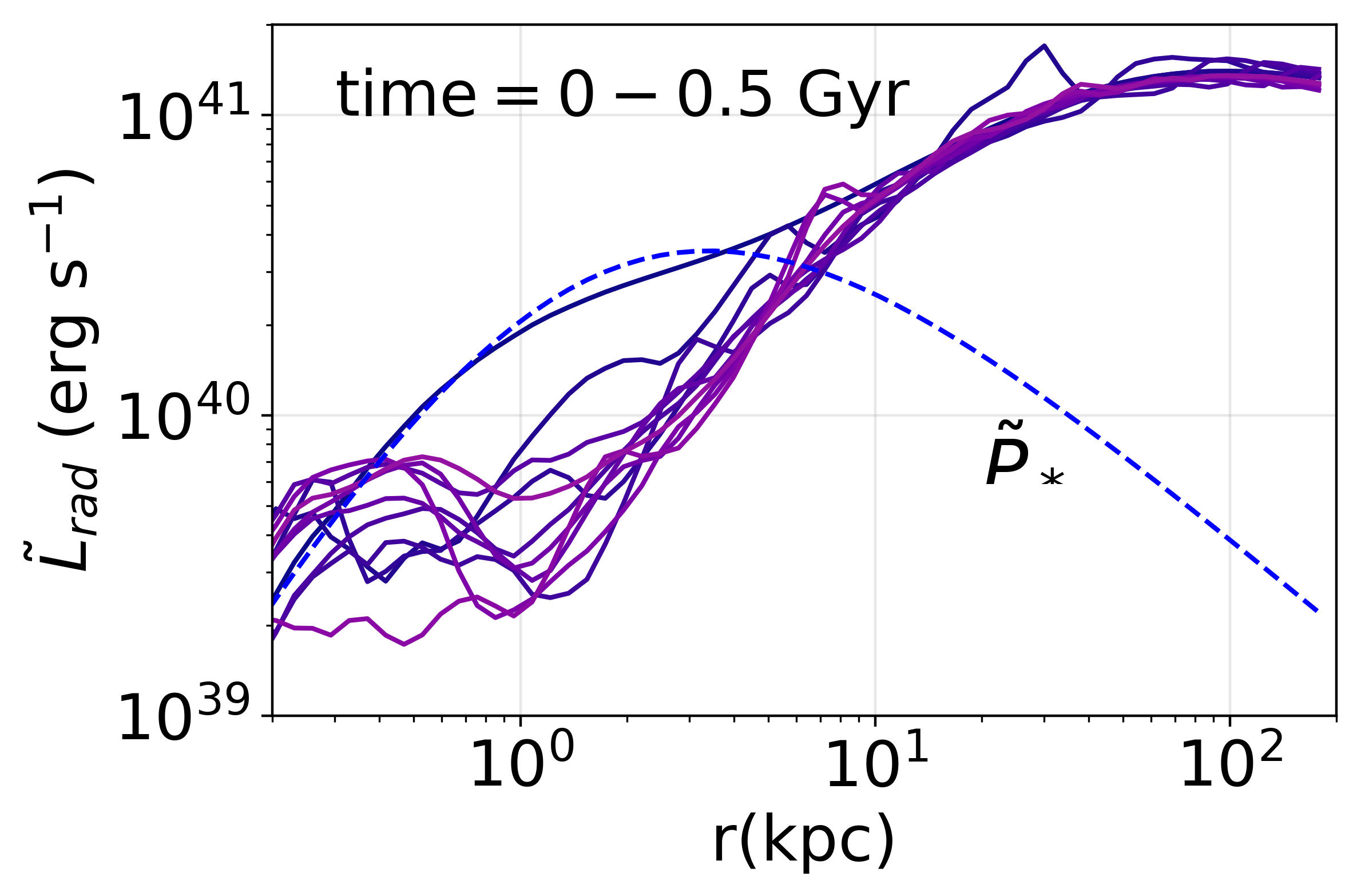}
 \includegraphics[width=2.0in,height=1.6in]{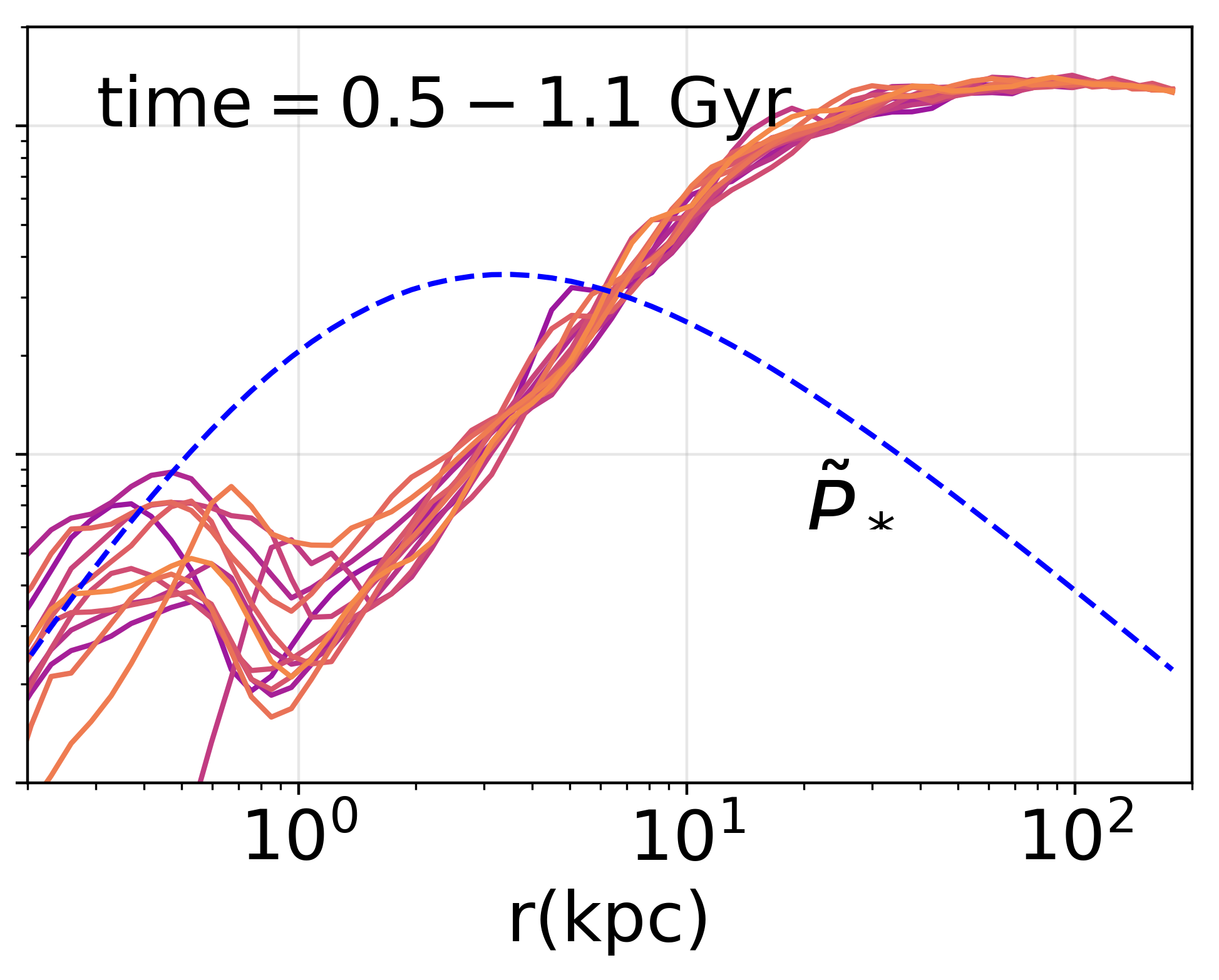}
 \includegraphics[width=2.5in,height=1.6in]{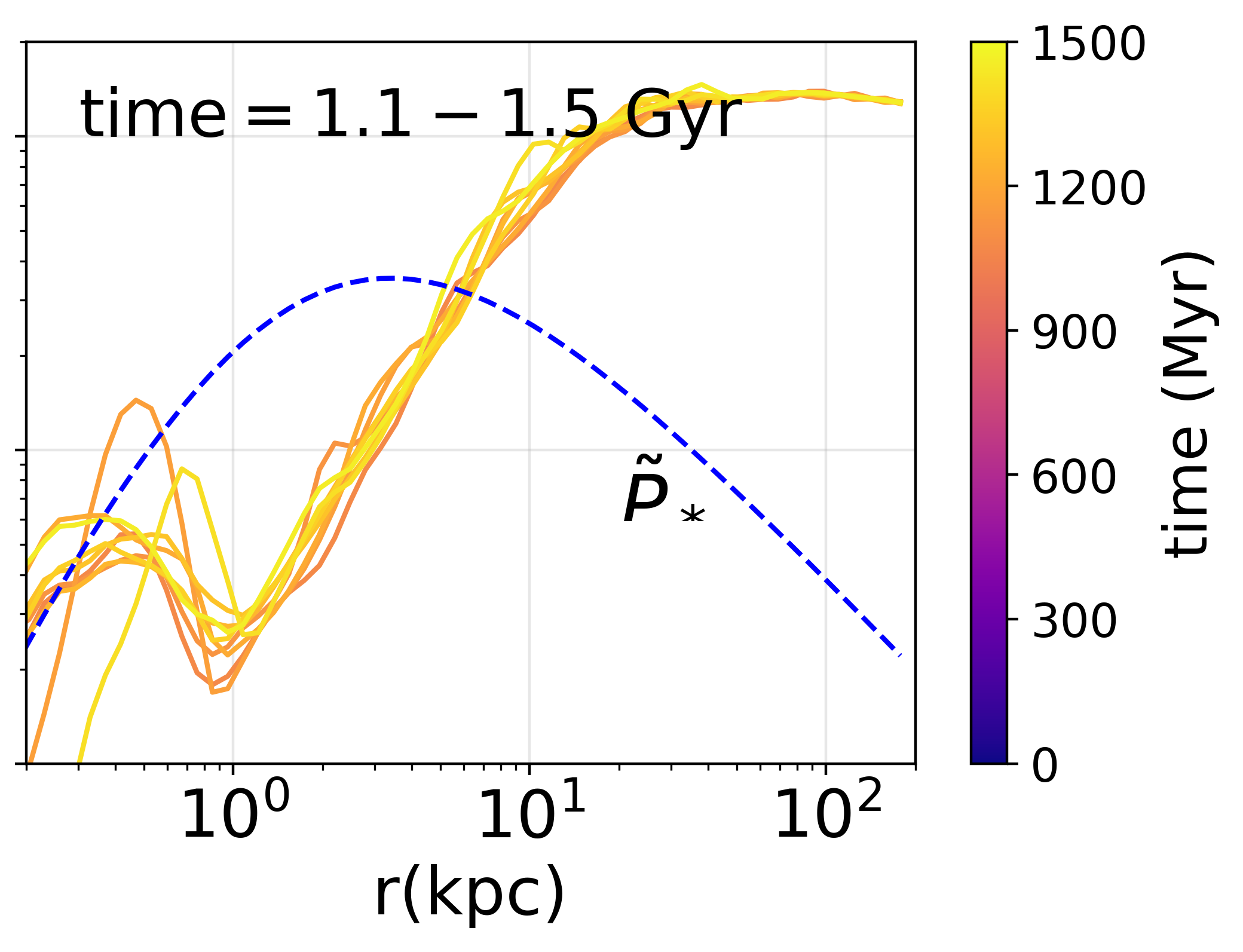}
 \includegraphics[width=6.7in,height=3.4in]{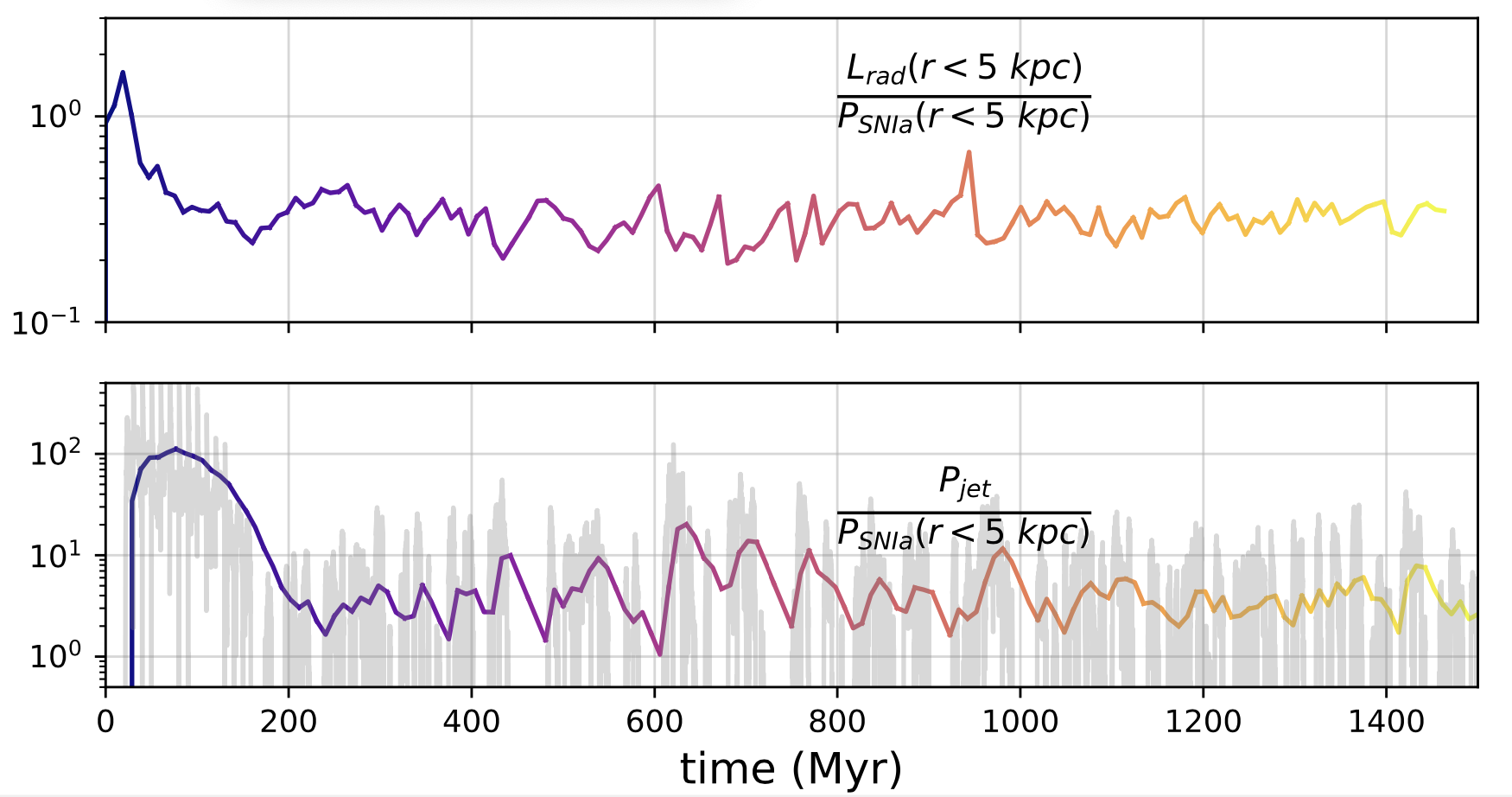}
 \caption{
 Regulation of AGN power in the fiducial SPG simulation.  The top row of panels shows how the varying radial distribution of radiative losses ($\tilde{L}_{\rm rad}$, colored solid lines) compares with the constant radial distribution of stellar power input ($\tilde{P}_*$, blue dashed lines). 
 For clarity, the 1.5~Gyr time period of the simulation is broken into three sub-periods as labeled, with lines showing the radial distribution of $\tilde{L}_{\rm rad}$ at 50~Myr intervals.  The lower two panels show how both radiative losses from the central 5~kpc and jet power compare with stellar power input within 5~kpc (P$_{\rm snia}\sim6\times10^{40} $ erg s$^{-1}$) during the time period of the simulation.  A strong initial outburst of AGN power lowers the atmospheric density at $r < 10$~kpc during the first 50~Myr, until SNIa power input exceeds radiative losses from $\sim$0.5--5~kpc.  Thereafter, the system remains in a nearly steady state, with a slow rise in time-averaged radiative losses and time-averaged jet power exceeding stellar power by a factor of several. 
 }  
 \label{fig:lum_spg}
\end{figure*}

Figure \ref{fig:lum_spg} shows how AGN self-regulation proceeds in the fiducial SPG simulation. Three panels across the top of the figure illustrate how the radial distribution of local radiative losses ($\tilde{L}_{\rm rad}$) compares with the radial distribution of local stellar power input ($\tilde{P}_{\rm rad}$). By $t = 50$~Myr, AGN feedback has reduced the gas density at $< 10$~kpc enough for stellar power to exceed radiative cooling at all radii from $\sim 0.5$--5~kpc. Thereafter, the atmosphere remains in a quasi-steady configuration, maintained by a combination of stellar heating and AGN feedback.

The two bottom panels show how both jet power and radiative losses from within 5~kpc compare with SNIa power from within 5~kpc (which is $\approx 6\times10^{40} \, {\rm erg \, s^{-1}}$).  Radiative losses from that inner region initially exceed stellar power by a factor $\sim 2$ but drop below stellar power and remain there after the first 50~Myr.  Time-averaged jet power, on the other hand, exceeds stellar power by nearly an order of magnitude during the entire simulation period. It therefore exceeds radiative losses from $r < 5$~kpc by at least an order of magnitude after the first 100~Myr.  Yet, the AGN power is not substantially altering the structure of the galaxy's atmosphere in that region.  

\subsubsection{Fiducial MPG}

\begin{figure*}[!t]
 \includegraphics[width=2.3in,height=1.6in]{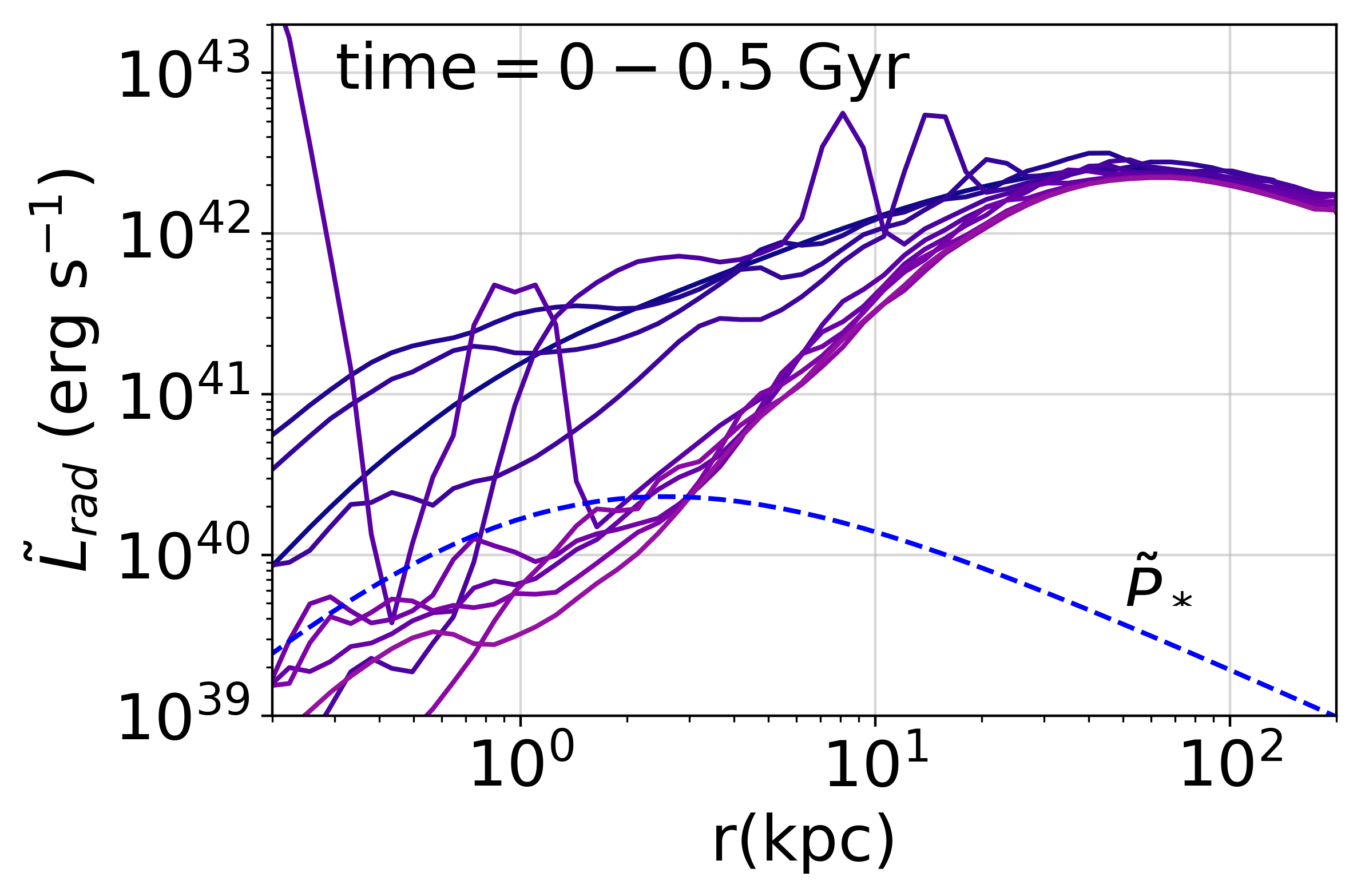}
 \includegraphics[width=2.0in,height=1.6in]{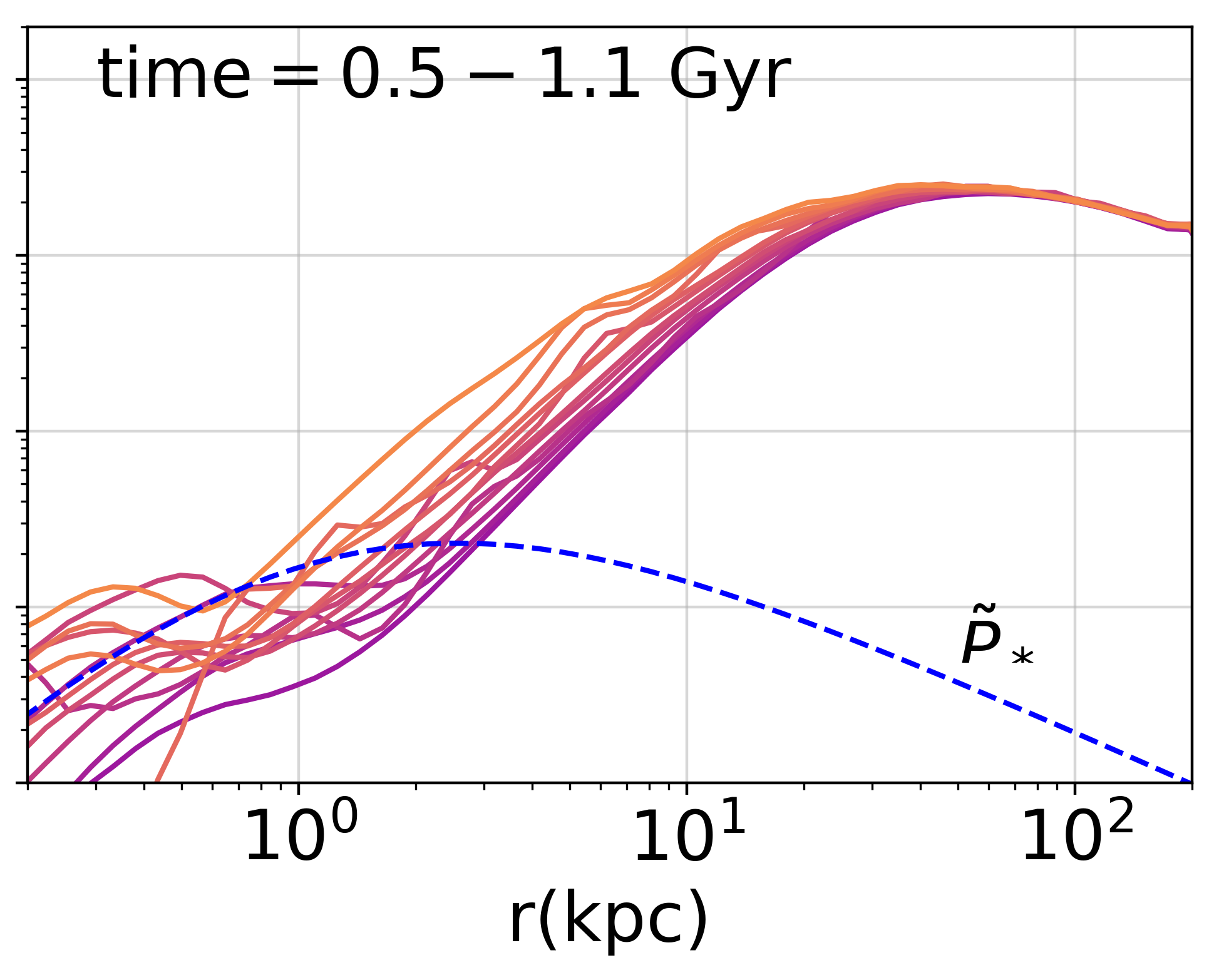}
 \includegraphics[width=2.5in,height=1.6in]{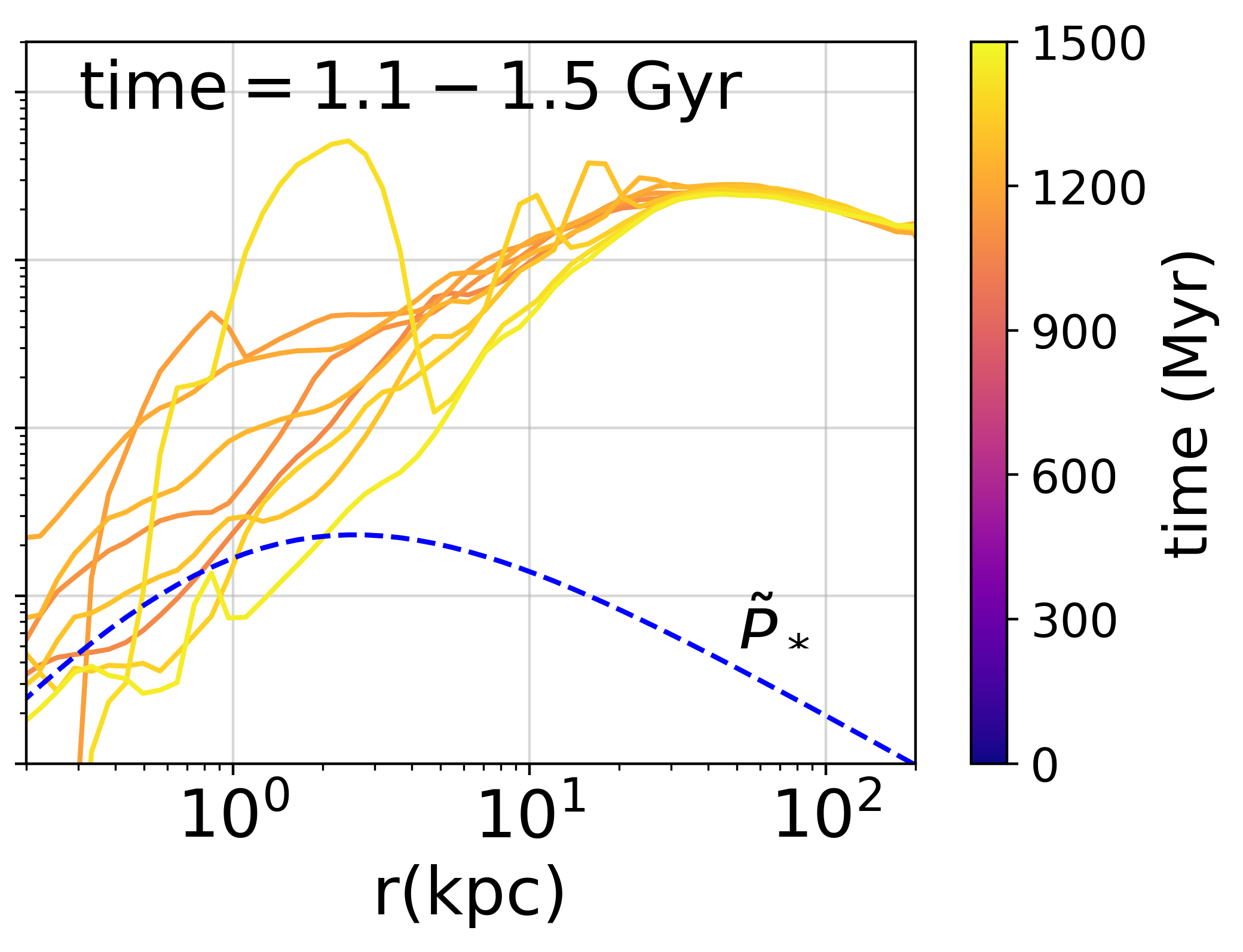}
 \includegraphics[width=6.8in,height=3.4in]{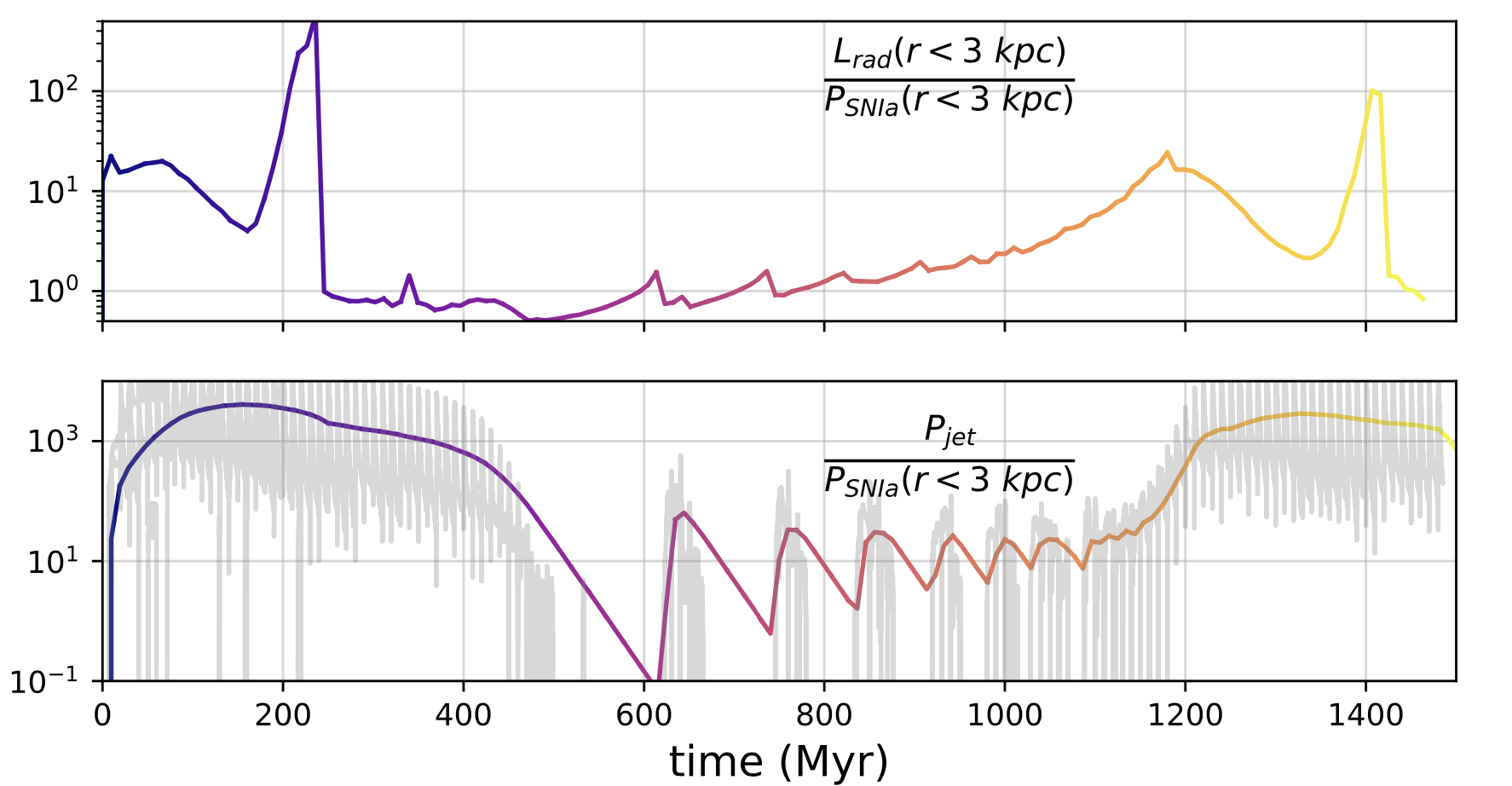}
 \caption{
 Regulation of AGN power in the fiducial MPG simulation.  The top row of panels shows how the varying radial distribution of radiative losses ($\tilde{L}_{\rm rad}$, colored lines) compares with the constant radial distribution of stellar power input ($\tilde{P}_*$, blue dashed line). 
 For clarity, the 1.5~Gyr time period of the simulation is broken into three sub-periods as labeled, with lines showing the radial distribution of $\tilde{L}_{\rm rad}$ at 50~Myr intervals. The lower two panels show how both radiative losses ($L_{\rm rad}$) from the central 3~kpc and jet power ($P_{\rm jet}$) compare with stellar power input within 
 3~kpc {($P_{\rm SNIa}\sim 4.34\times 10^{40}$ erg s$^{-1}$)} during the time period of the simulation.  A strong initial outburst of AGN power lowers the atmospheric density at $r < 15$~kpc during the first 250~Myr, until SNIa power input exceeds radiative losses from $\sim$0.3--3~kpc.  During the middle sub-period ($0.5 - 1.1$~Gyr), radiative losses slowly rise as atmospheric density within 10~kpc slowly increases. Shortly after 1.1~Gyr, radiative losses once again exceed stellar power input, resulting in another powerful AGN outburst that again pushes radiative losses below stellar power input at $r \lesssim 3$~kpc by $t = 1.5$~Gyr.  
 }  
 \label{fig:lum_mpg}
\end{figure*}

Figure \ref{fig:lum_mpg} shows how AGN self-regulation proceeds in the fiducial MPG simulation.  As in Figure \ref{fig:lum_spg}, three panels across the top of the figure illustrate how local radiative losses compare with local stellar power, but the story they tell is different.  It takes 270~Myr for strong AGN feedback ($P_{\rm jet} \gtrsim 10^{44} \, {\rm erg \, s^{-1}}$) to lower the inner gas density enough for stellar power to exceed radiative cooling at $\sim 0.3$--3~kpc.  Stellar power remains competitive with radiative cooling at small radii for $t = 0.25$--$1.1$~Gyr.  During that time, AGN power output is relatively low and fails to compensate for radiative losses at larger radii. Circumgalactic pressure rises, as does the gas density at $< 10$~kpc.  Eventually, radiative losses exceed stellar power at all radii, ushering in a second period of high-power AGN feedback at $t \approx 1.2$~Gyr. By $t = 1.5$~Gyr, this second powerful outburst ($> 10^{44} \, {\rm erg \, s^{-1}}$) has again reduced the inner atmosphere's density enough for stellar feedback to exceed radiative losses at $< 3$~kpc. 

The $L_{\rm rad}/P_{\rm SNIa}$ ratio plot shows a sharp spike in X-ray luminosity following each major jet event. Much of the X-ray power in those spikes comes from dense shock regions at 1--3~kpc in which the outflows impact infalling low-entropy gas.

\subsection{Atmospheric Reconfiguration}
\label{sec:config}

In all of our simulations, most of the AGN's energy output ultimately goes into \textit{lifting} of circumgalactic gas rather than heating of atmospheric gas within the galaxy.  As a result, AGN jets reconfigure the circumgalactic medium (CGM) during our simulations. Large scale (tens of kpc) circulation of the CGM on $\sim 10$--100~kpc scales therefore plays a critical role in preventing overcooling of gas in these simulated galaxies. This section discusses how AGN feedback reconfigures the atmospheres of the simulated galaxies in our fiducial SPG and MPG runs.  

\subsubsection{Fiducial SPG}

\begin{figure*}[!t]
 \includegraphics[width=3.6in,height=2.4in]{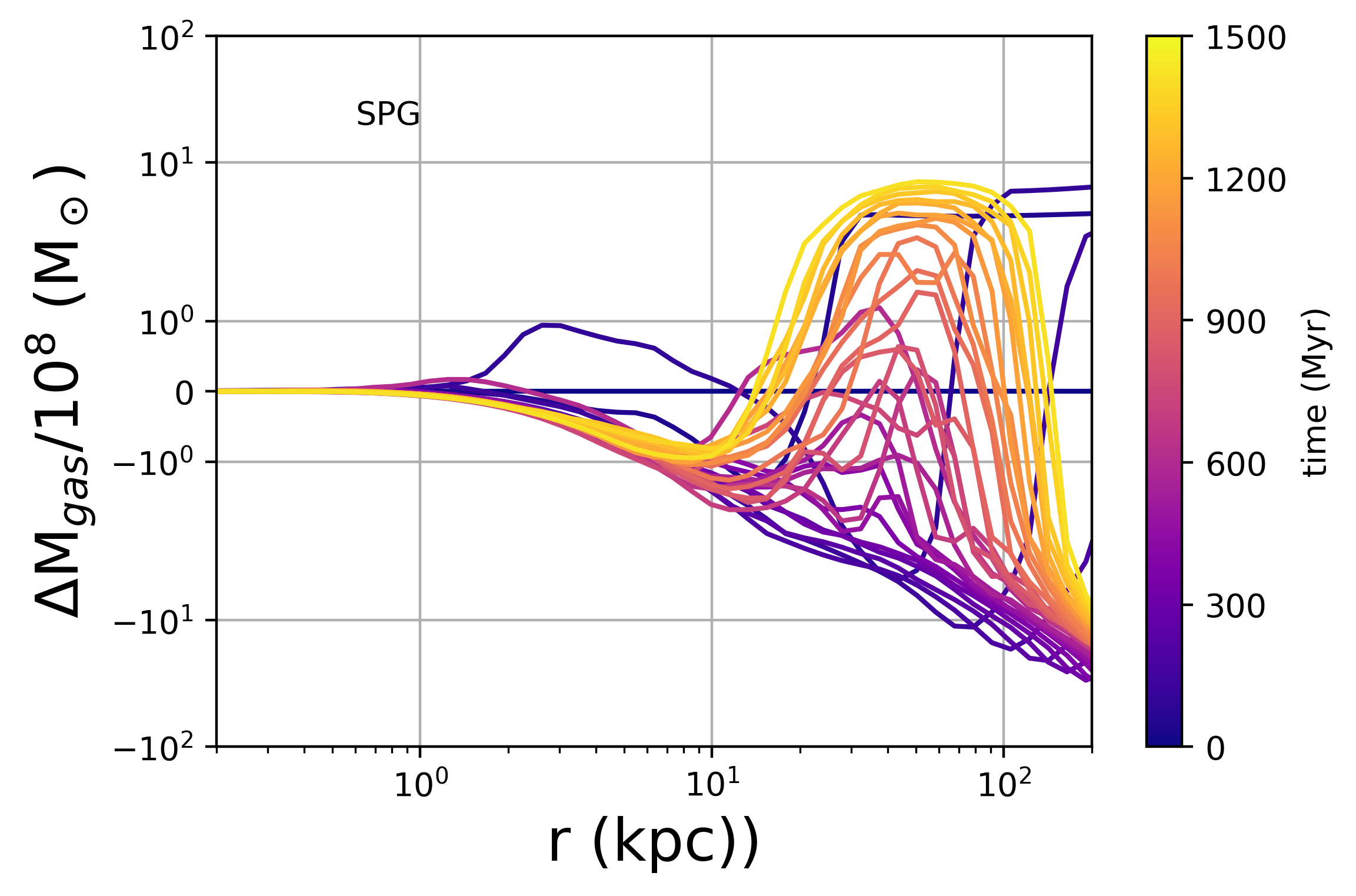}
 \includegraphics[width=3.6in,height=2.4in]{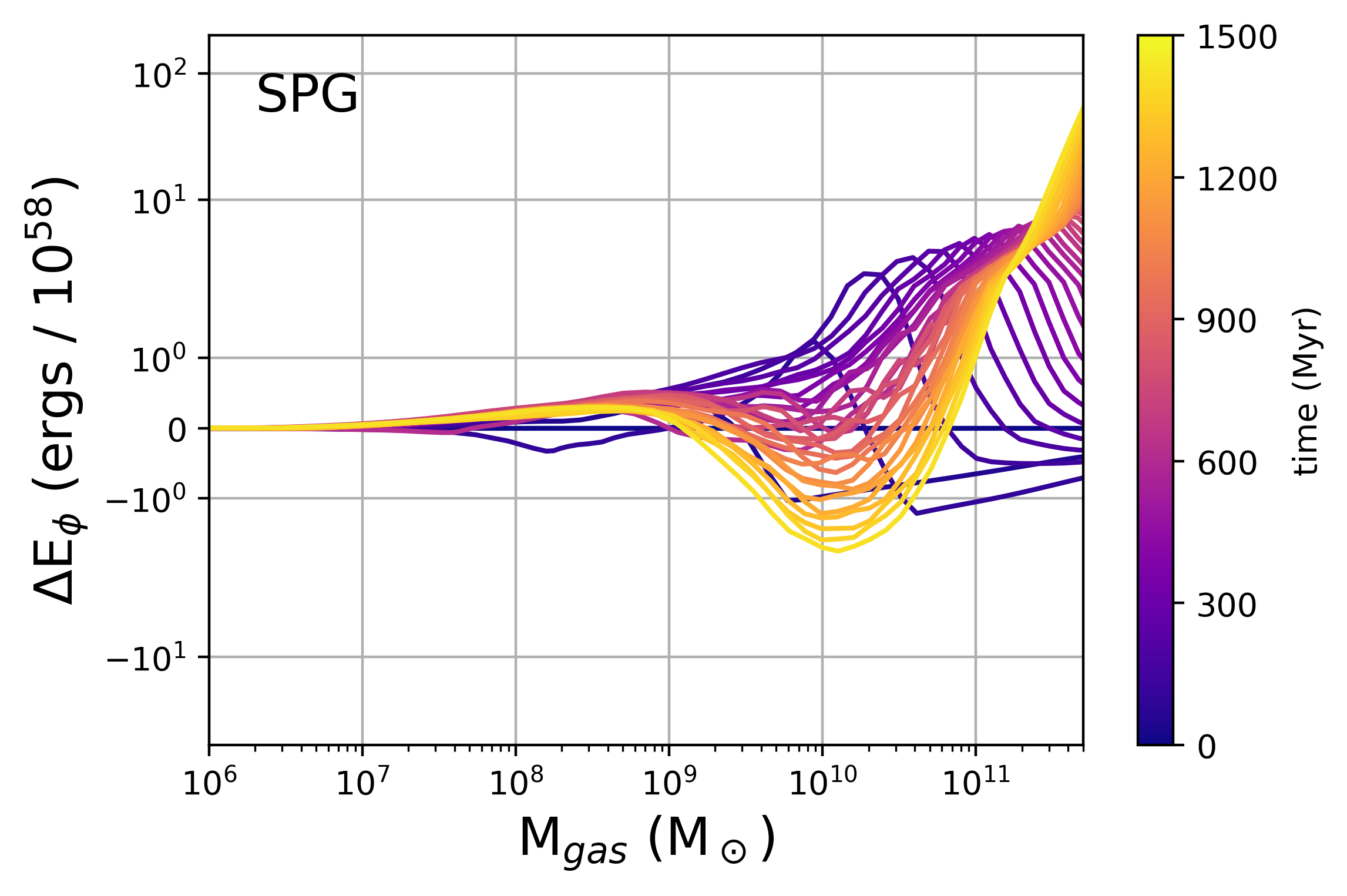}
 \caption{Atmospheric reconfiguration in the fiducial SPG simulation.  \textit{Left:} Changes in the gas mass ($\Delta M_{\rm gas}$) within radius $r$, relative to the initial configuration. \textit{Right:} Changes in atmospheric gravitational energy ($\Delta E_\phi$), relative to the initial configuration, as a function of the gas mass $M_{\rm gas}(r)$ enclosed within radius $r$.  Colored lines show intervals of 50~Myr.  The initial AGN outburst reduces the gas mass within 10~kpc by $\sim 10^8 \, M_\odot$ and reduces the gas mass within 100~kpc by $\sim 10^9 \, M_\odot$.  Most of the AGN's energy output ($\sim 10^{59} \, {\rm ergs}$) therefore goes into lifting of circumgalactic gas out to distances exceeding 100~kpc.  Meanwhile, during the AGN's low-power state ($\sim 300$--1500~Myr), gas ejected by the galaxy's stars flows out of the galaxy and accumulates at $\sim 30$--100~kpc, causing a steady rise in CGM pressure at smaller radii. 
}
 \label{fig:dmde_SPG}
\end{figure*}

Figure \ref{fig:dmde_SPG} shows atmospheric reconfiguration
in the fiducial SPG simulation.  By $t = 100$~Myr, the initial outburst has lowered the gas mass within $10$~kpc by $\sim 10^8 \, M_\odot$, relative to the initial state.  It has also lowered the gas mass within $ 100$~kpc by $\sim 10^9 \, M_\odot$.  This reconfiguration has increased the gravitational potential energy of the atmosphere.  The potential energy of the innermost $10^9 \, M_\odot$, corresponding roughly to $r < 10$~kpc, has increased by less than $10^{58} \, {\rm erg}$.  However, the total potential energy increase at larger radii is an order of magnitude greater.  After the first 300~Myr, the innermost $10^{11} M_\odot$ has gained $\sim 5 \times 10^{58} \, {\rm erg}$ of potential energy.  That amount of energy is similar to the integrated feedback power during that same period of time and is far larger than the integrated radiative losses from the atmosphere (several times $10^{57} \, {\rm erg}$). 
This finding implies that only a small fraction of the AGN power actually couples to the central region of the galactic atmosphere.  A larger fraction propagates out to several tens of kpc, leading to large scale cicrulation resulting in CGM reconfiguration. The low entropy gas initially at small radii essentially gets lifted to larger radii while high entropy gas rushes in to the center.

\subsubsection{Fiducial MPG}

\begin{figure*}[!t]
 \includegraphics[width=3.6in,height=2.4in]{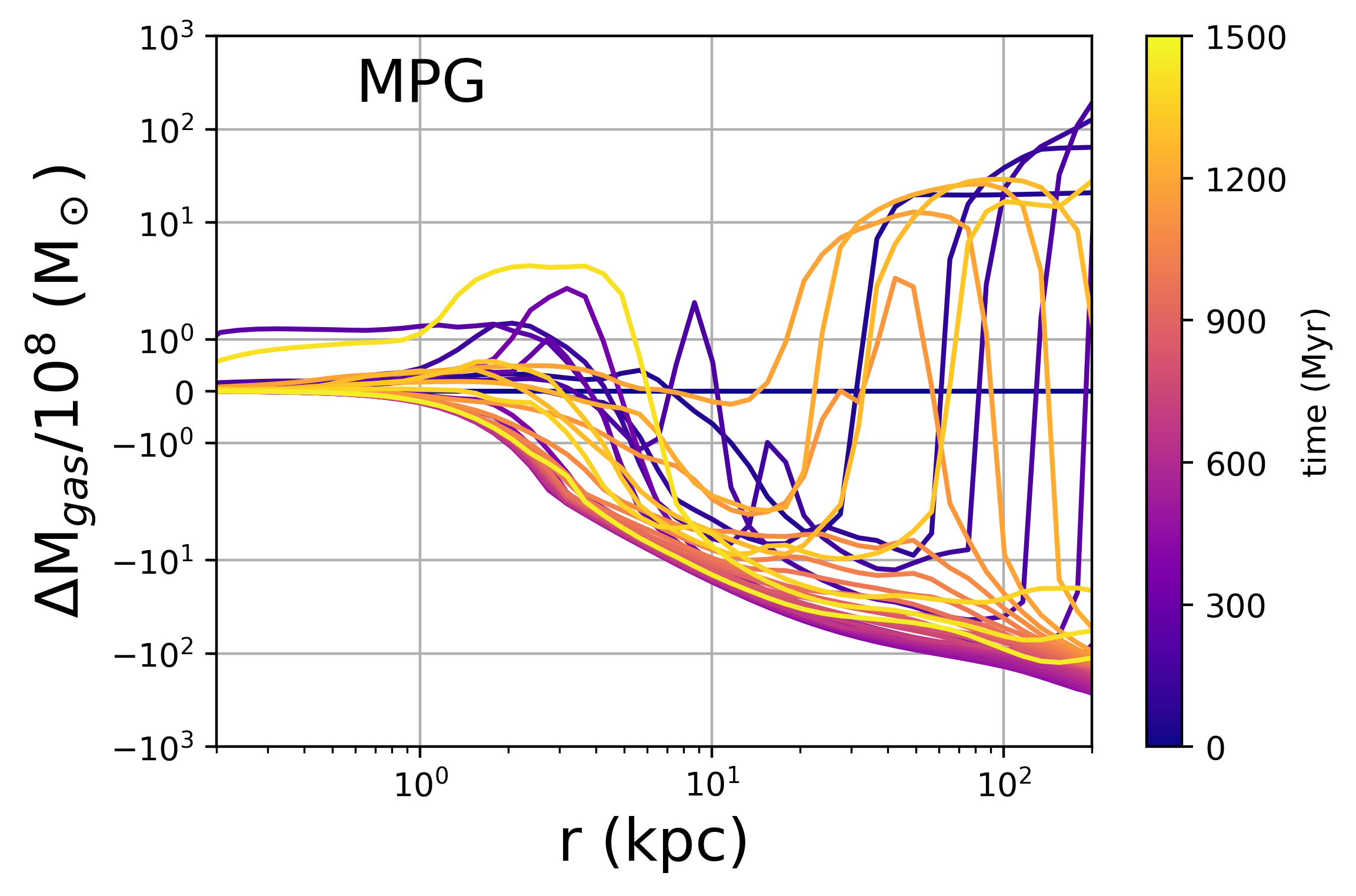}
 \includegraphics[width=3.6in,height=2.4in]{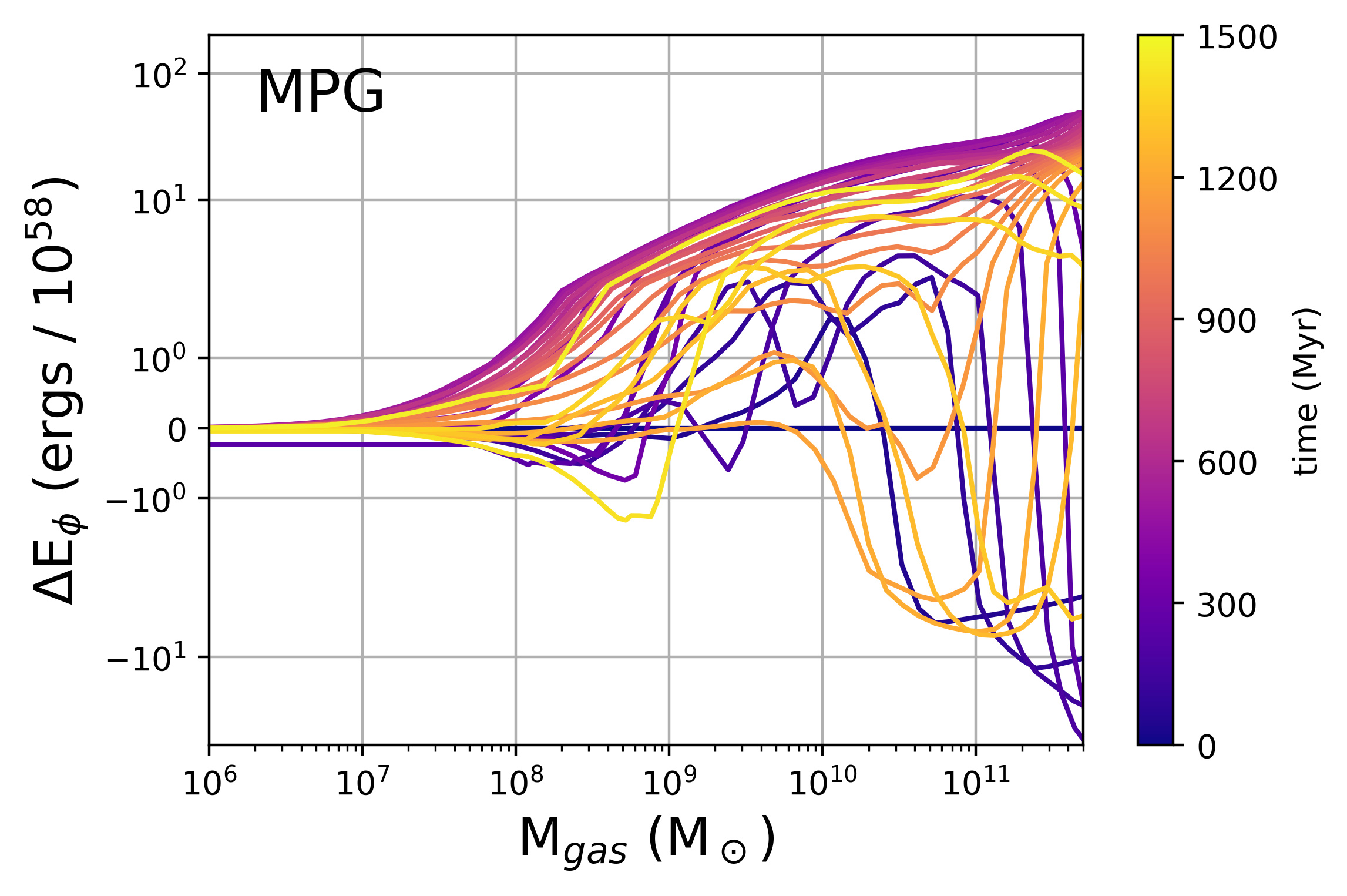}
 \caption{Atmospheric reconfiguration in the fiducial MPG simulation.  \textit{Left:} Changes in the gas mass ($\Delta M_{\rm gas}$) within radius $r$, relative to the initial configuration. \textit{Right:} Changes in atmospheric gravitational energy ($\Delta E_\phi$), relative to the initial configuration, as a function of the gas mass $M_{\rm gas}(r)$ enclosed within radius $r$.  Colored lines show intervals of 50~Myr.  The initial AGN outburst reduces the gas mass within 10~kpc by $\sim 10^9 \, M_\odot$ and reduces the gas mass within 100~kpc by $\sim 10^{10} \, M_\odot$.  Most of the AGN's energy output (several times $10^{59} \, {\rm ergs}$) therefore goes into lifting circumgalactic gas out to distances exceeding 100~kpc.  Meanwhile, gas ejected by the galaxy's stars during the AGN's low-power state ($\sim 600$--1100~Myr) flows out of the galaxy and accumulates at $\sim 30$--100~kpc, causing a steady rise in circumgalactic pressure. That rise in pressure confinement of the galaxy's central atmosphere initiates a second powerful AGN outburst when radiative losses again exceed stellar heating within the central few kiloparsecs (see Figure \ref{fig:dmde_MPG}). 
 }
 \label{fig:dmde_MPG}
\end{figure*}

Figure \ref{fig:dmde_MPG} shows how AGN feedback reconfigures the galactic atmosphere in the MPG simulation.  The initial outburst lowers the gas mass at $< 10$~kpc by 
$\sim 10^{9} \, M_\odot$, relative to the initial state. It also lowers the gas mass at $< 100$~kpc by $\sim 10^{10} \, M_\odot$.
The potential energy of the innermost $10^9 \, M_\odot$ has increased by several times $10^{58} \, {\rm erg}$, and the total potential energy increase at larger radii is several times $10^{59} \, {\rm erg}$ as the CGM has been pushed outwards by powerful AGN activity during the first $t\lesssim 0.5$ Gyr. Subsequent small jet outbursts help to maintain this state until $t\sim1$ Gyr. \\

However, this reconfiguration of the MPG atmosphere is unsustainable, because radiative losses exceed the sum of both stellar power and AGN feedback power.  At $t \approx 1.1$--1.2~Gyr, the atmosphere at $r < 20$~kpc approaches the initial cooling-dominated configuration, and uncompensated cooling then fuels another episode of strong AGN feedback.  That outburst increases the atmosphere's gravitational potential energy by more than $10^{59} \, {\rm erg}$ over the course of a couple hundred Myr, because it greatly exceeds radiative losses. Like the first strong outburst, the second one again lowers the mass within $r<100$ kpc by $\sim 10^{10}$ M$_\odot$ relative to the initial state and restarts the low-power jet activity phase.   \\

\subsection{Accumulation of Stellar Ejecta}
\label{sec:accumulation}

During the low-power periods of AGN feedback in both fiducial simulations, gas ejected by the central galaxy's stars is continually swept out of the galaxy by stellar heat input.  But stellar heating cannot push the gas very far out of the galaxy. The ejected gas slowly builds up in the CGM, resulting in higher CGM density and pressure. 

\subsubsection{Fiducial SPG}
Figure \ref{fig:dmde_SPG} shows that ejected stellar gas in the fiducial SPG simulation accumulates in the CGM at $\sim 30$--100~kpc from the central galaxy.  This slow rise in CGM gas mass causes a slow rise in atmospheric pressure and density rise at smaller radii, reflected by the slow increase in radiative losses depicted in the $L_{\rm rad} / P_{\rm SNIa}$ panel of Figure \ref{fig:lum_spg}. Given that the buildup of the ejected matter is at $r\sim20-100$ kpc, where $t_{\rm cool}\gtrsim 2$ Gyr, it does not have time to become extended cold gas in our simulation. Only central ($r<1$ kpc) cooling can fuel AGN feedback while SNIa feedback sweeps the matter ejected by stars at larger radii into the CGM. Therefore, there is essentially no star formation in the fiducial SPG simulation.

\subsubsection{Fiducial MPG}
Accumulation of ejected stellar gas has swifter consequences in the fiducial MPG simulation. Stellar heating sweeps the ejected gas out of the galaxy into the CGM ($r\sim 20-10$ kpc) 
during the AGN's low-power phase ($t\sim 0.5-1.1$ Gyr). As the gas density there rises, radiative losses eventually exceed the stellar and AGN heating, leading to cooling in the region at $r\sim3-10$ kpc. This gas then flows into the center, fuelling the next massive AGN outburst by $t\sim 1.1$ Gyr.  That large outburst then lowers the CGM density and pressure, again allowing stellar heating to become competitive with radiative cooling, causing AGN feedback to revert to a low-power state. 

\subsection{Origin of Excess Entropy}
\label{sec:ent}

\begin{figure}[b]
 \includegraphics[width=0.48\textwidth]{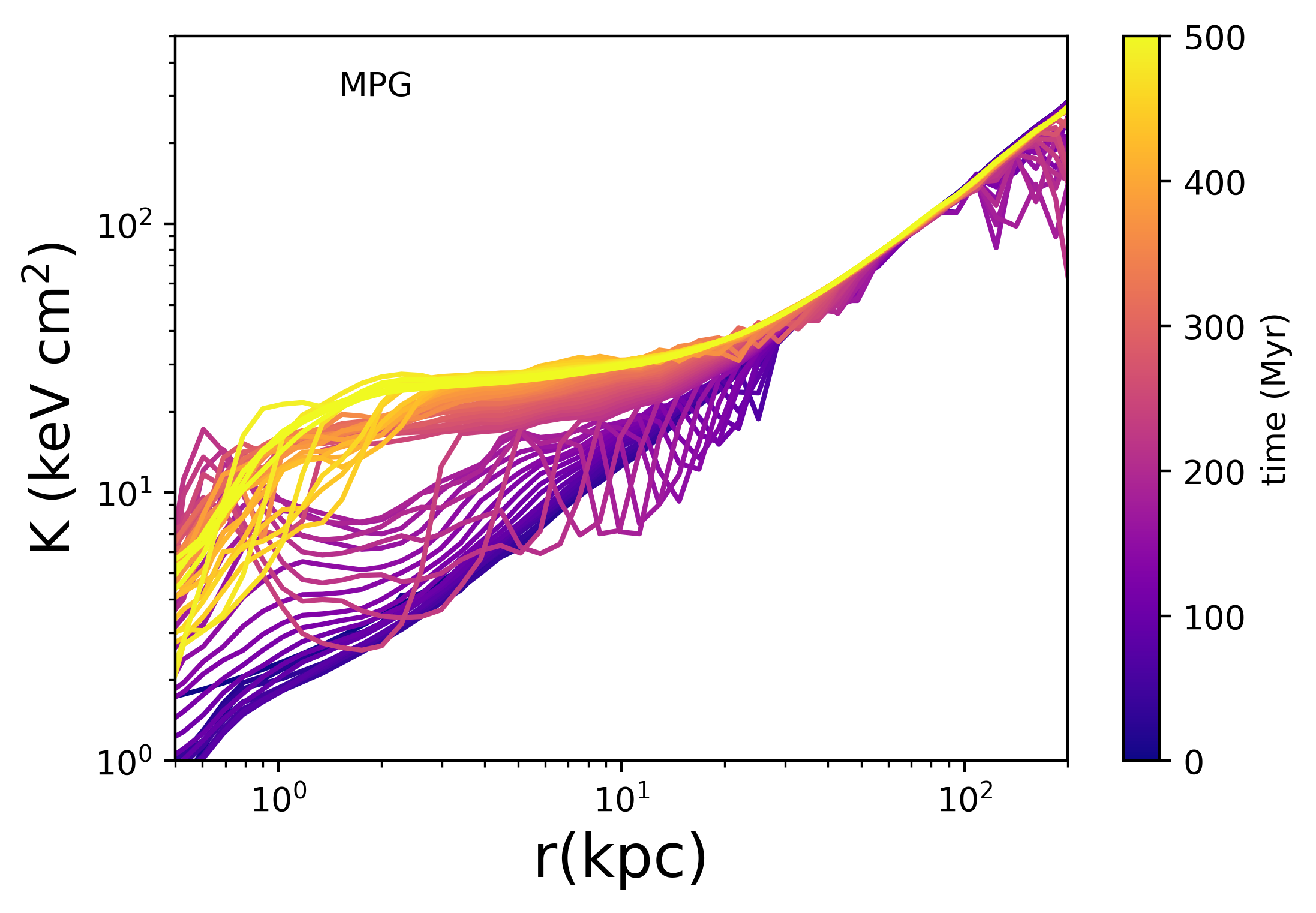}
 \caption{Evolution of the radial entropy profile during the first 0.5~Gyr of the fiducial MPG simulation.  Median entropy at $< 10$~kpc continually rises during that time period, resulting in an entropy plateau exceeding $20 \, {\rm keV \, cm^2}$ by the end of it. 
 }
 \label{fig:entropy_MPG}
\end{figure}

Now we return to the excess entropy issue spotlighted in \S \ref{sec:introduction} and examine it in light of the findings of \S \ref{sec:regulation} and \S \ref{sec:config}. Paper~I hypothesized that the entropy increases within 10~kpc, evident in Figures \ref{fig:entropy} and \ref{fig:entropy_MPG}, arise from excessive thermalization of jet energy in the central few kpc.  However, Figure \ref{fig:init_spg} shows that the initial AGN outburst, which is by far the strongest in the fiducial SPG simulation, does not significantly raise the entropy of much of the gas within the central 10~kpc.  Instead, the bipolar outflow drives a shock that heats atmospheric gas asymmetrically as it passes through that region.  By $t = 40$~Myr, the shock front has propagated beyond the central 10~kpc and much of the central gas remains at $\lesssim 10$~keV~cm$^2$.  Thereafter, the AGN outflow travels through the central 10~kpc without dissipating much of its energy in that region but the median entropy level at $< 10$~kpc continues to rise.  The entropy excess developing there in the fiducial SPG simulation must therefore have an origin other than direct AGN heating. Section \ref{sec:circ} will show that much of the excess arises from atmospheric circulation driven by the AGN. Here we assess the amount of circulation needed to produce the entropy excesses shown in Figures \ref{fig:entropy} and \ref{fig:entropy_MPG}.

\subsubsection{Fiducial SPG}

\begin{figure}[b]
 \includegraphics[width=0.48\textwidth]{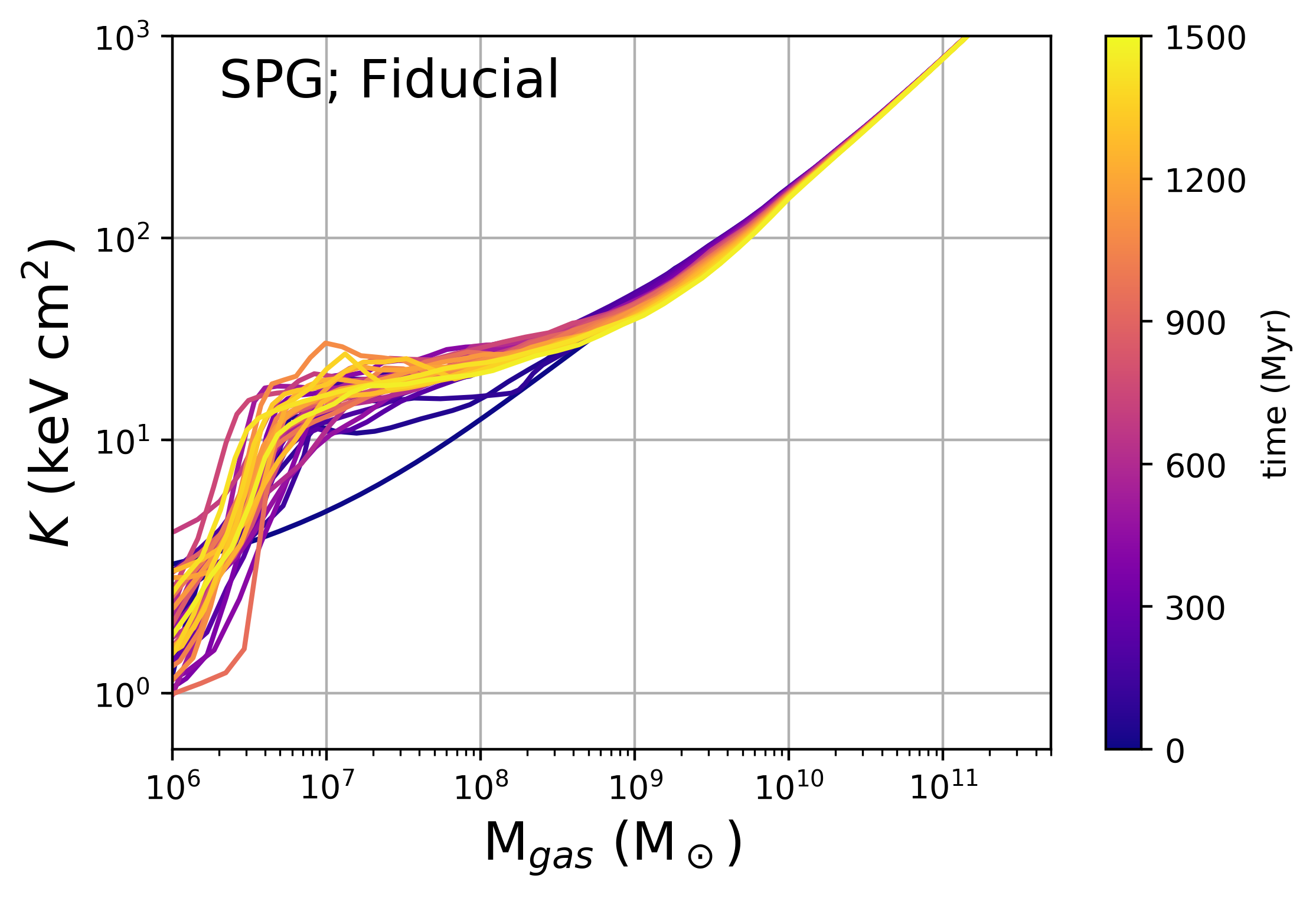}
 \includegraphics[width=0.48\textwidth]{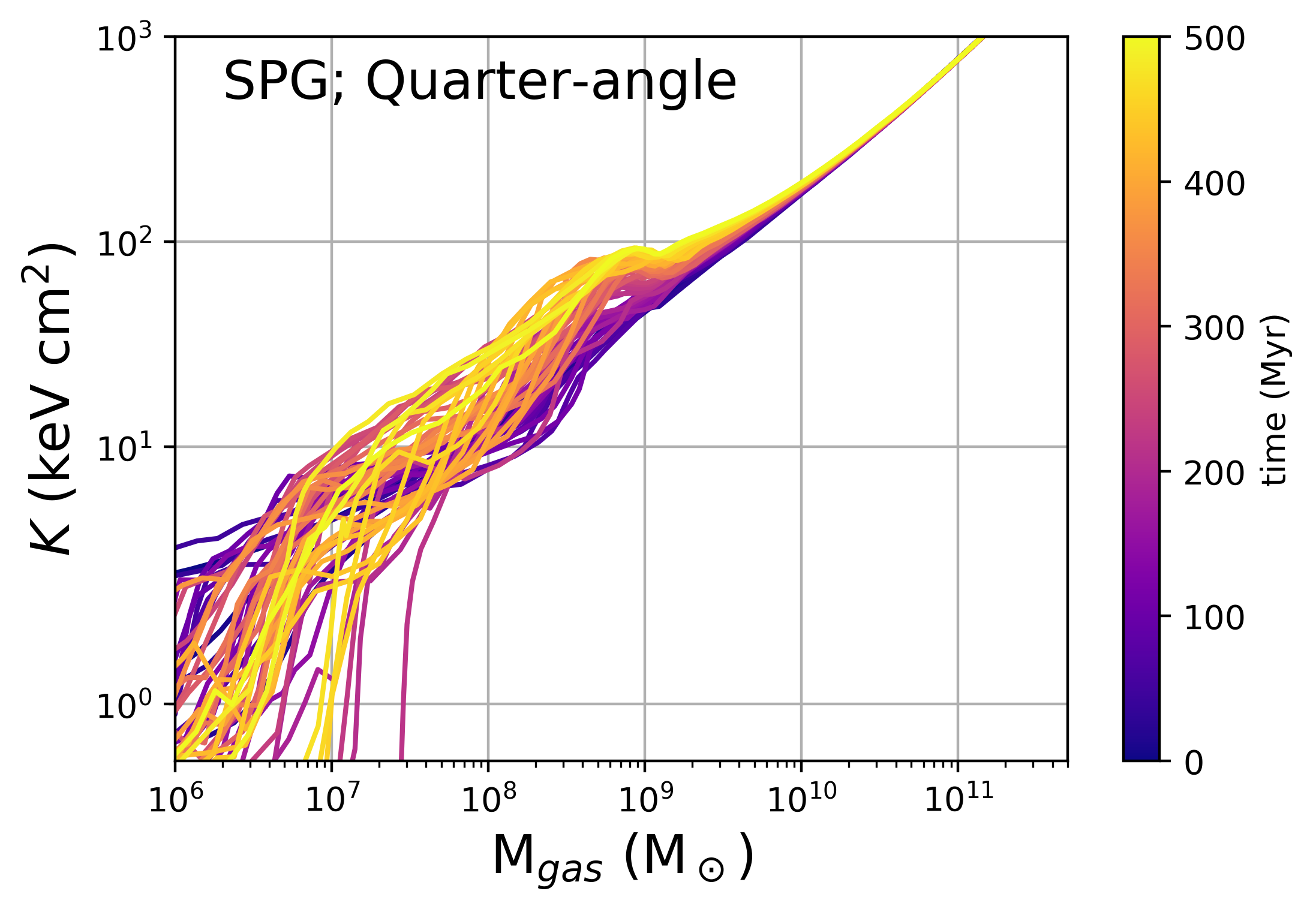}
 \includegraphics[width=0.48\textwidth]{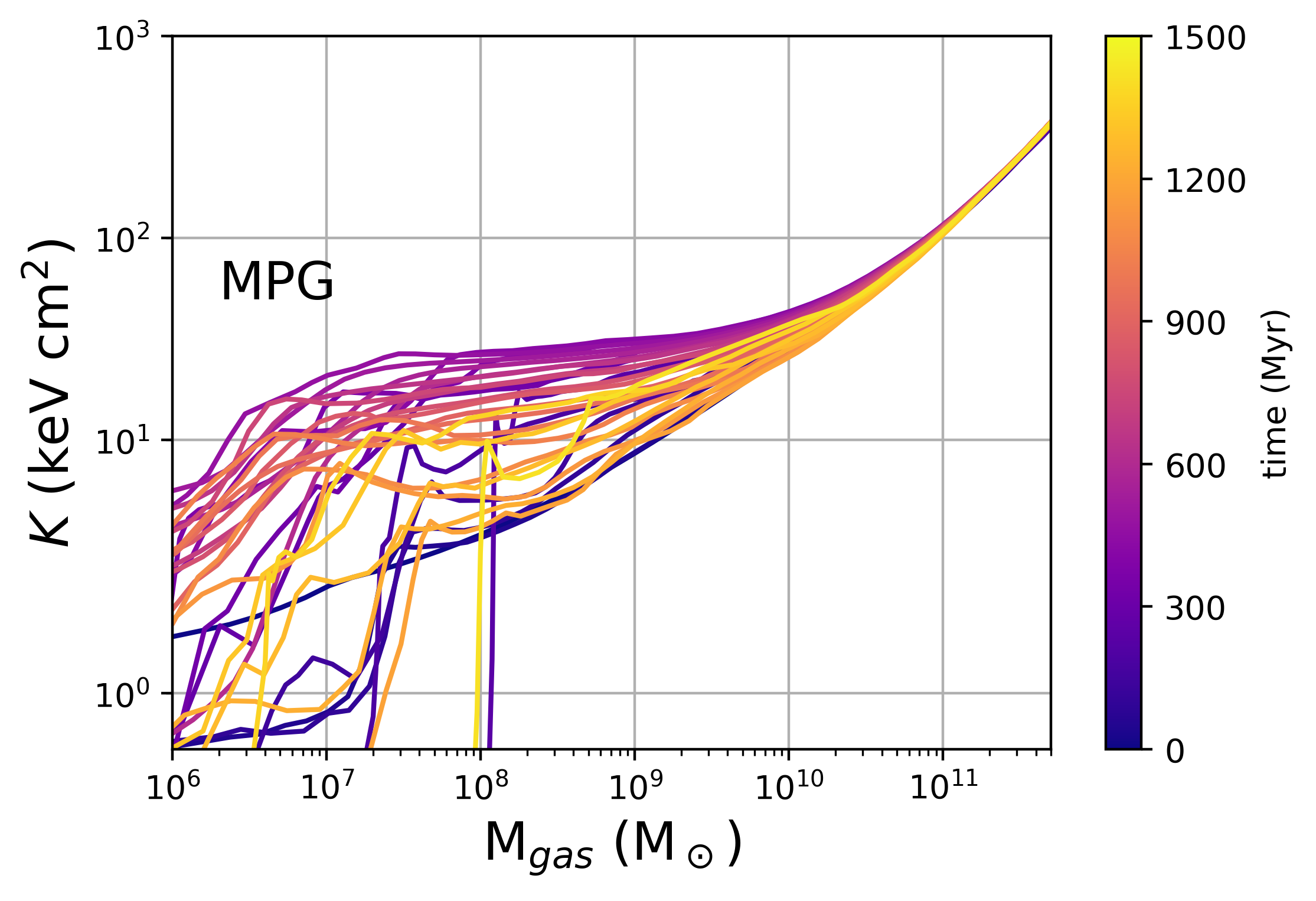}
\caption{
Distribution of entropy as a function of gas mass and its evolution with time in the fiducial SPG (top), quarter-angle SPG, and fiducial MPG simulations.
 }
 \label{fig:ent_mgas}
\end{figure}

In the fiducial SPG simulation, the entropy level at 1--3~kpc gradually climbs to $\sim 15 \, {\rm keV \, cm^2}$ during the first $\sim 200$~Myr  (see Figure \ref{fig:entropy}).  Gas that begins the simulation at $r \approx 6$~kpc has a similar
entropy level, and Figure \ref{fig:ent_mgas} shows that $\sim 2 \times 10^8 \, M_\odot$ of gas initially has lower entropy.  Therefore, removing that lower-entropy gas from the inner 6~kpc can allow gas with $K \sim 15 \,{\rm keV \, cm^2}$ to sink inward from larger radii.  That inflow can produce an entropy excess at smaller radii, as long as the timescale for inflow is shorter than the cooling timescale.  In this case, the cooling time of the gas with $K \approx 15 \, {\rm keV \, cm^2}$ is $\sim 300$~Myr, meaning that inflowing higher-entropy gas can indeed produce at least a temporary entropy excess at smaller radii.

There are two ways to remove lower-entropy gas from within 6~kpc without heating it.  One is consumption by either the black hole or star formation.  The other is expulsion from the central region either by direct momentum transfer from the AGN outflow or in the trailing plume behind a buoyantly rising bubble. Note that our AGN feedback algorithm technically does not consume any gas, because all of the cold gas mass removed from the central 0.5~kpc is re-introduced by the jets, which can propel it to large radii.

In our fiducial simulations, the AGN feedback algorithm converts accreted rest-mass energy to AGN feedback energy with an efficiency of $10^{-4}$, meaning that accretion of $2 \times 10^8 \, M_\odot$ of gas releases $4 \times 10^{58} \, {\rm erg}$.  AGN feedback does indeed produce a comparable amount of energy during the first $\sim 200$~Myr of the fiducial SPG simulation, meaning that gas consumption (followed by ejection via jets) at a rate $\lesssim 1 \, M_\odot \, {\rm yr}^{-1}$ is a plausible origin for the entropy excess that arises at $\sim 1$--3~kpc. However, it is not clear how much of the entropy excess at 1--3~kpc in the fiducial SPG simulation is due to circulation and how much is due to direct AGN heating. Section \ref{sec:circ} will show that circulation plays a larger role in our other simulations. 

\subsubsection{Quarter-Angle SPG}

Changing the jet opening angle in the SPG simulation changes the nature of the entropy excess that develops.  Figure \ref{fig:entropy} shows that the mass-weighted mean entropy at $\sim 1$~kpc remains at $\lesssim 10 \, {\rm keV \, cm^2}$, but an excess reaching nearly $100 \, {\rm keV \, cm^2}$ arises at $\sim 5$--10~kpc.  That entropy level corresponds to $r \sim 30$~kpc in the atmosphere's initial configuration (see Figure \ref{fig:entropy}), which encloses a gas mass $\sim 3 \times 10^9 \, M_\odot$ (see Figure \ref{fig:ent_mgas}).  Consequently, circulation exceeding $6 \, M_\odot \, {\rm yr}^{-1}$ at $\sim 10$~kpc is necessary to produce such an entropy increase within 500~Myr.  However, the entropy of gas at $\lesssim 3$~kpc does not become excessive, indicating that large-scale circulation does not transport gas with $K \gg 10 \, {\rm keV \, cm^2}$ into the SPG galaxy's central region.

\subsubsection{Fiducial MPG}

Figure~\ref{fig:entropy_MPG} shows that an entropy plateau near $25 \, {\rm keV \rm cm^2}$ develops at $\sim 2$--10~kpc during the first 500~Myr of the fiducial MPG simulation. That entropy level corresponds to $r \sim 25$~kpc in the atmosphere's initial configuration (see Figure \ref{fig:entropy}) and encloses a gas mass $\sim 10^{10} \, M_\odot$ (see Figure \ref{fig:ent_mgas}).  The amount of circulation needed to produce that entropy excess over 500~Myr is therefore $\gtrsim 20 \, M_\odot \, {\rm yr}$.   

\section{Atmospheric Circulation}
\label{sec:circ}
This section looks more closely at the role of circulation in the AGN feedback loop, as realized in the simulations of Paper~I. Section \ref{sec:mass_circ} explores the large scale gas mass circulation in the circumgalactic medium due to AGN feedback. This is followed with a discussion on the resultant  entropy circulation due to gas mass circulation in Section \ref{sec:ent_circ}. Sections \ref{sec:circ_SPG} and \ref{sec:circ_MPG} present the mass and entropy circulation analysis for the fiducial SPG and MPG simulations, respectively. Finally in Section \ref{sec:opang} we analyse the role of narrow jet opening angle on the AGN-CGM coupling and its consequences. \\

\subsection{Gas Mass Circulation}
\label{sec:mass_circ}
 
\begin{figure*}[!t]
 \includegraphics[width=0.45\textwidth]{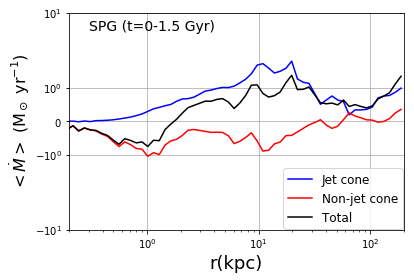}
 \includegraphics[width=0.45\textwidth]{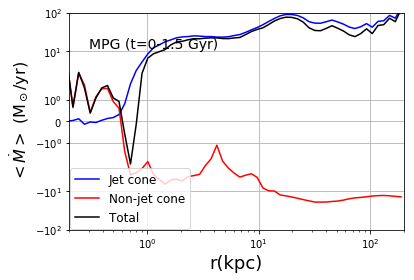}
 \caption{Time-averaged mass flow rates inside and outside of the jet-driven bipolar outflow during the full 1.5~Gyr simulation period for the fiducial SPG (left) and MPG (right) simulations.  For making this comparison, we define the jet cone to be the region within $\theta = 30^\circ$ around the jet axis.  Blue lines show the net gas mass that flows through radius $r$ within the jet cone.  Red lines show the net gas mass that flows through radius $r$ outside of the jet cone.  AGN feedback in those simulations drives large-scale circulation that generally flows outward along the jet axis and inward along directions far from the jet axis.  The mean circulation rate at $\sim 10$~kpc is $\sim 1 \, M_\odot \, {\rm yr}^{-1}$ in the SPG simulation and $\sim 30 \, M_\odot \, {\rm yr}^{-1}$ in the MPG simulation.  
 }
 \label{fig:mass_flux}
\end{figure*}

Bipolar AGN feedback in our simulations generally drives gas near the jet axis outward, allowing gas farther from the jet axis to sink inward.  Figure \ref{fig:mass_flux} shows mean flow rates in those directions during the fiducial SPG and MPG simulations.  Blue lines illustrate the net flow rate of gas mass through radius $r$ within $30^\circ$ of the jet axis, a region we refer to as the ``jet cone.''  Red lines illustrate the net flow rate of gas mass through radius $r$ outside of the jet cone. Black lines show the net flow rate of gas mass through the entire spherical shell at each radius $r$.  
 
In the fiducial SPG simulation, the mean circulation rate at 10~kpc is $\sim 1 \, M_\odot \, {\rm yr}^{-1}$, implying that $\sim 10^9 \, M_\odot$ circulates through that radius during the 1.5~Gyr simulation period.  The polar circulation pattern extends to $\sim 50$~kpc.  Outside of that radius, the net flow is outward both inside and outside of the jet cone, because the primary outcome of AGN feedback at those radii is atmospheric reconfiguration via expansion in all directions (see \S \ref{sec:config}).  Atmospheric expansion also happens at smaller radii, but the net mass flow associated with expansion is smaller than the circulation flow.  

The mean circulation rate in the fiducial MPG simulation is much greater. Outward gas flow within the jet cone happens at a mean rate of several tens of $M_\odot \, {\rm yr}^{-1}$ at 10--100~kpc during the simulation period, while the mean inward gas flow outside of the jet cone is $\sim 10 \, M_\odot \, {\rm yr}^{-1}$.  At smaller radii (1--10~kpc) the overall pattern is similar, but the amounts of gas displaced are smaller by a factor of a few.


\subsection{Entropy Circulation}
\label{sec:ent_circ}

Large scale circulation of gas in the fiducial SPG and MPG simulations plays a major role in the AGN feedback loop because of how it alters the entropy structure of the inner atmosphere.  In an entropy-stratified atmosphere, circulation that transports gas inward raises $K$ at small radii. Bipolar circulation that happens on a timescale shorter than the cooling time of the circulating gas can raise the entropy level of the inner atmosphere if the typical specific entropy of gas flowing outward through the jet cone is lower than the typical specific entropy of the gas that replaces it by flowing inward through the (non-jet cone) equatorial regions. Such a circulation pattern can regulate cooling of atmospheric gas within a galaxy without any heat input
by the AGN, as long as the low-entropy gas pushed outward along the jet axis mixes with higher-entropy gas at greater altitudes before falling back into the galaxy (\citealt{hillsok}).
This section illustrates how that circulation pattern plays out in our fiducial simulations.

\subsubsection{Fiducial SPG}
\label{sec:circ_SPG}

\begin{figure*}[!t]
\centering
 \includegraphics[width=2.2in,height=2.2in]{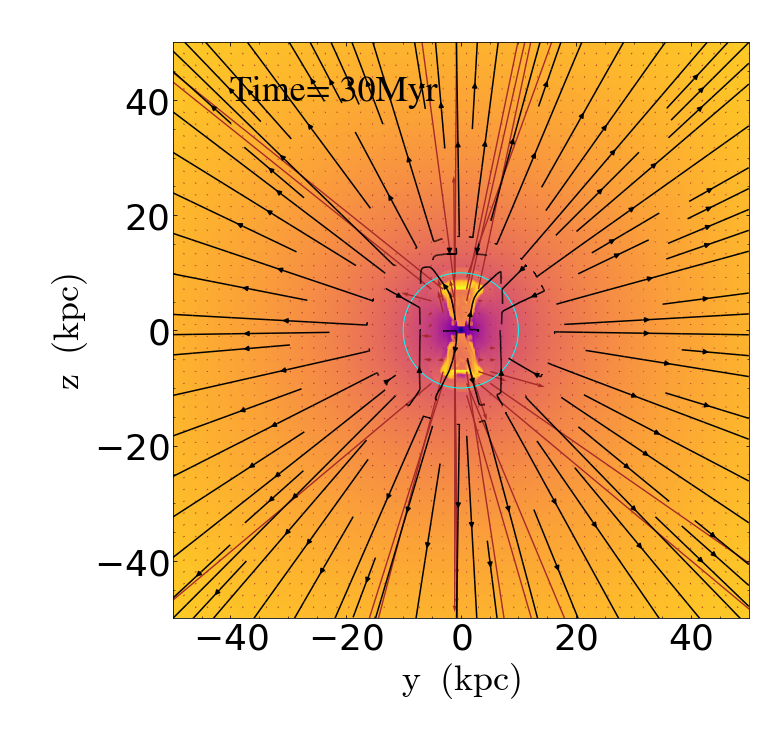}
  \includegraphics[width=2.2in,height=2.2in]{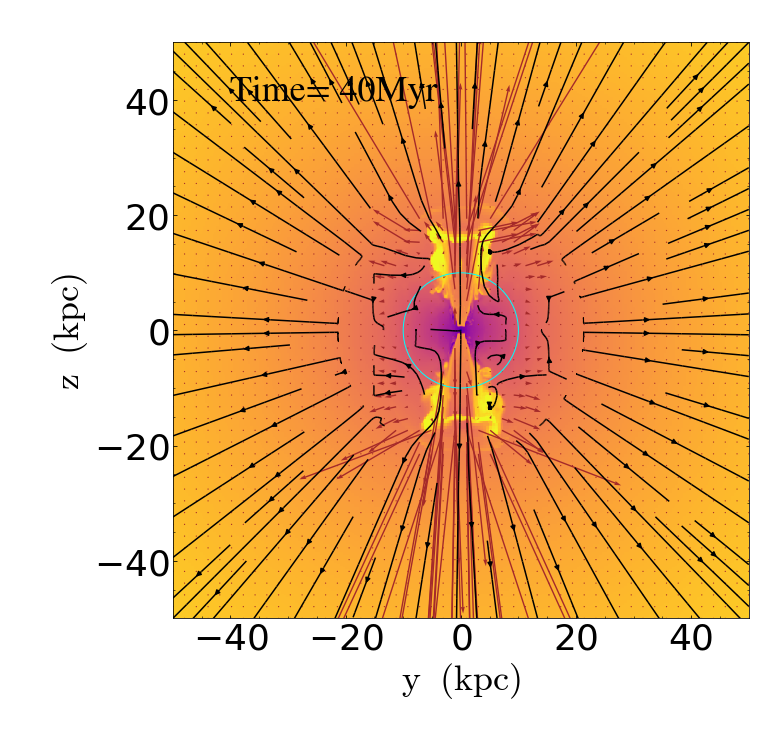}
   \includegraphics[width=2.4in,height=2.2in]{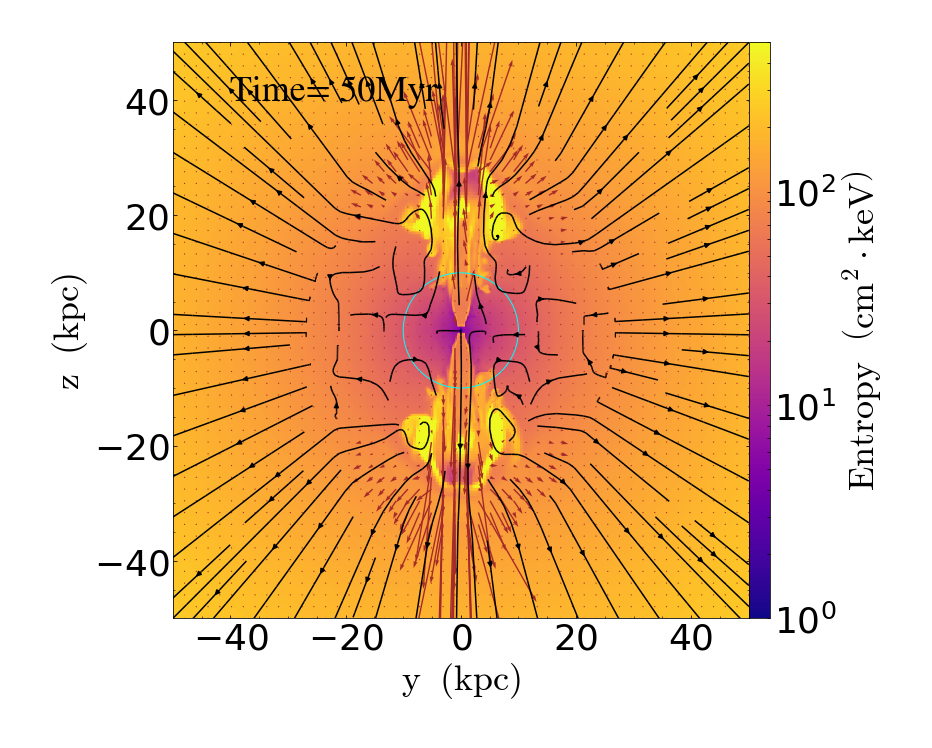}
    \includegraphics[width=2.2in,height=2.2in]{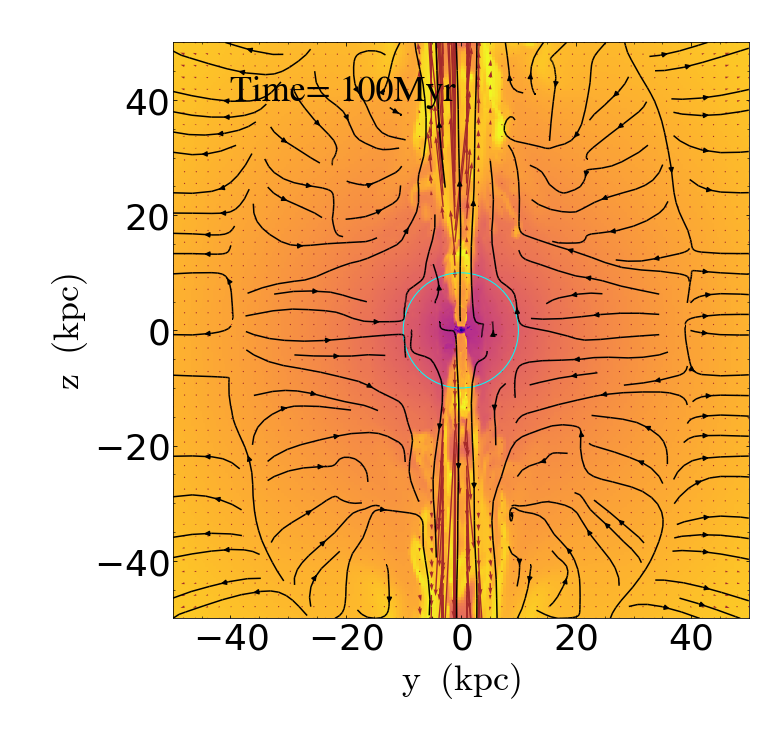}
  \includegraphics[width=2.2in,height=2.2in]{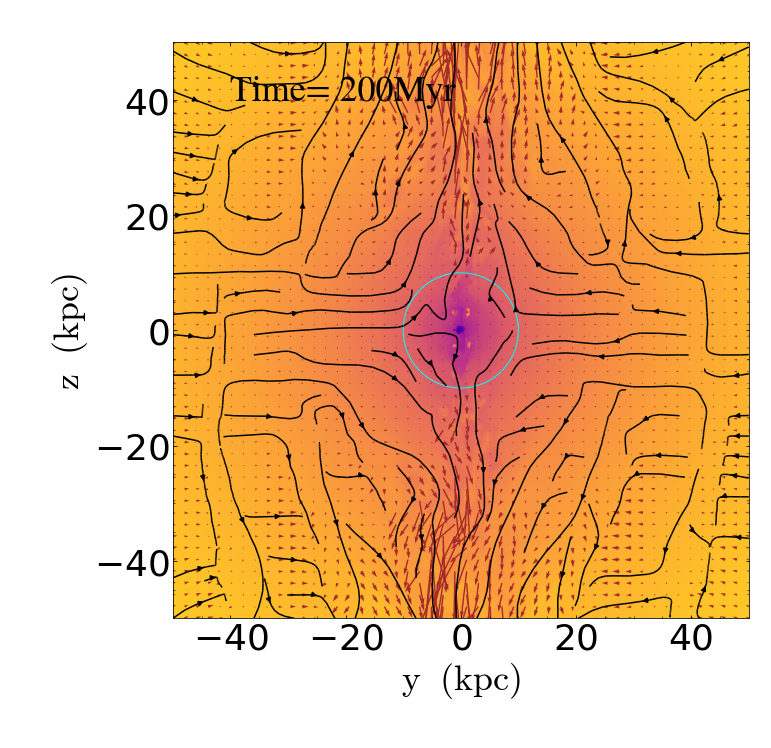}
   \includegraphics[width=2.4in,height=2.2in]{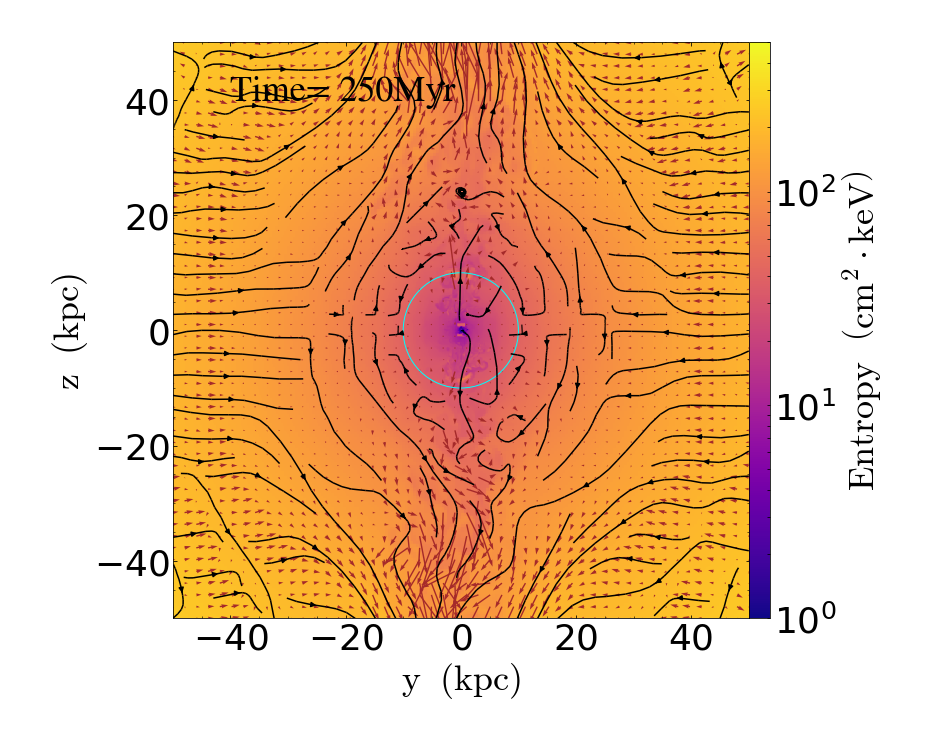}
    \includegraphics[width=2.2in,height=2.2in]{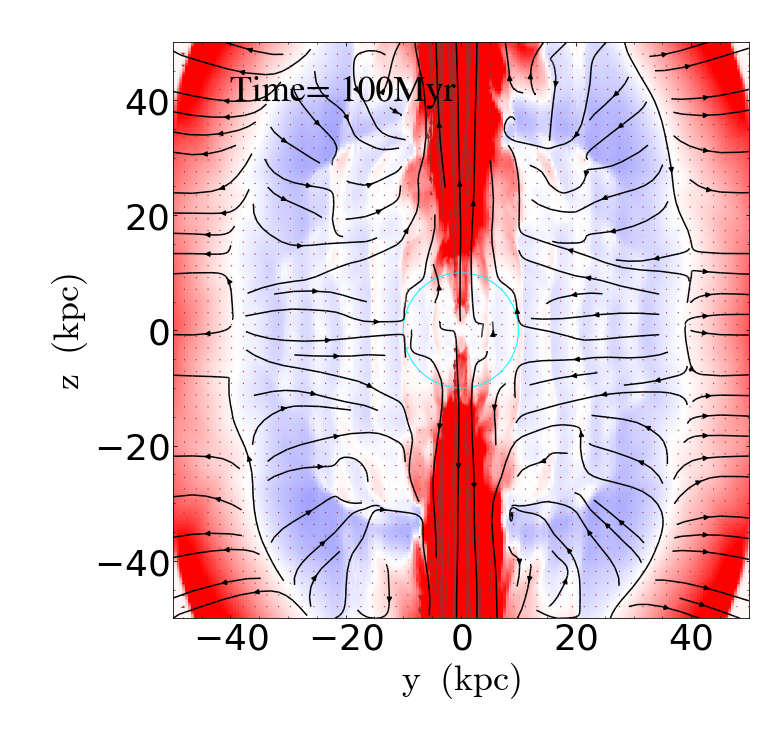}
  \includegraphics[width=2.2in,height=2.2in]{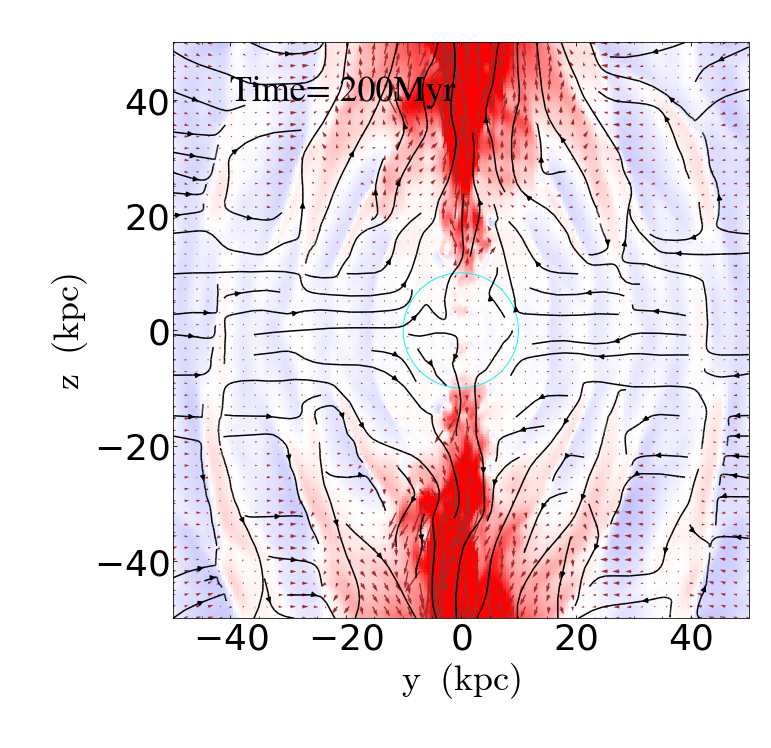}
   \includegraphics[width=2.4in,height=2.2in]{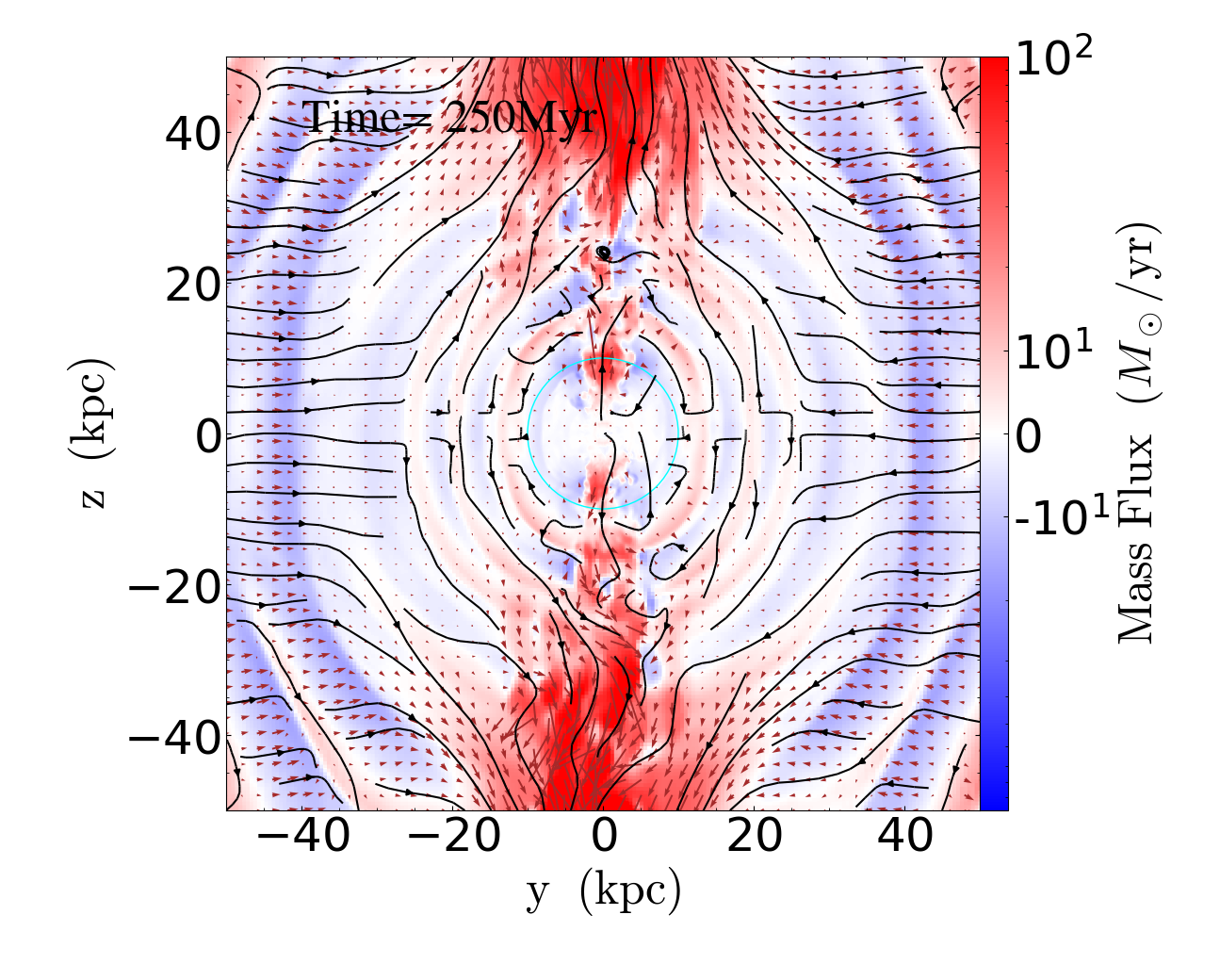}
 \caption{Circulation of entropy and gas mass during the initial outburst of the fiducial SPG simulation.  \textit{Top two rows:}  Azimuthally averaged entropy at $t=30$, 40, 50, 100, 200, and 250~Myr. Black arrows show streamlines while brown arrows show relative mass flux.  \textit{Bottom row:} Azimuthally averaged radial mass flux (shown as $4 \pi r^2 \rho v_r$) at $t=100$, 200, and~250 Myr.  Black arrows again show streamlines while brown arrows show the relative magnitude of the Bernoulli energy flux.  Cyan circles in all panels show a radius of $r\lesssim10$ kpc around the galaxy. The top row shows how the outburst produces high-entropy bubbles that form $\sim 20$~kpc from the AGN.  At 100~Myr, an acoustic impulse from the initial outburst is approaching $r \sim 60$~kpc, and the AGN is driving large-scale circulation within $r \sim 30$~kpc. At 200~Myr, AGN power has declined by nearly an order of magnitude, and circulation is slowing down.  At 250~Myr, circulation within the central 20~kpc is minimal.  
}
 \label{fig:SPG_fid_circ}
\end{figure*}

Figure \ref{fig:SPG_fid_circ} shows how bipolar circulation alters atmospheric entropy structure during the first 250~Myr of the fiducial SPG simulation.  Across the top of the figure are three panels corresponding to propagation of the initial feedback outburst to beyond $\sim 30$~kpc.  At 30~Myr, some high-entropy gas is beginning to accumulate at $\sim 10$~kpc inside the jet cone.  That outflowing gas has reached high entropy by passing through the reverse shock produced
as the bipolar outflow encounters resistance from the CGM.  By 50~Myr a high-entropy gas bubble is developing at $\sim 20$~kpc.  However, much of the gas outside of the jet cone and within 10~kpc still has relatively low entropy ($\lesssim 10 \, {\rm keV \, cm^2}$).

After the first 50~Myr, a large-scale circulation pattern develops, as shown by the streamlines in the 100~Myr and 200~Myr panels.  During that time period, the AGN's power is declining (see Figure \ref{fig:lum_spg}). Simultaneously, the mean entropy level at $\lesssim 10$~kpc is rising.  Some of that entropy rise can come from SNIa heating of the outflowing gas ejected by stars. In that case, the resulting entropy level depends on CGM pressure confinement of the heated outflow \citep[e.g.,][]{voit2020}.  A continual rise in entropy therefore implies a continual decrease in CGM pressure confinement, which comes about through the circulation pattern.

Across the bottom of Figure \ref{fig:SPG_fid_circ} are three panels showing the bipolar mass flow during the time period from 100~Myr through 250~Myr.  Red and blue colors show the quantity $4 \pi r^2 \rho v_r$, with red indicating outflow and blue indicating inflow. At 100~Myr there is a strong bipolar outflow along the jet cone, extending down to within 5~kpc.  Outside of the jet cone, the light blue color at $< 10$~kpc indicates a weak inflow consistent with the mean circulation rate $\sim 1 \, M_\odot \, {\rm yr^{-1}}$ indicated by Figure \ref{fig:mass_flux}.  At larger radii, the initial feedback impulse has reached $r > 40$~kpc in all directions, and sound waves are reverberating through the net inflow outside of the jet cone. By 200~Myr the initial feedback impulse has propagated outside the frame, and sound waves are still reverberating.  Even though the AGN has significantly declined in power, bipolar circulation is continuing because of its inertia. Circulation streamlines indicate that equatorial gas is still flowing inward through the shell at 10~kpc.  However, inflow of gas to within 10~kpc has largely ceased by 250~Myr, by which time the AGN has entered its low-power state.

\subsubsection{Fiducial MPG}
\label{sec:circ_MPG}
 
\begin{figure*}[!t]
\centering
 \includegraphics[width=2.2in,height=2.2in]{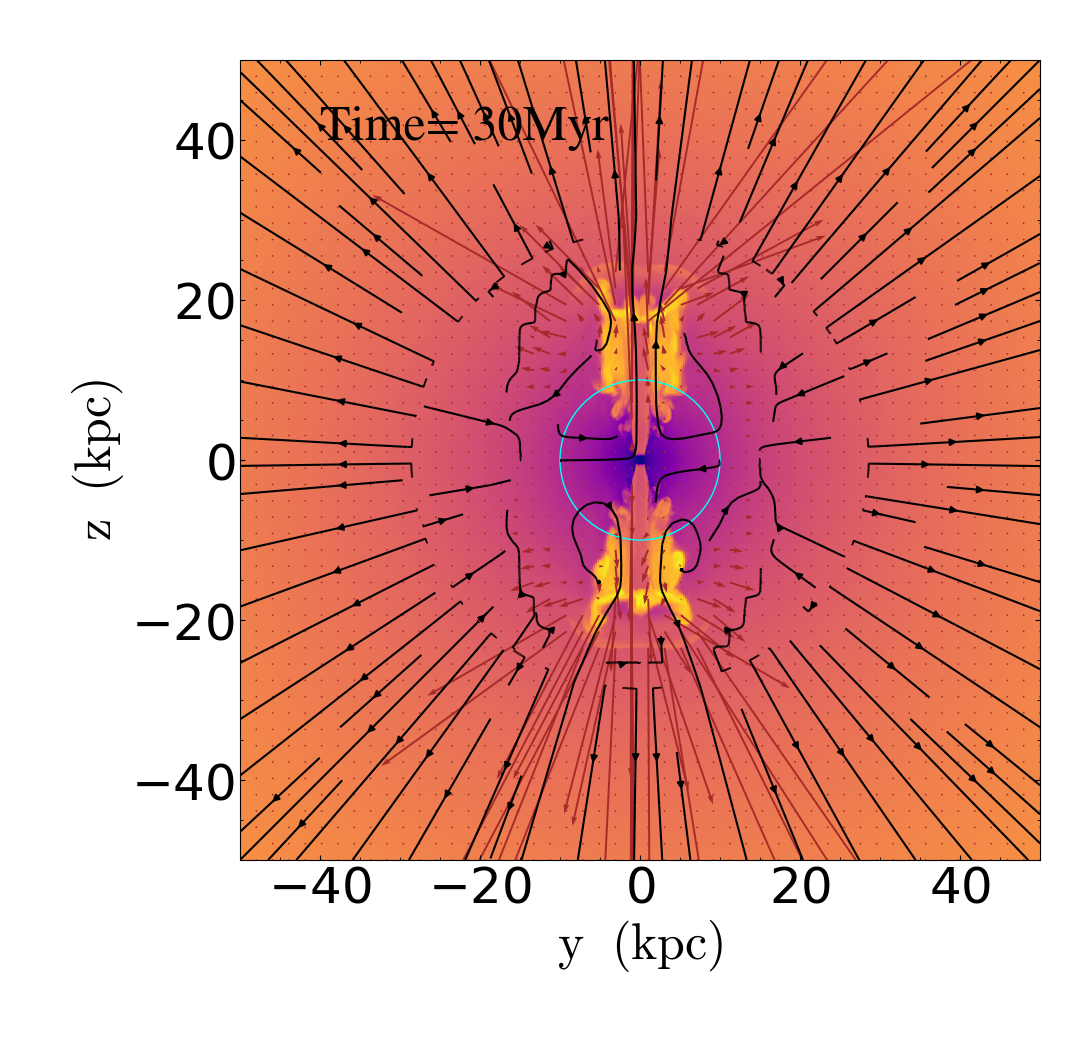}
  \includegraphics[width=2.2in,height=2.2in]{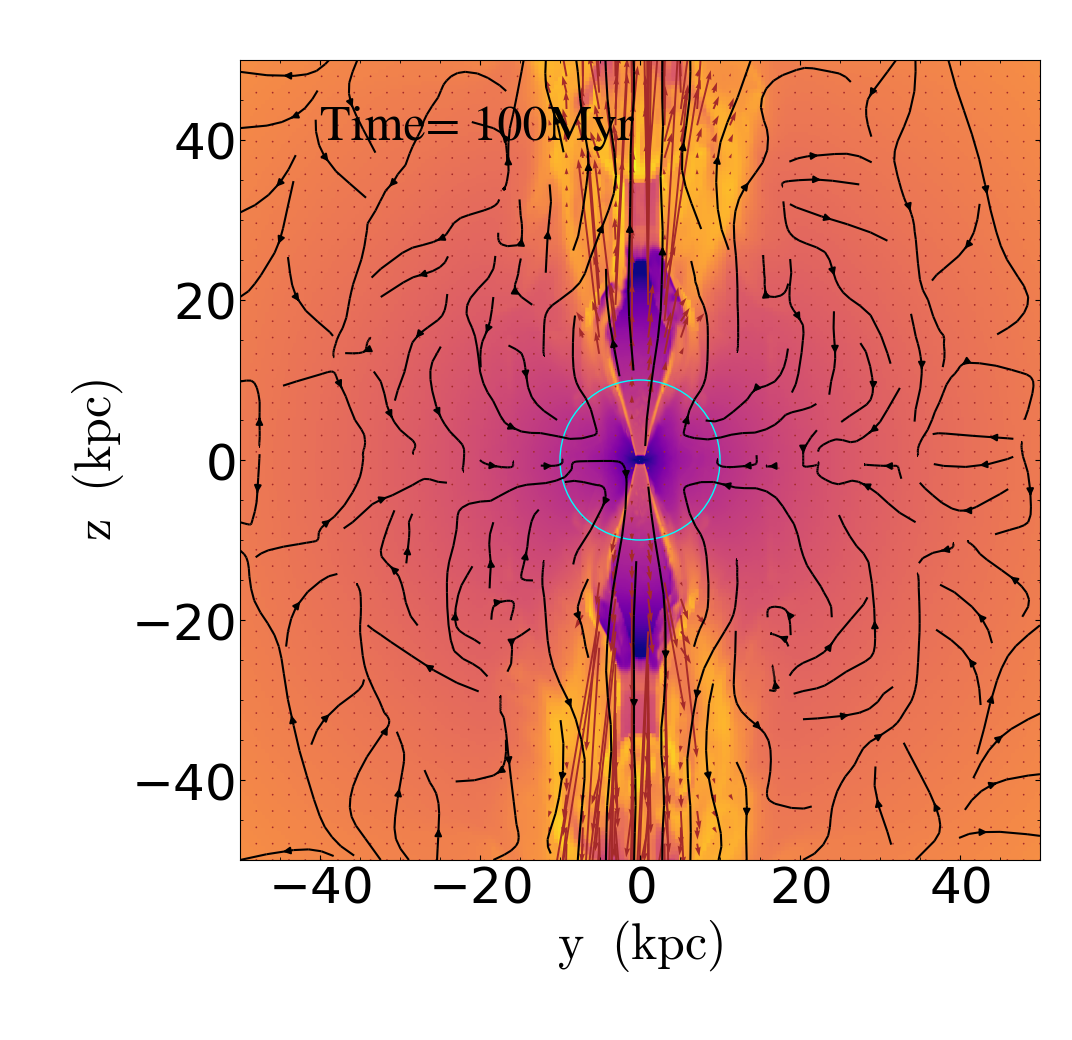}
   \includegraphics[width=2.6in,height=2.2in]{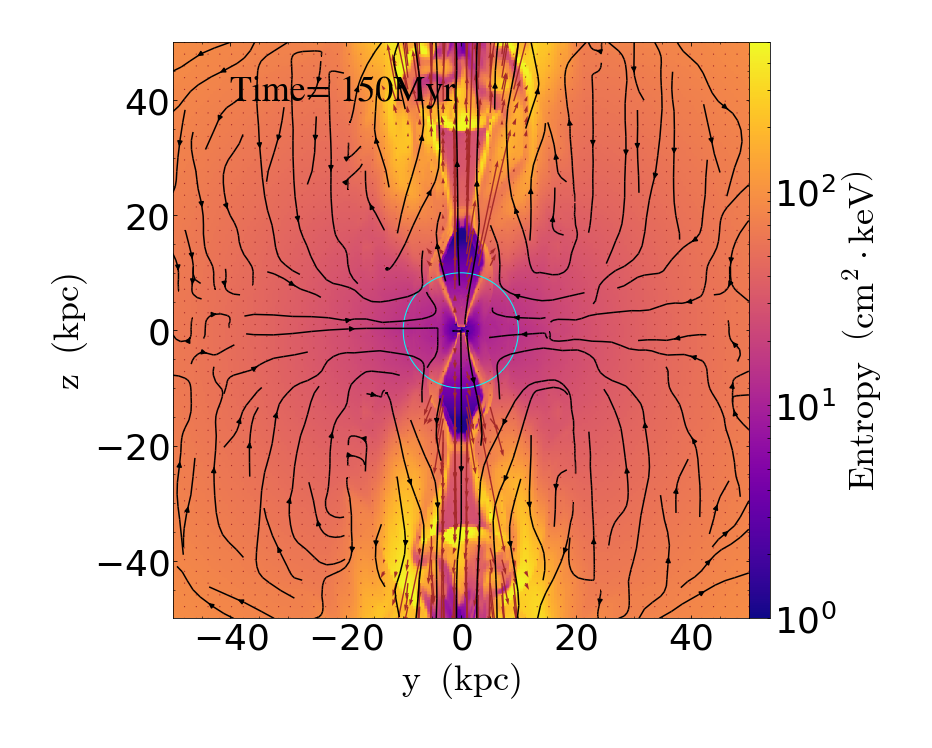}
  \includegraphics[width=2.2in,height=2.2in]{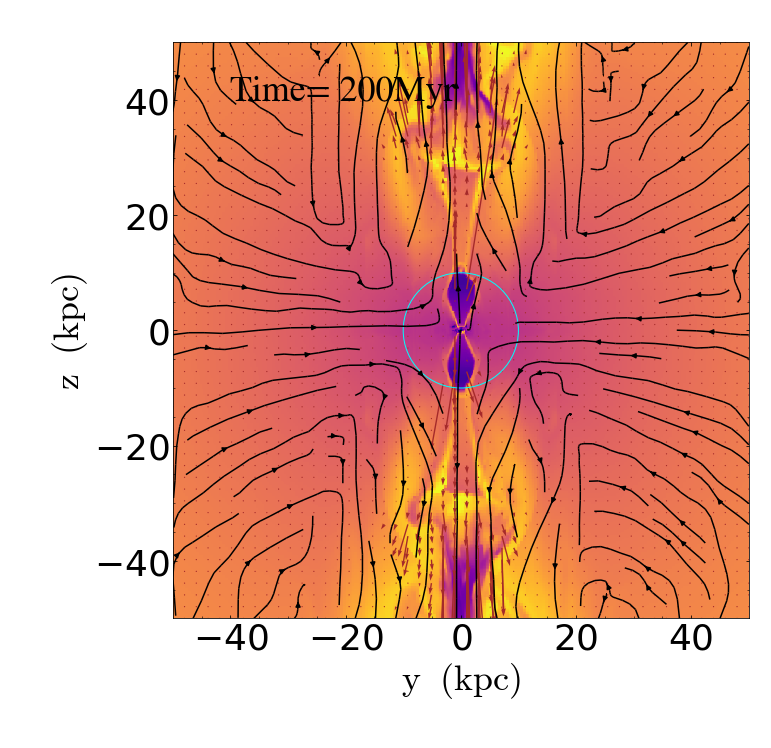}
  \includegraphics[width=2.2in,height=2.2in]{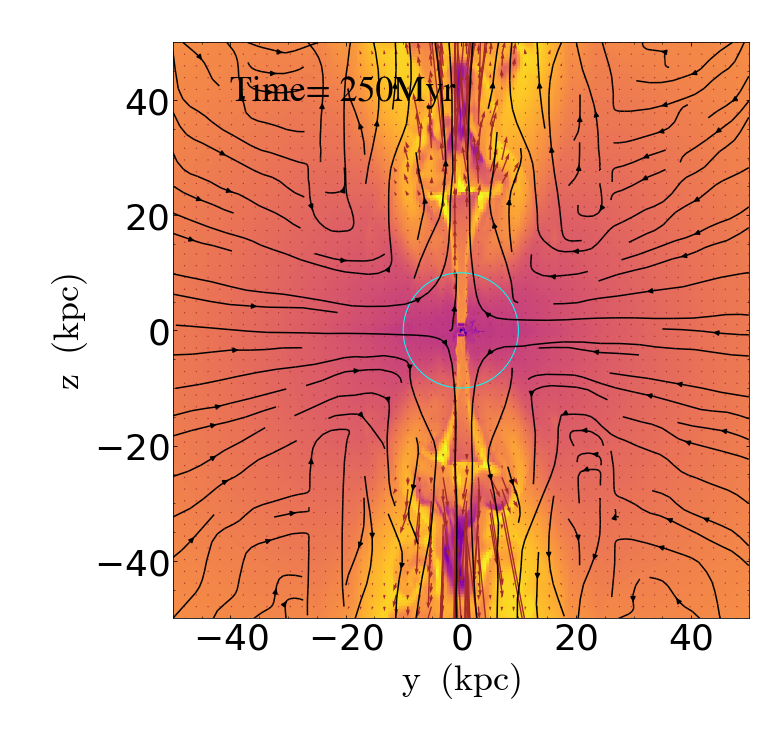}
   \includegraphics[width=2.6in,height=2.2in]{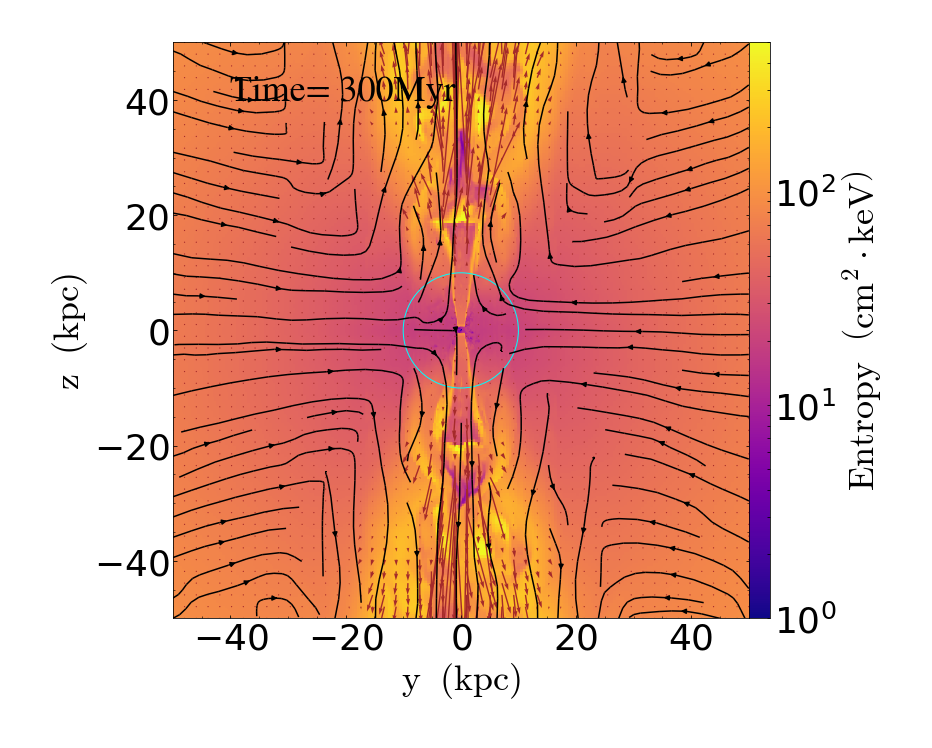}
\caption{Circulation of entropy during the initial outburst of the fiducial MPG simulation at $t=30$, 100, 150, 200, 250, and 300~Myr. Black arrows show streamlines while brown arrows show relative mass flux.  All figure elements are as in the top two rows of Figure \ref{fig:SPG_fid_circ}.  At 40~Myr, the initial AGN impulse has reached $\sim 30$~kpc.  However, much of the off-axis gas within 10~kpc still has $K < 10 \, {\rm keV \, cm^2}$.  The median entropy level within 10~kpc then gradually rises as AGN feedback starts to drive large-scale circulation.  Polar outflow lifts gas out of the central region, and equatorial inflow transports higher entropy gas to within $\sim 10$~kpc.  (As discussed in \S \ref{sec:circ_MPG}, prominent regions of outflowing gas within the jet cone at 100--200~Myr correspond to episodes of unusually large AGN accretion rates.  They arise because the AGN feedback algorithm re-injects all of the accreted mass, resulting in a dense outflow.  However, strong internal shocks raise the entropy of that gas before significant cooling can happen.)   
}
 \label{fig:MPG_fid_circ_1}
\end{figure*}

Circulation plays a much bigger role in the fiducial MPG simulation, because the high-power mode of its AGN is considerably more powerful in its high-power state.  Figure \ref{fig:MPG_fid_circ_1} shows how atmospheric entropy structure in the fiducial MPG evolves during $t = 30$--300~Myr.  As in the fiducial MPG simulation, the initial AGN outburst produces high-entropy bubbles at $\sim 20$~kpc while leaving much of the gas outside of the jet cone in a low-entropy state (see the 30~Myr panel).  By 150~Myr, a large scale bipolar circulation flow has developed.  The typical entropy of gas within 10~kpc has increased somewhat but there is still plenty of low-entropy gas accreting onto the AGN, making this the epoch of maximum AGN power.

An anomalous feature of the fiducial MPG simulation during this time period is the very low entropy of gas being fed into the jets.  Its low entropy during this period of high AGN power results from assumptions built into the feedback algorithm, which specifies that the total energy of ejected gas is $10^{-4}$ times the rest-mass energy, with 90\% in kinetic form and 10\% in thermal form.  When AGN power is high, the mass flow of the jets (equal to $\dot{M}_{\rm acc}$ by design) is also high, resulting in
\begin{equation}
    K_{\rm jet} \approx 140 \, {\rm keV \, cm^2}
            \left( \frac {\dot{M}_{\rm acc}} 
                         {1 \, M_\odot \, {\rm yr^{-1}}} \right)^{-2/3}
                         \; \; .
\end{equation}
When AGN power is particularly high, the initial entropy of the jet material is
\begin{equation}
    K_{\rm jet} \approx 5 \, {\rm keV \, cm^2}
            \left( \frac {P_{\rm jet}} 
                         {10^{45} \, {\rm erg s^{-1}}} \right)^{-2/3}
                         \; \; .
\end{equation}
Temporary excursions to very high jet power (see the gray line in Figure \ref{fig:lum_mpg}) therefore can cause the jet material to have a low entropy level. The entropy of that outflowing gas does not remain low for very long, because it soon passes at high speed through a reverse shock, gaining more entropy as shown in Figure \ref{fig:MPG_fid_circ_1}.  By 250~Myr, those low-entropy outflows are gone.  But before they go, they produce a prominent luminosity spike peaking near 220~Myr in Figure \ref{fig:lum_mpg}.  The luminosity peaks when the density of gas that has just passed through the reverse shock is greatest.  However, that peak is a feature specific to this feedback implementation and thus we do not expect something similar to happen in real galaxies.

\begin{figure*}[!t]
\centering
 \includegraphics[width=2.2in,height=2.2in]{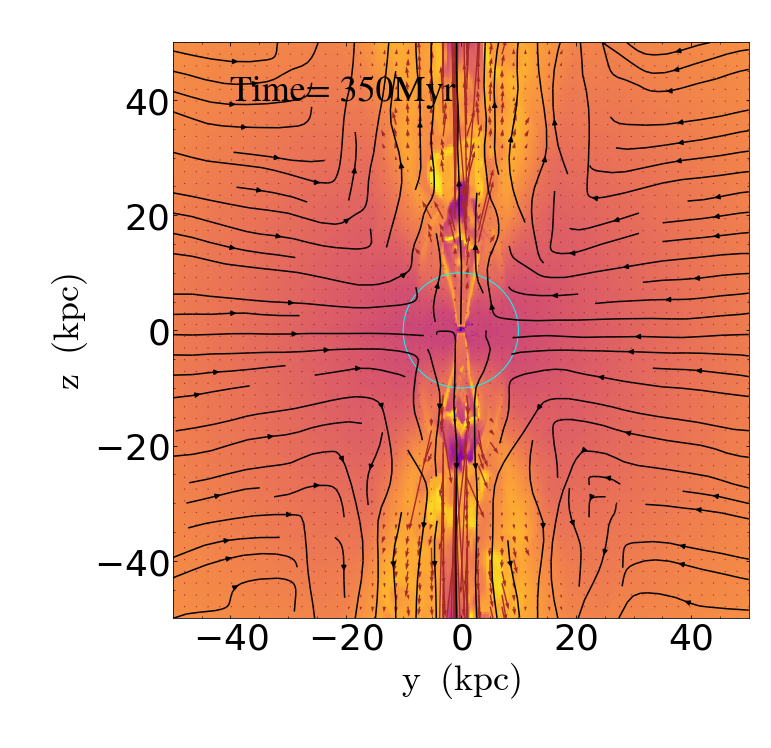}
  \includegraphics[width=2.2in,height=2.2in]{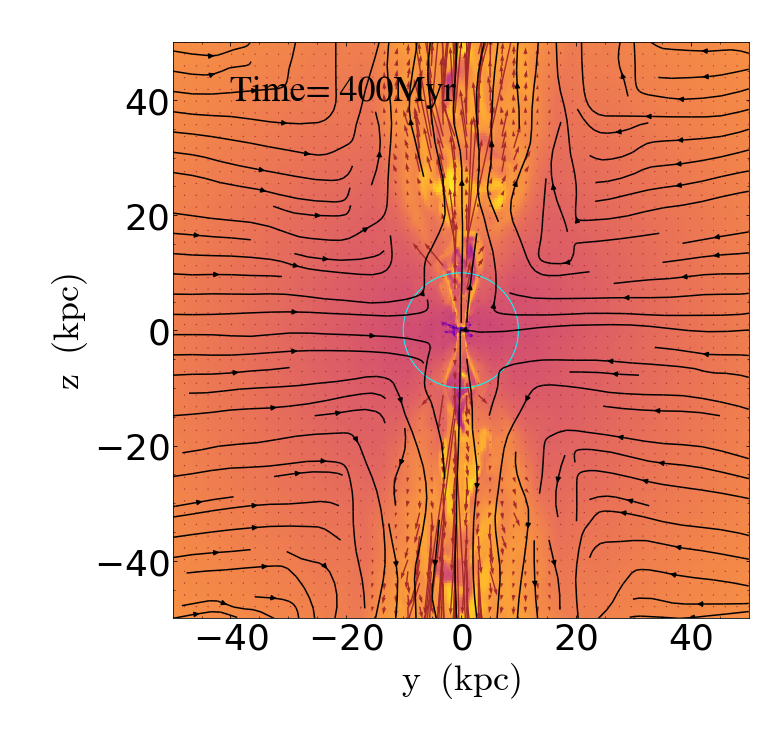}
   \includegraphics[width=2.6in,height=2.2in]{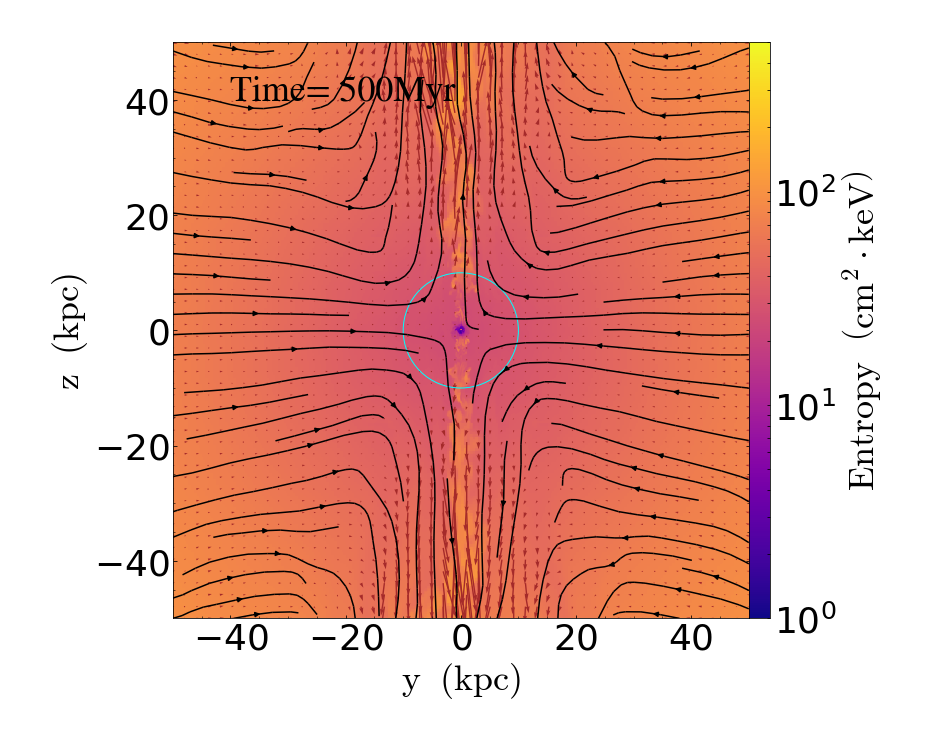}
  \includegraphics[width=2.2in,height=2.2in]{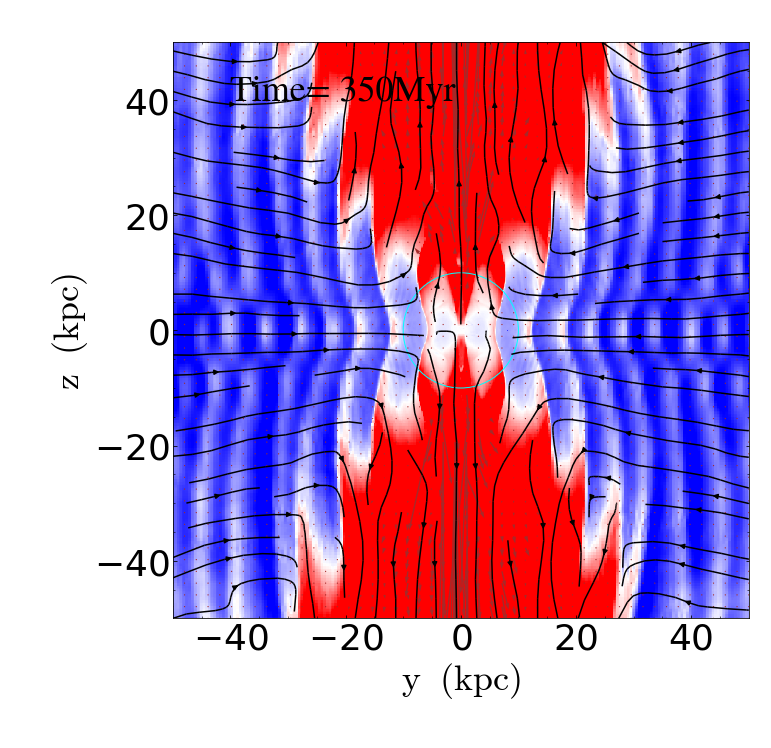}
  \includegraphics[width=2.2in,height=2.2in]{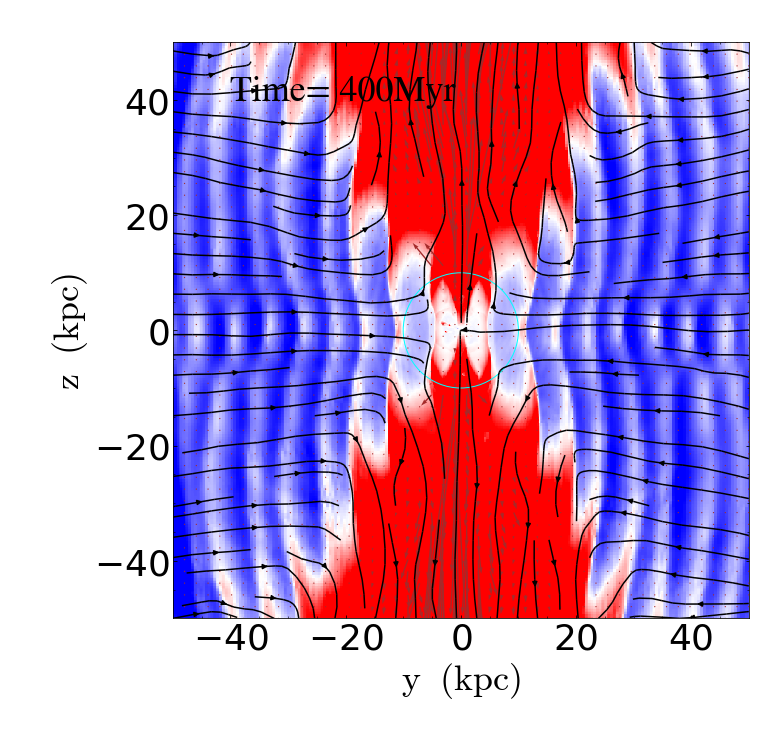}
   \includegraphics[width=2.6in,height=2.2in]{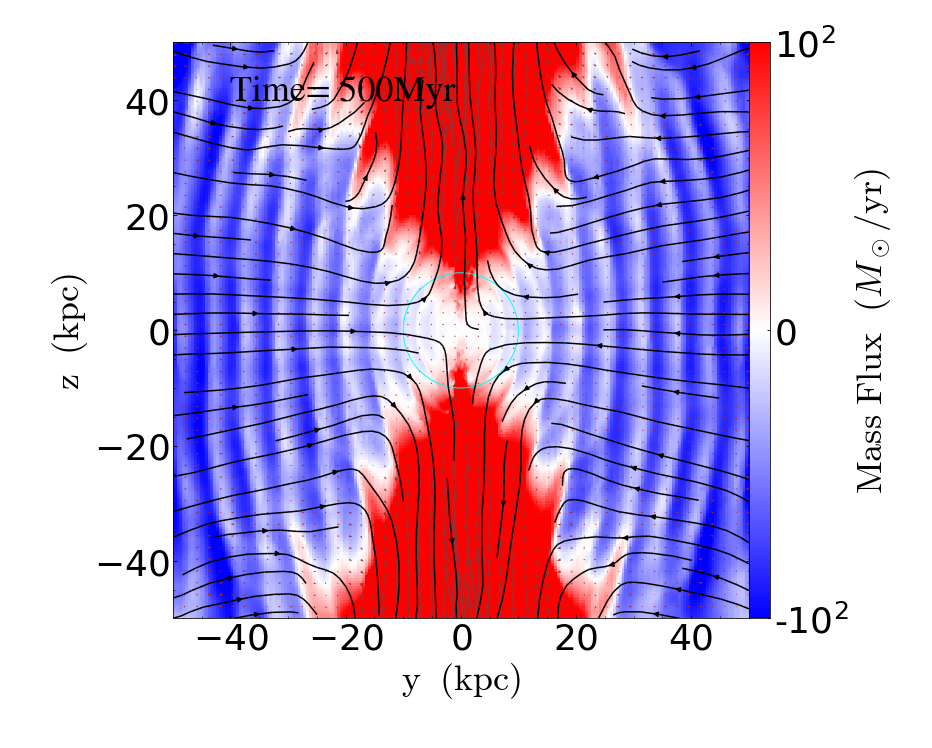}
\caption{Circulation of entropy and gas mass as the initial outburst of the fiducial MPG simulation winds down at $t=350$, 400, and 500~Myr.  All figure elements are as in Figure \ref{fig:SPG_fid_circ}.  Strong large-scale circulation continues to drive equatorial inflow during this time period, until $t \approx 500$~Myr.  (Note the whiteness within $\sim 10$~kpc of the mass-flow panel at $t = 500$~Myr.)
}
 \label{fig:MPG_fid_circ_2}
\end{figure*}

In the fiducial MPG simulation, AGN power gradually subsides between 400~Myr and 500~Myr as the AGN enters its low-power state. Figure \ref{fig:MPG_fid_circ_2} illustrates the circulation pattern that persists during this transitional time period.  Gas mass flux into the central 10~kpc along directions outside of the jet cone is significant during this time period but drops off at 500~Myr (resulting in a white region within the 10~kpc circle in the lower right panel).

Figure \ref{fig:entropy_MPG} shows that the atmospheric entropy level at 10~kpc gradually increases from 300~Myr to 500~Myr.  This is the time period during which large scale circulation transports higher entropy gas from larger radii into the central 10~kpc along directions approximately perpendicular to the jet axis.  That rise in entropy lowers the atmospheric density and pressure surrounding the galaxy, enabling SNIa heating to exceed radiative cooling within the central 3~kpc during the same time period (see Figure \ref{fig:lum_mpg}).  As in the fiducial SPG simulation, not much shock heating is evident outside of the jet cone. Instead the entropy change at $\sim 10$~kpc appears to arise primarily from atmospheric reconfiguration via circulation.

\begin{figure*}[!t]
\centering
 \includegraphics[width=2.2in,height=2.2in]{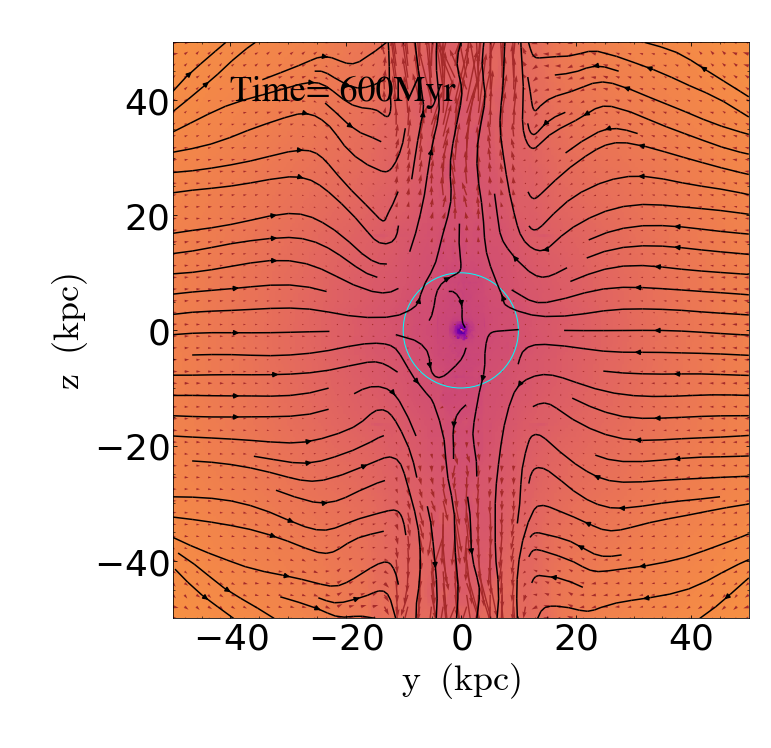}
  \includegraphics[width=2.2in,height=2.2in]{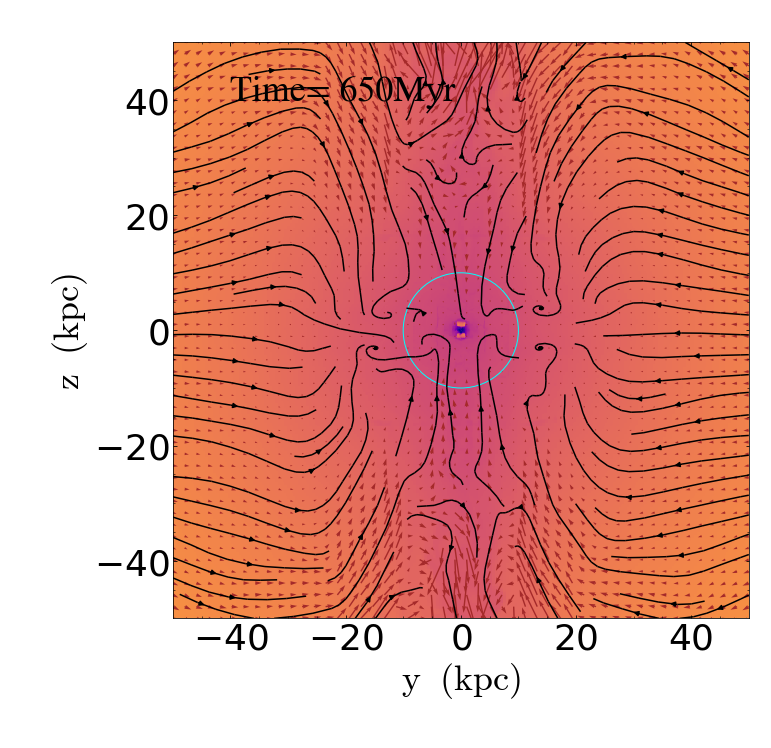}
   \includegraphics[width=2.4in,height=2.2in]{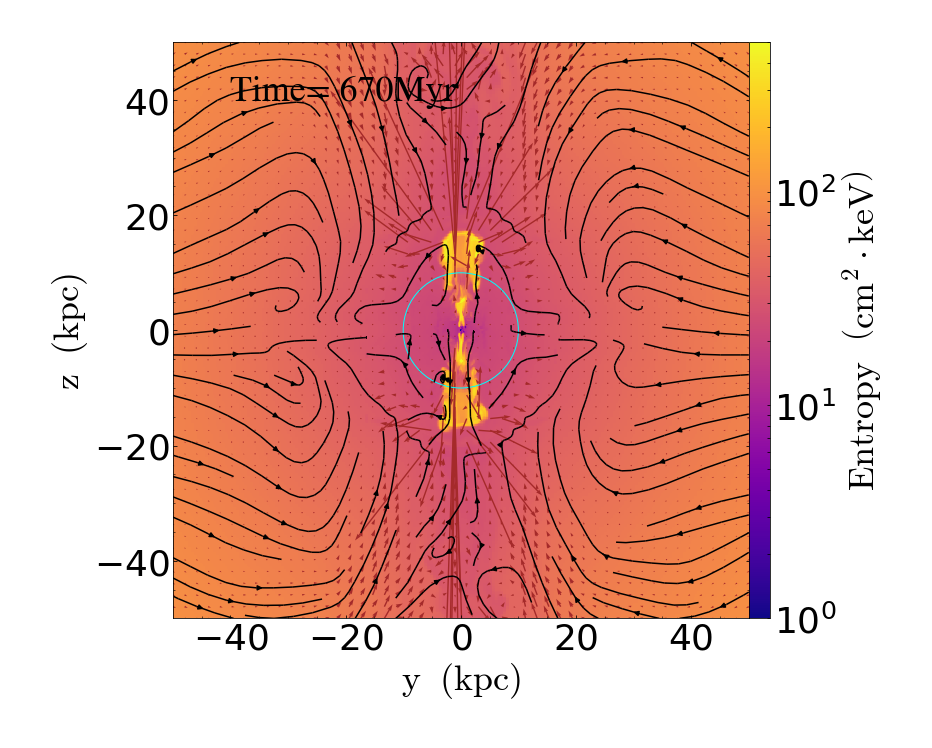}
  \includegraphics[width=2.2in,height=2.2in]{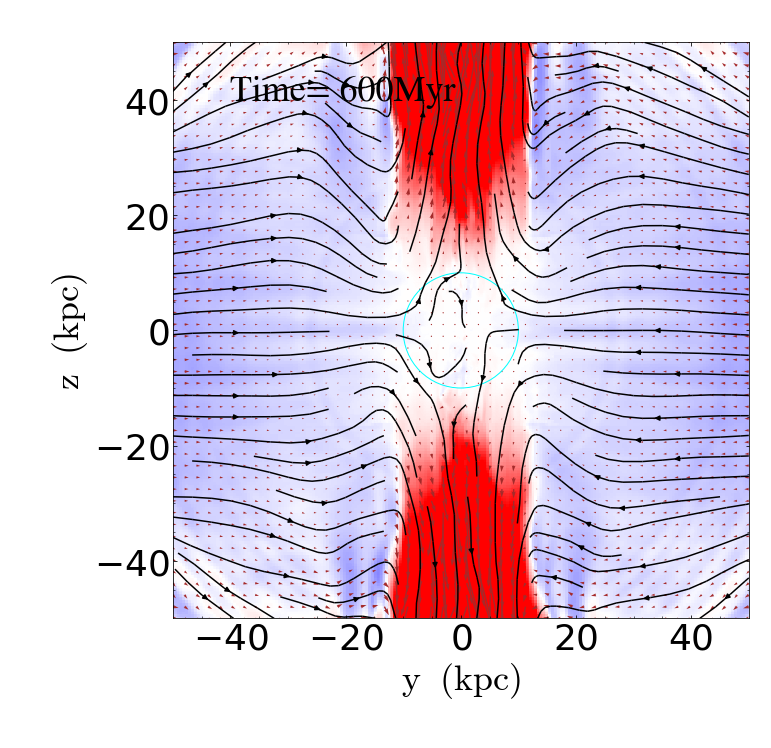}
  \includegraphics[width=2.2in,height=2.2in]{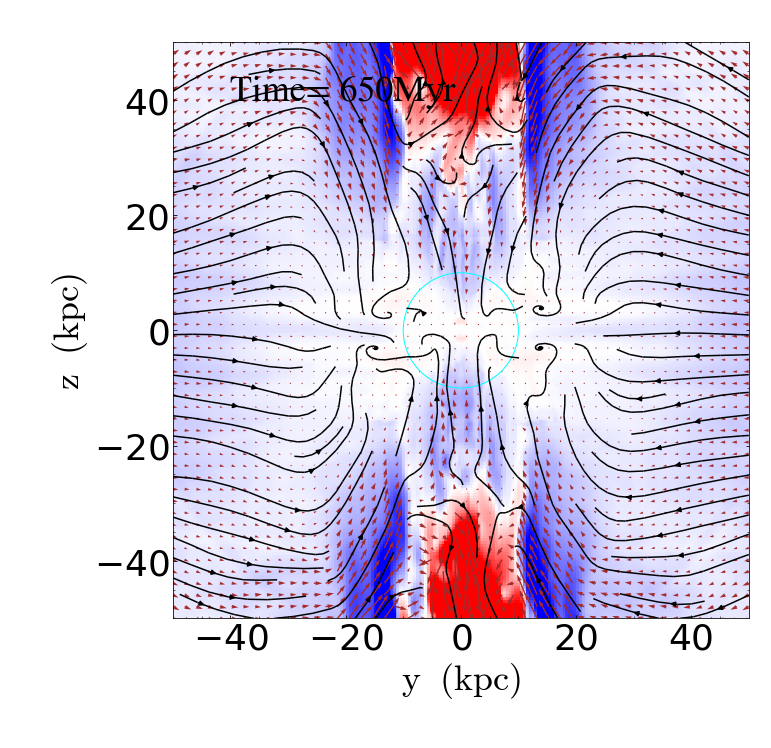}
   \includegraphics[width=2.4in,height=2.2in]{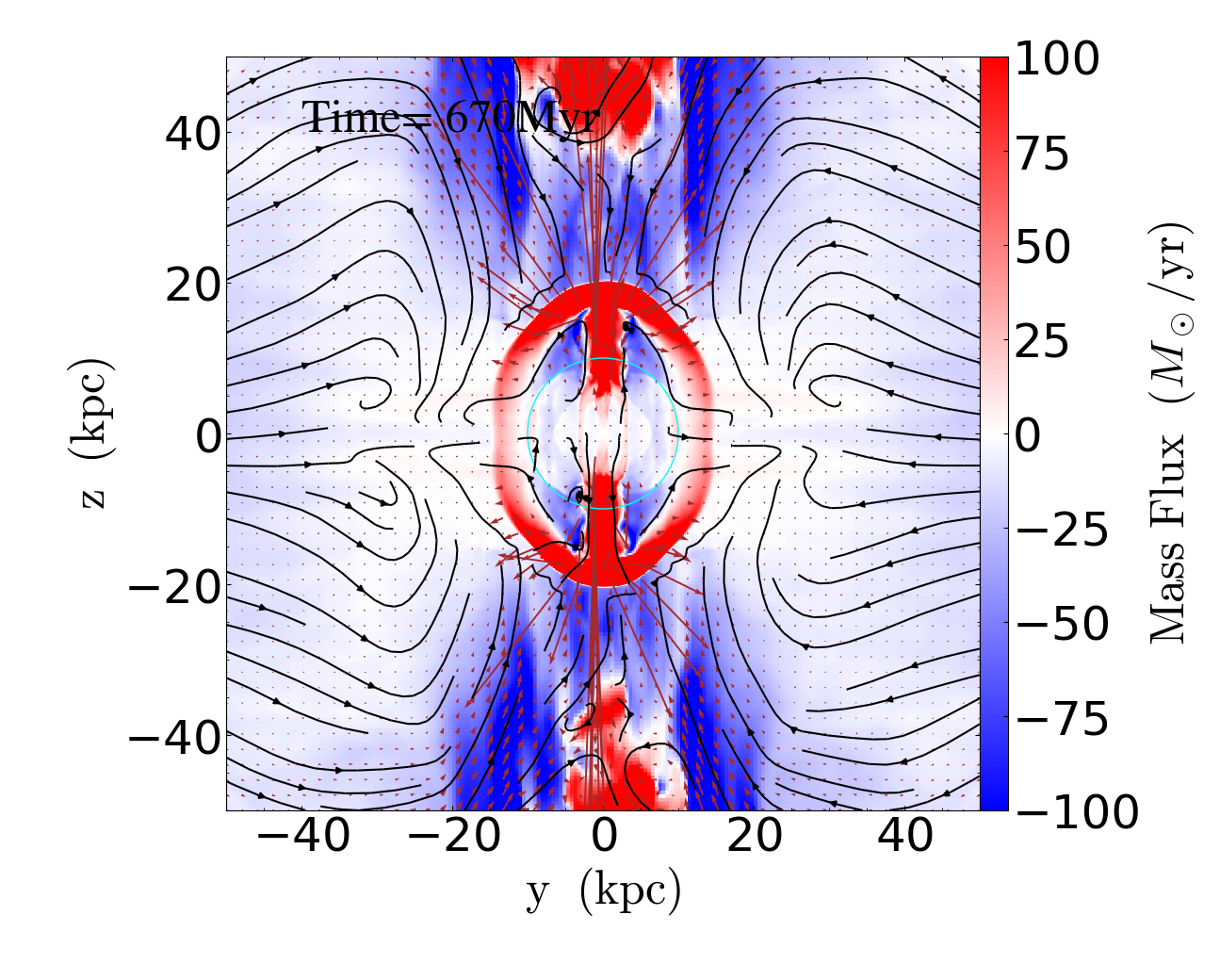}
\caption{Circulation of entropy and gas mass in the fiducial MPG simulation as it enters its low-power state at $t=600$, 650, and 670~Myr.  All figure elements are as in Figure \ref{fig:SPG_fid_circ}. The large-scale circulation pattern fundamentally changes when AGN feedback weakens. Lower-entropy gas lifted along the jet axis to altitudes that exceed 50~kpc starts to sink back toward the center.  At 650~Myr, inflow is both equatorial and polar.  The AGN then propels a weak outburst, driving an acoustic impulse that reaches $\sim 15$~kpc at 
670~Myr.}
 \label{fig:MPG_fid_circ_3}
\end{figure*}

\begin{figure*}[!t]
\centering
 \includegraphics[width=2.2in,height=2.2in]{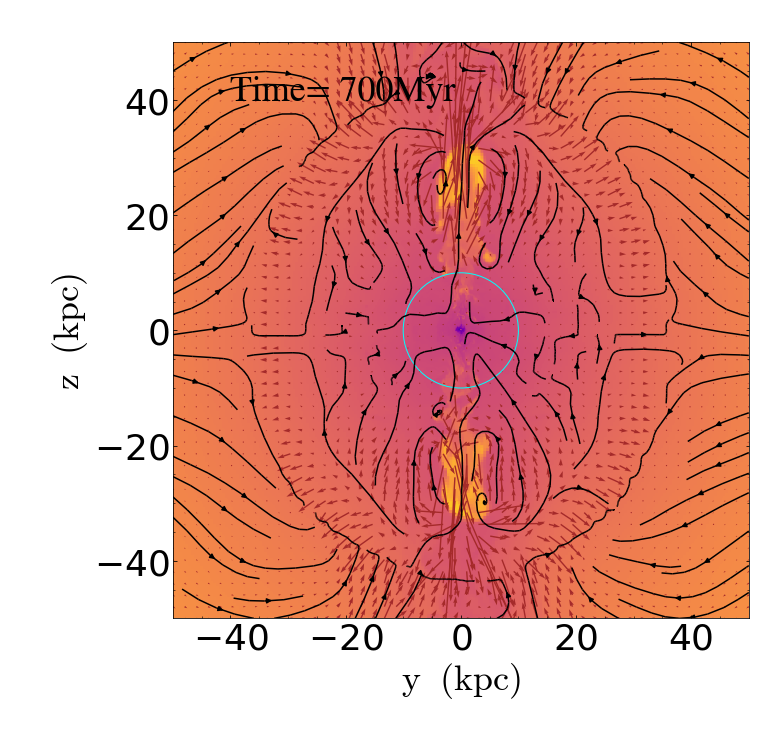}
  \includegraphics[width=2.2in,height=2.2in]{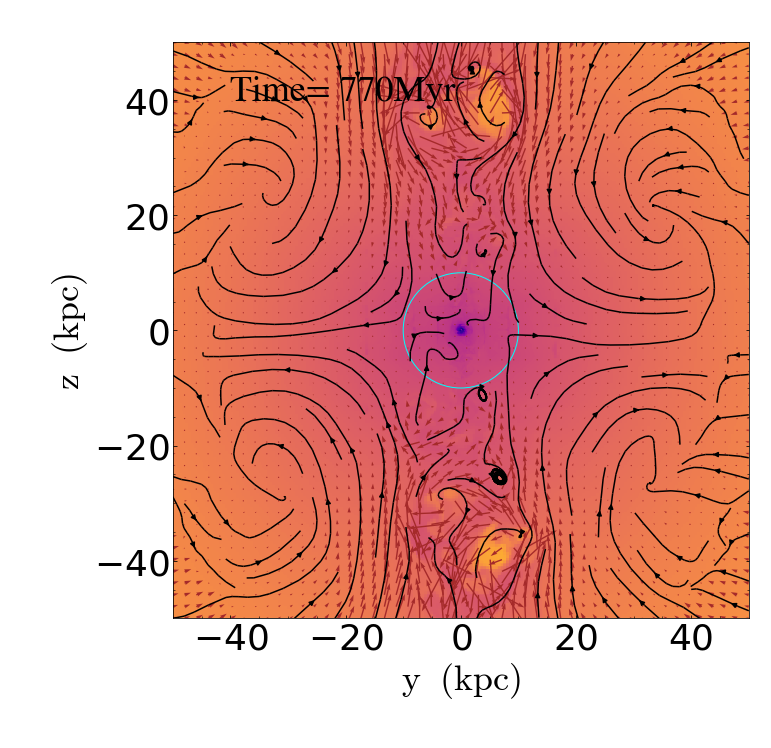}
   \includegraphics[width=2.4in,height=2.2in]{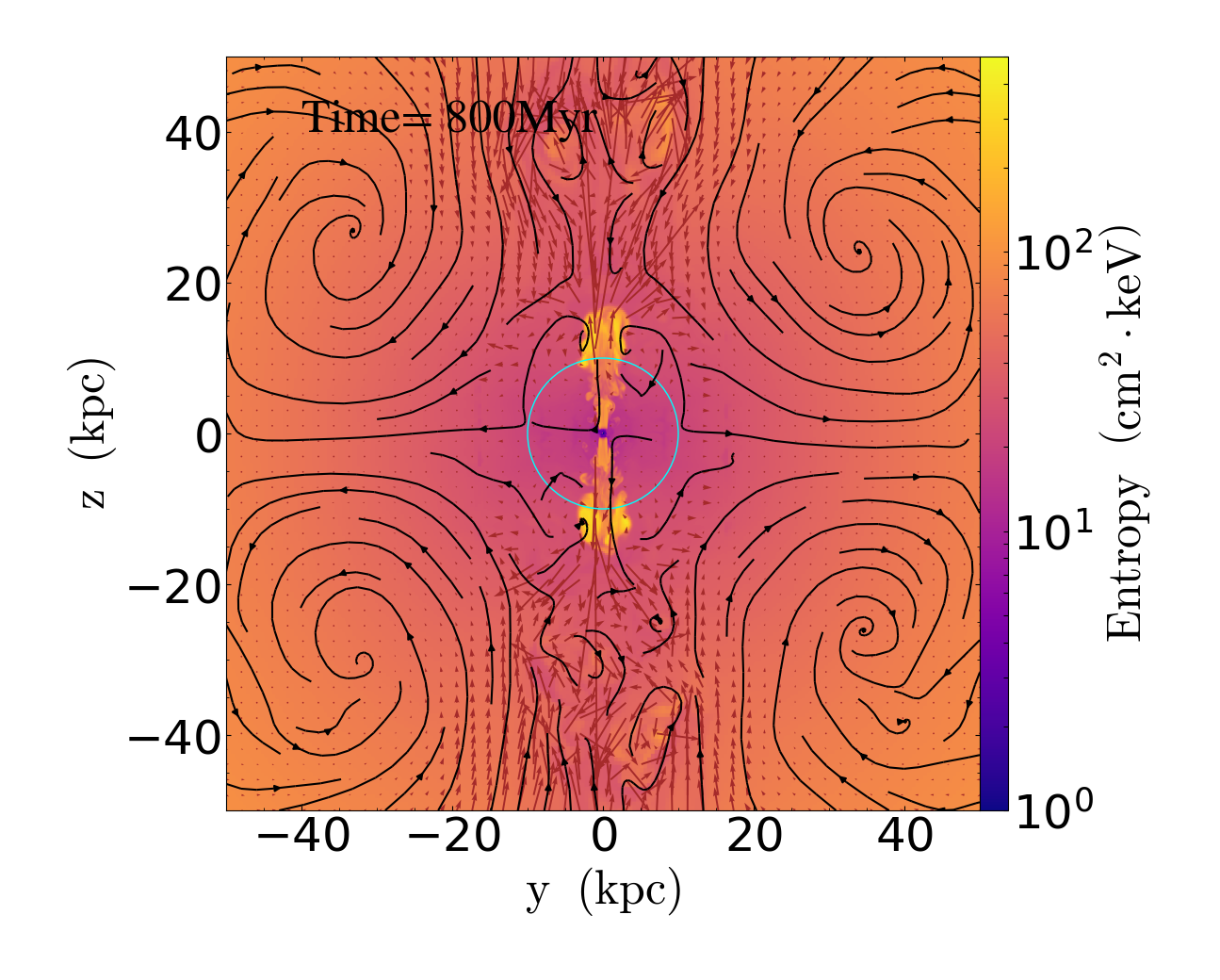}
  \includegraphics[width=2.2in,height=2.2in]{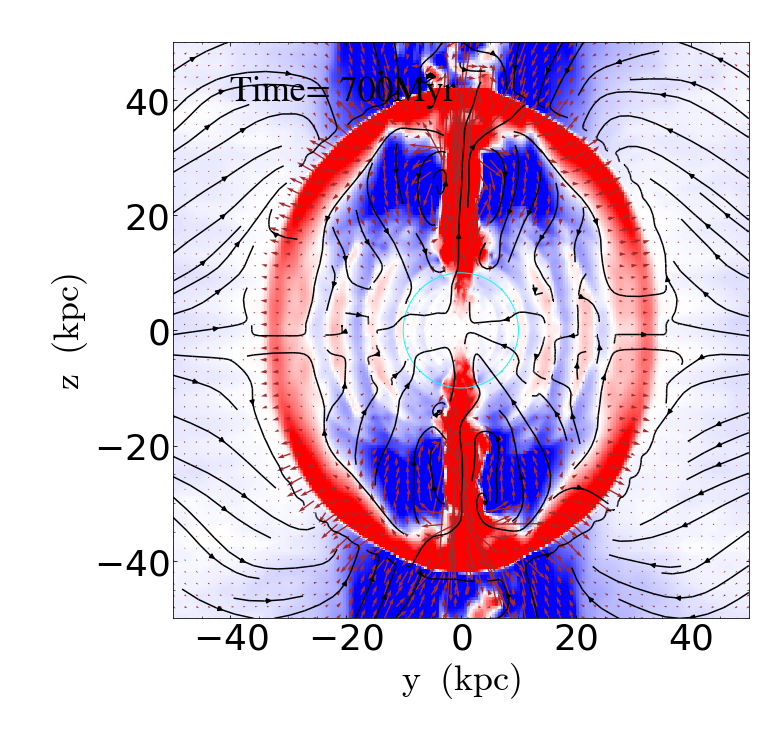}
  \includegraphics[width=2.2in,height=2.2in]{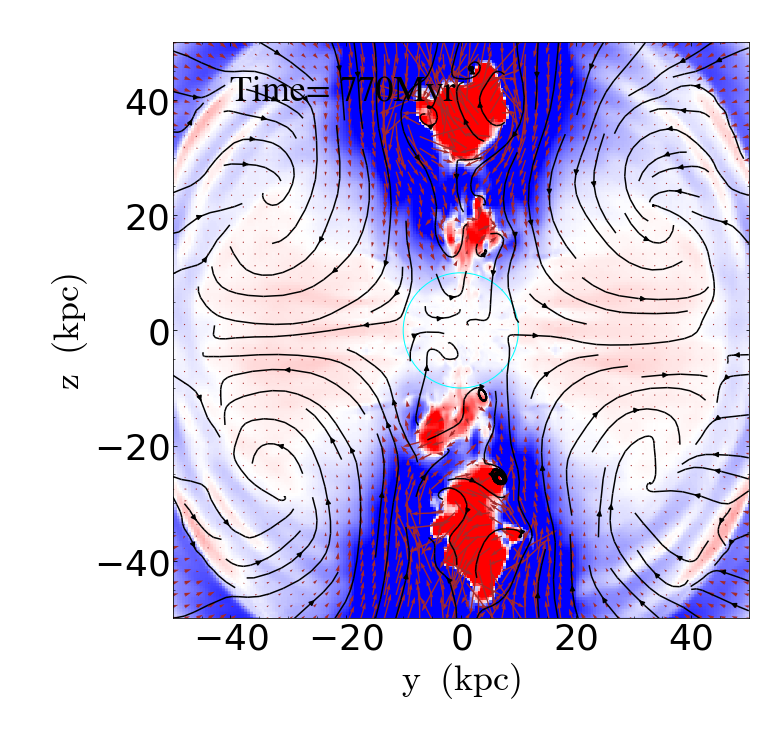}
   \includegraphics[width=2.4in,height=2.2in]{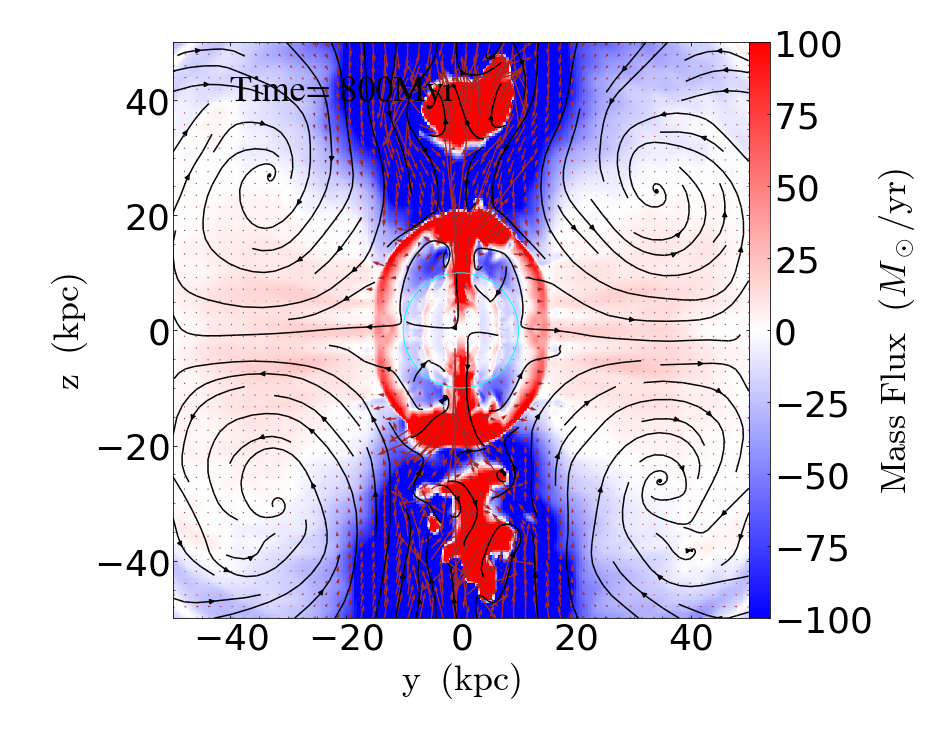}
\caption{Circulation of entropy and gas mass in the fiducial MPG simulation at $t=700$, 770, and 800~Myr during its low-power state.  All figure elements are as in Figure \ref{fig:SPG_fid_circ}. The AGN impulse propagating outward at 700~Myr fails to stop infall of lower-entropy gas along the jet axis.  High-entropy bubbles can be see rising outward at 770~Myr, but by that time an equatorial outflow has begun, as low-entropy gas falling inward along the poles squeezes out higher-entropy gas.  At 800~Myr, another weak AGN outburst is propating outward.
}
 \label{fig:MPG_fid_circ_4}
\end{figure*}

After 500~Myr, while AGN power is low, the atmospheric circulation pattern changes as shown in Figures \ref{fig:MPG_fid_circ_3} and \ref{fig:MPG_fid_circ_4}.  Lower-entropy gas that circulation has transported upward along the polar axis is no longer being suspended there by a strong outflow.  It therefore sinks back toward the center of the potential well, reversing the sense of circulation.  Between 650~Myr and 670~Myr, condensation of gas near the center feeds a minor AGN outburst that produces high-entropy bubbles at $\sim 10$~kpc, but those bubbles rise through the low-entropy gas that is falling in along the polar axis.  By 770~Myr, the flow pattern outside of 10~kpc is primarily inward along the polar axis and outward along the equator.

The simulation's behavior at 500~Myr through 800~Myr shows that the strong initial outburst has excited a large-scale internal gravity wave with considerable power in the quadrupole mode. By $\sim 800$~Myr, infall of low-entropy gas perpendicular to the jet axis has produced symmetric eddies with a size scale $\sim 30$~kpc. Detections of similar eddies in real galactic atmospheres would provide evidence for prior uplift episodes and stimulation of large-scale circulation by strong jet outbursts.

\subsection{Dependence on Jet Width}
\label{sec:opang}

Significant narrowing of the jet opening angle in our simulations affects how AGN feedback couples with the CGM.  Figure~\ref{fig:entropy} has already shown how the radial entropy profile changes when jet width is reduced by a factor of four in the SPG simulations (see also Paper I).  Entropy near 1~kpc remains closer to the inital state, in better agreement with observations, but a large entropy excess builds up at $\sim 5$--20 kpc.  This section discusses the role of circulation in producing that entropy excess.

\begin{figure*}[!t]
\centering
  \includegraphics[width=2.2in,height=2.2in]{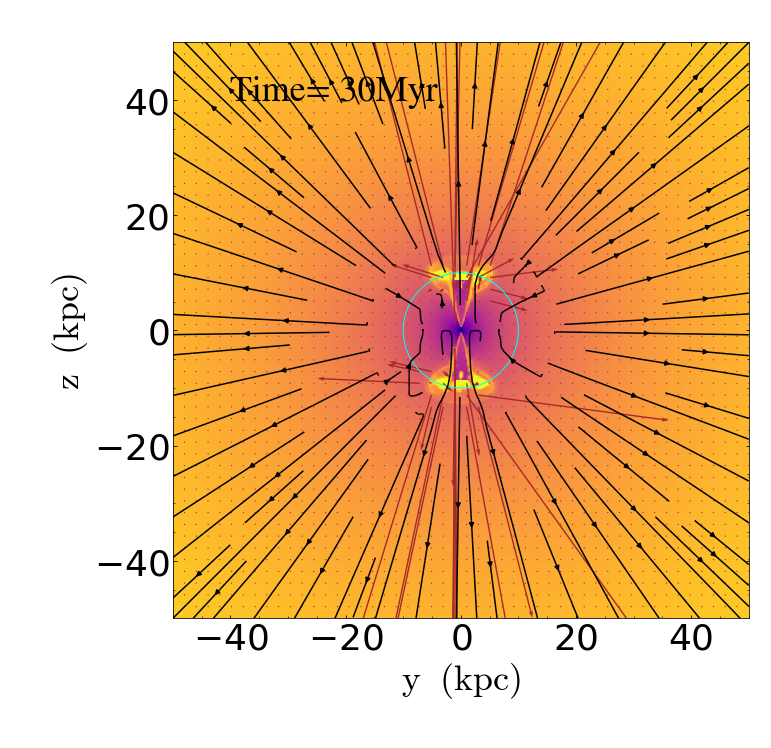}
  \includegraphics[width=2.2in,height=2.2in]{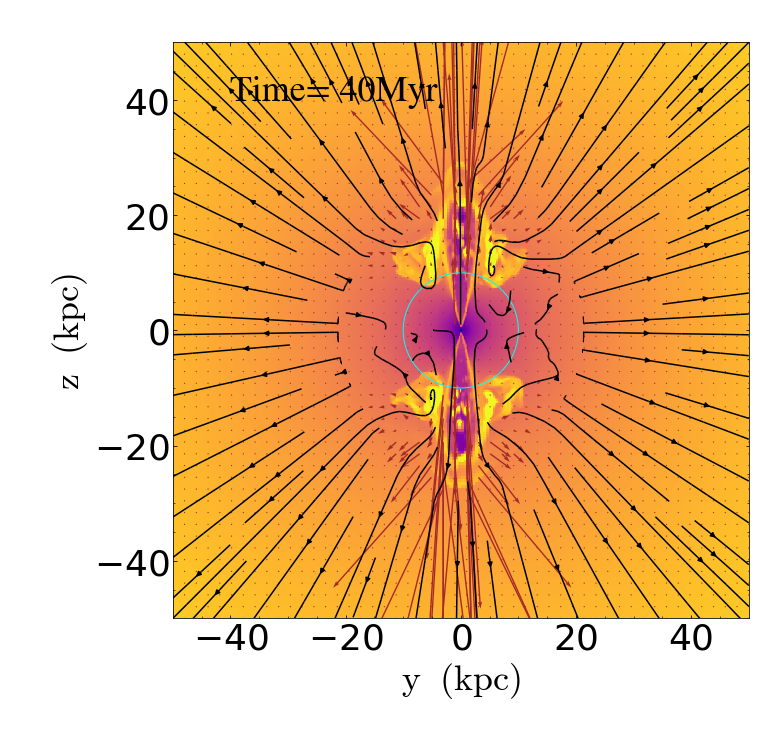}
  \includegraphics[width=2.4in,height=2.2in]{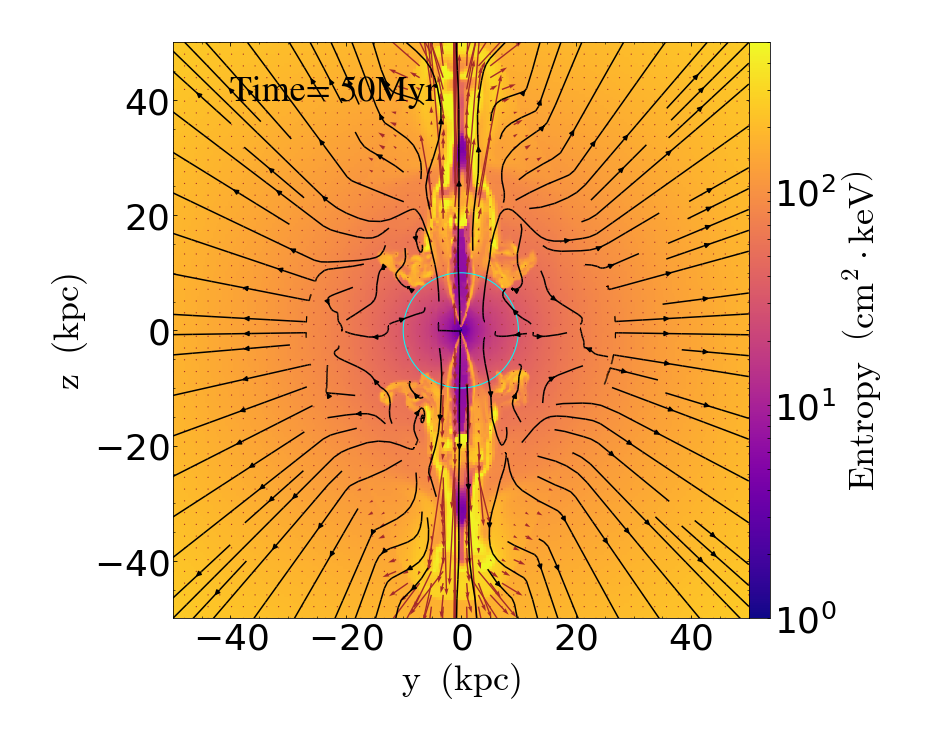}
  \includegraphics[width=2.2in,height=2.2inh]{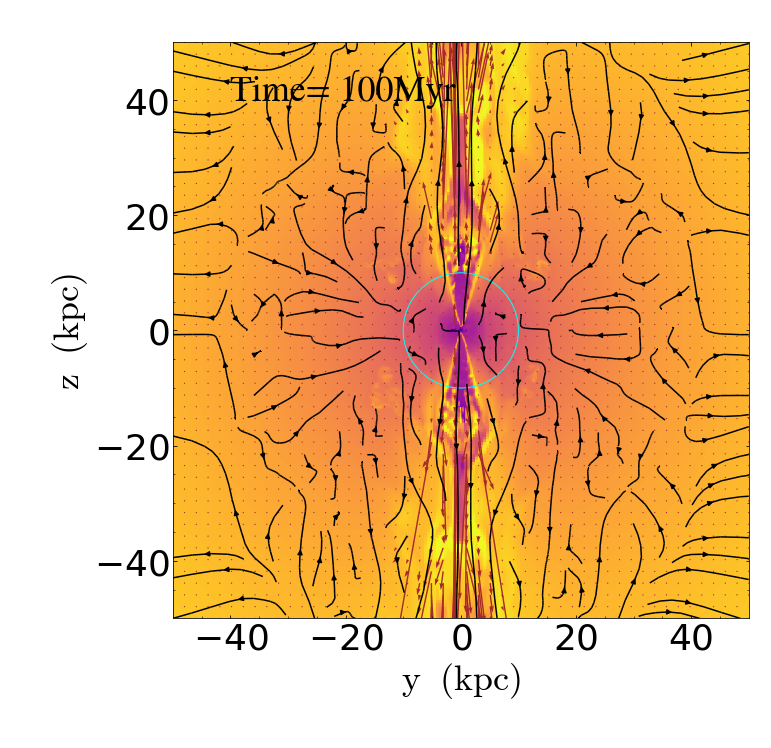}
  \includegraphics[width=2.2in,height=2.2in]{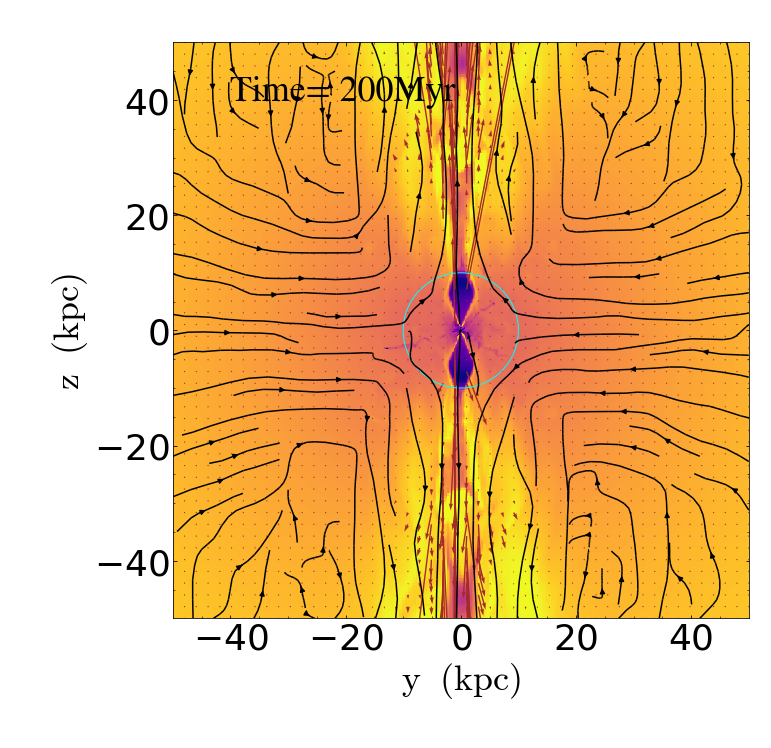}
  \includegraphics[width=2.4in,height=2.2in]{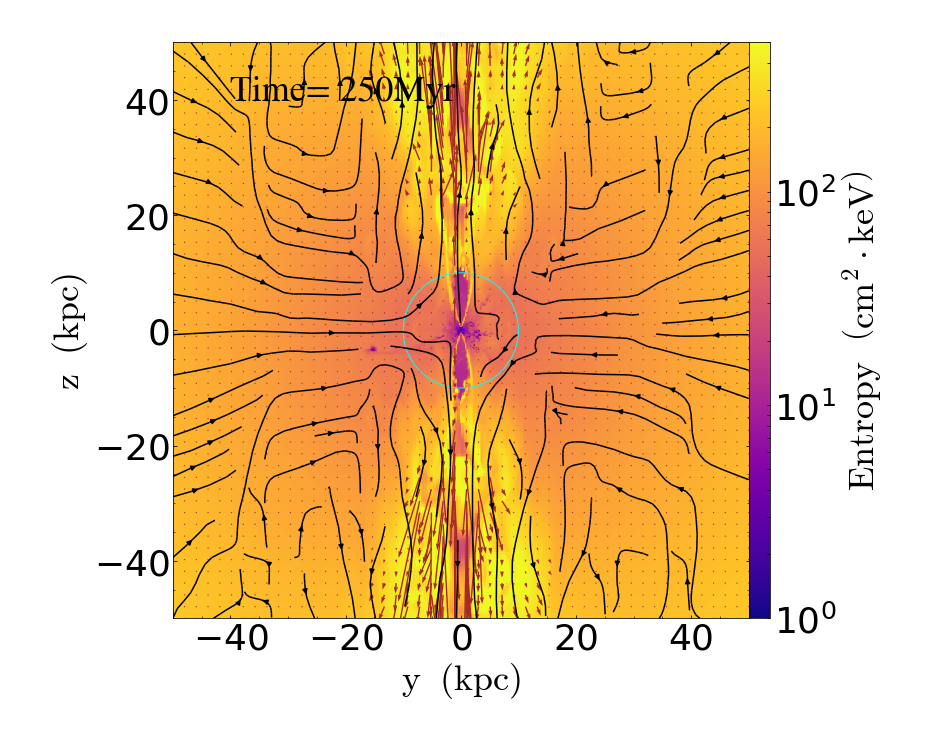}
  \includegraphics[width=2.2in,height=2.2in]{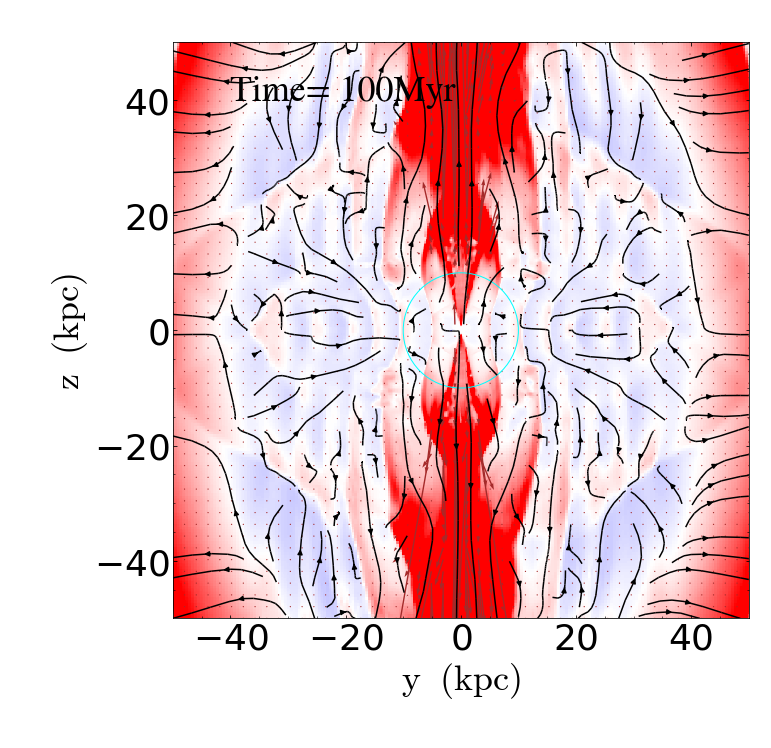}
  \includegraphics[width=2.2in,height=2.2in]{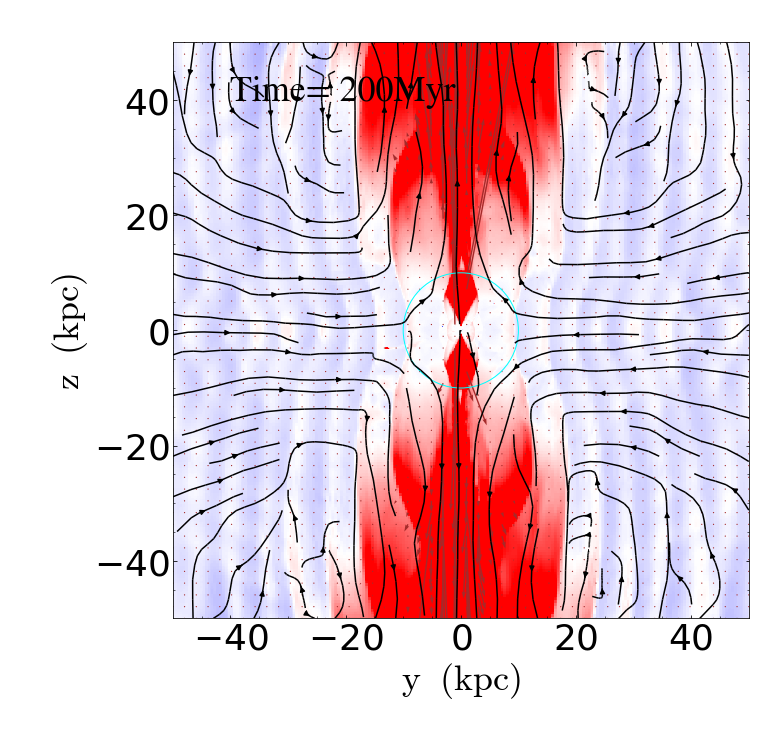}
  \includegraphics[width=2.4in,height=2.2in]{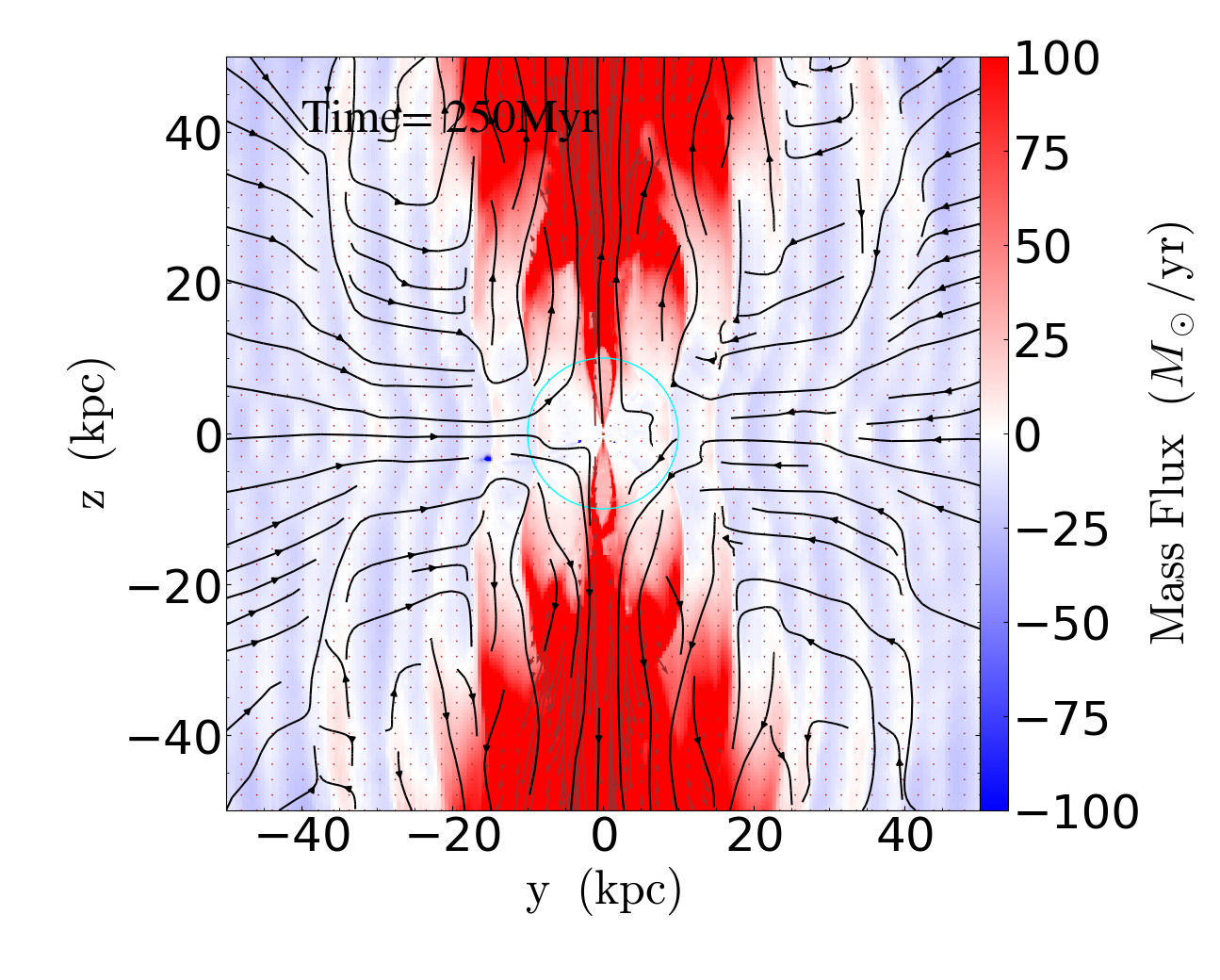}
\caption{Circulation of entropy and gas mass in the quarter-angle SPG simulation.  \textit{Top two rows:}  Azimuthally averaged entropy at t=30, 40, 50, 100, 200, and 250~Myr. Black arrows show streamlines while brown arrows show relative mass flux.  Cyan circles show a radius of $r\lesssim10$ kpc around the galaxy. \textit{Bottom row:} Azimuthally averaged radial mass flux (shown as $4 \pi r^2 \rho v_r$) at t=100, 200, and~250 Myr.  Black arrows again show streamlines while brown arrows show the relative magnitude of the Bernoulli energy flux. Comparison with Figure~\ref{fig:SPG_fid_circ} shows that high-entropy bubbles form at greater distances because gas in the jet propagates farther before being shock-heated to higher entropy.  Also, the large-scale circulation pattern is more persistent, continuing through $t=250$~Myr.
}
 \label{fig:SPG_quarter_circ}
\end{figure*}

Figure~\ref{fig:SPG_quarter_circ} illustrates entropy and mass circulation during the first 250~Myr of the quarter-angle SPG simulation.  Comparing the top three panels in that figure to the top three in Figure~\ref{fig:SPG_fid_circ} shows that the narrower jets unsurprisingly drill more effectively through the CGM, propagating to a greater distance and thermalizing less of their energy at smaller radii.  The middle row of panels, depicting the period from 100~Myr through 250~Myr, shows that large-scale circulation then develops and transports higher-entropy gas inward to within 10~kpc along directions perpendicular to the jet axis. However, the circulation pattern does not lead to a large enough entropy rise at smaller radii to promptly flip the AGN into a low-power state.  There is still enough low-entropy gas fueling the AGN at 250~Myr to maintain it in a high-power state.  As a result, large-scale bipolar circulation continues beyond 250~Myr without forcing AGN feedback into a low-power state. Instead of settling into a low-power state, AGN feedback in this simulation suppresses star formation by continually consuming low-entropy gas and ejecting it to large radii through the jets.

\section{Discussion}
\label{sec:disc}

Sections \ref{sec:AGN} and \ref{sec:circ} have shown that AGN feedback in this paper's simulations sometimes self-regulates by causing large-scale circulation. During periods of strong AGN feedback, powerful outflows lift low-entropy gas out of the central region along the jet axis, allowing higher-entropy gas to flow inward perpendicular to the jet axis. Those periods of strong feedback come to an end when the central gas density has dropped enough for stellar heating to exceed radiative cooling within the central few kiloparsecs. Then the simulations enter a low-power mode in which less powerful AGN outbursts allow stellar heating to maintain a nearly steady state within the galaxy (see Figures \ref{fig:lum_spg} and \ref{fig:lum_mpg}), consistent with the black-hole feedback valve scenario of \citet{voit2020}.

Should the large-scale circulation produced in our simulations be considered realistic? The answer to that question depends critically on features of the AGN's radial momentum output that are currently poorly constrained by observations. Two key assumptions of our AGN feedback implementation determine the momentum flux of our outflows: the mass-energy conversion efficiency $\epsilon_{\rm AGN}$ and the jet opening angle $\theta_{\rm jet}$. The resulting momentum flux along the jet axis is then
\begin{equation}
    \langle \rho v_r^2 \rangle_{\rm jet} \: \sim \: 
      \frac {1} {\pi \theta_{\rm jet}^2 (2 \epsilon_{\rm AGN})^{1/2}}  
      \frac {\dot{E}_{\rm AGN}} {cr^2} 
      \; \; ,
\end{equation}
given $v_r \approx (2 \epsilon_{\rm AGN})^{1/2} c$ at the base of the jets. 

In our AGN feedback algorithm, the assumption of mass conservation (i.e., a jet mass output equal to $\dot{M}_{\rm acc}$), paired with a relatively low efficiency factor ($\epsilon_{\rm AGN} = 10^{-4}$), leads to a large momentum output capable of driving large-scale circulation.  However, the mass-energy conversion efficiency of real AGN feedback is likely to be considerably greater than $10^{-4}$, and the mass output in an AGN's jets could be orders of magnitude different from $\dot{M}_{\rm acc}$. For example, comparisons of observed black hole masses with CGM binding energy suggest that a fraction $\sim 10^{-2.5}$ of the black hole rest-mass energy is necessary to lift the CGM (\citealt{kormendy2013,voit15L,davies2019}).
And using that value for $\epsilon_{\rm AGN}$ in our simulations would reduce the jet momentum flux by more than an order of magnitude. 

Our simulations employ a smaller efficiency factor to account for additional uncertainties in the accretion process. Primary among those is uncertainty in the fraction of cold gas within the central 0.5~kpc that would ultimately accrete onto the central black hole. In a more realistic model, much of that gas would form stars before reaching the black hole, and another large fraction may be driven out of the central 0.5~kpc by AGN feedback itself.

Ultimately, observational detections of CGM circulation patterns similar to what our simulations produce might provide evidence for AGN outflows with a large momentum flux.  However, there are other reasons to believe that the simulations presented here require further revision and improvement. As noted in \S \ref{sec:introduction}, the current implementation of AGN feedback leads to entropy-profile anomalies at 1--10~kpc. Therefore, we will focus the rest of this discussion on how those anomalous features depend on the jet momentum flux.

\subsection{Jet Momentum Flux}

The radial domain over which jet impulses drive large-scale circulation is set by the condition $\langle \rho v_r^2 \rangle_{\rm jet} \gg P_{\rm CGM}$. Jets with a momentum flux much greater than the ambient pressure tend to drill through the ambient gas while transferring radial momentum to it. But a momentum flux comparable to the ambient pressure corresponds to an outflow that is decelerating to subsonic speeds and no longer drills effectively to larger radii.

\begin{figure}
 \includegraphics[width=0.45\textwidth]{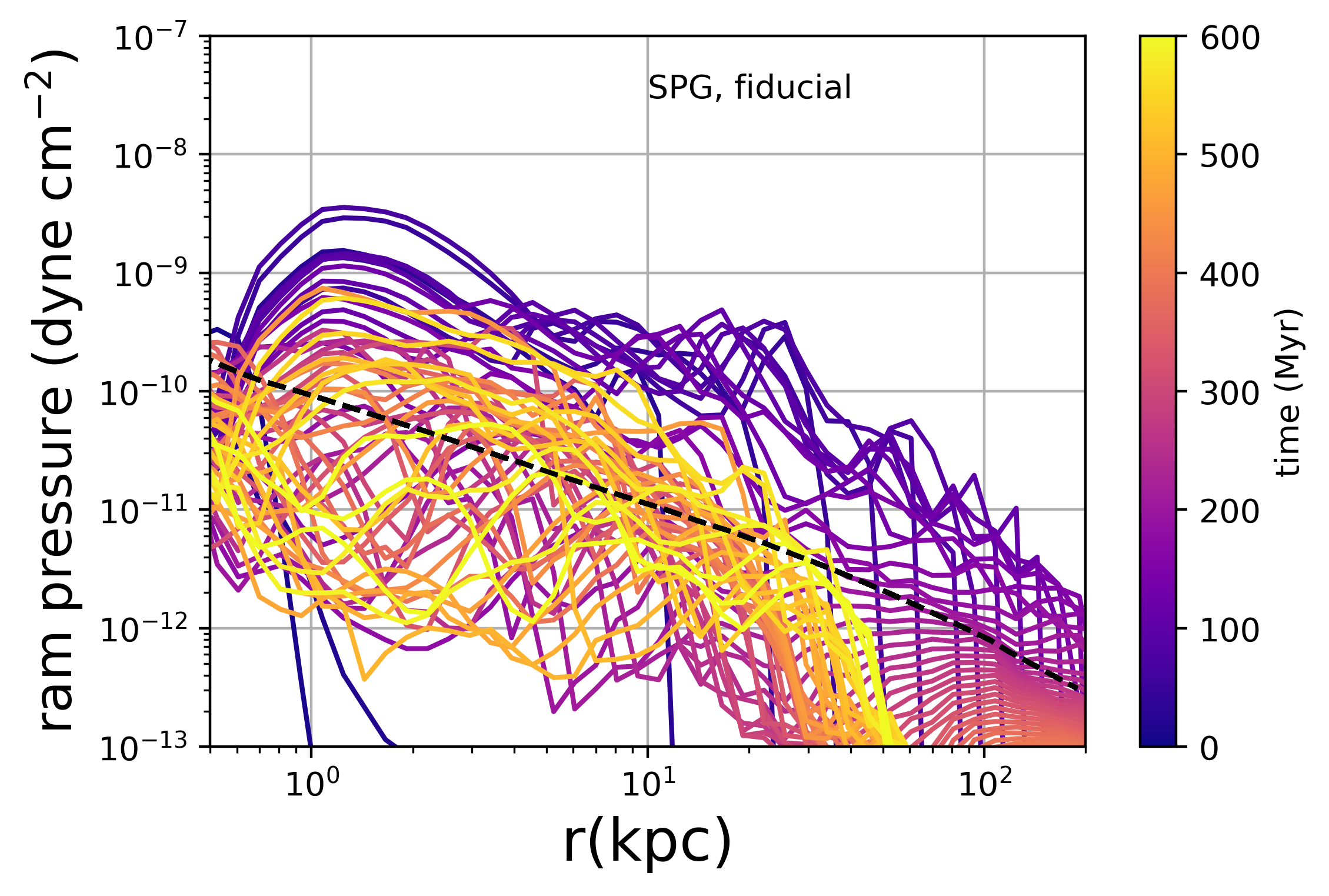}
 \includegraphics[width=0.45\textwidth]{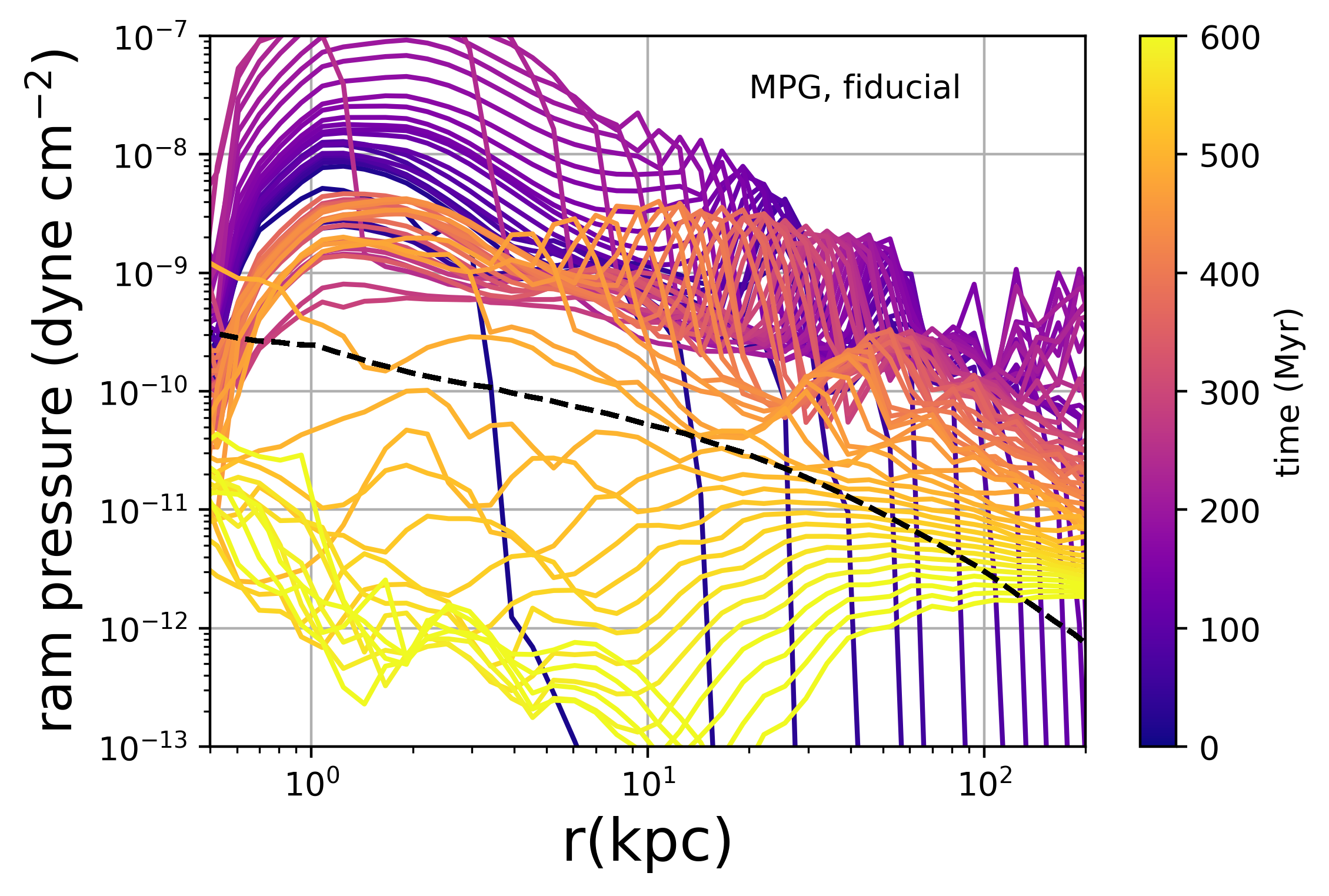}
 \includegraphics[width=0.45\textwidth]{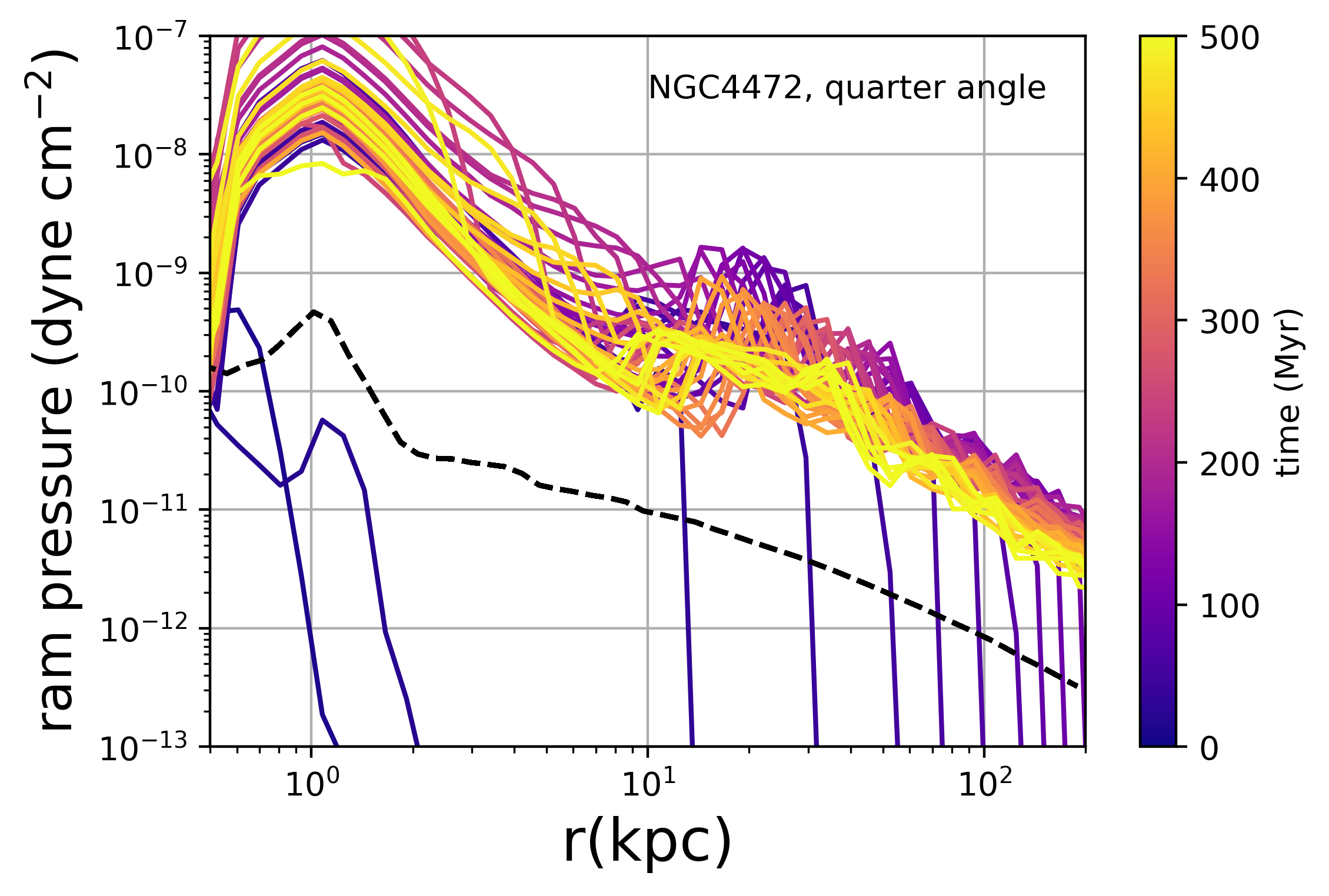}
 \caption{Comparisons between jet ram pressure (colored lines) and median thermal pressure profiles (dashed black line). In the fiducial SPG simulation (top panel) jet ram pressure exceeds the median thermal pressure profile out to beyond 100~kpc early in the strong initial jet outburst phase and later drops below it. In the fiducial MPG simulation (middle panel), jet ram pressure behaves similarly. But narrowing the jets, as in the quarter-angle SPG simulation, raises their ram pressure, enabling it to exceed thermal pressure out to beyond 100~kpc for at least 500~Myr.}
 \label{fig:Pram}
\end{figure}

Figure \ref{fig:Pram} shows how ram pressure along the jet axis compares with the ambient gas pressure during the initial outbursts in three of our simulations. Dashed black lines show the median pressure profile in each simulation, and temporal deviations from the median profile are much smaller than the temporal changes in ram pressure. All three panels show that the initial period of strong AGN feedback produces jets capable of drilling well beyond 100~kpc before slowing to subsonic speeds. This feature of strong kinetic feedback is consistent with the circulation patterns evident in Figures \ref{fig:SPG_fid_circ}, \ref{fig:MPG_fid_circ_1}, \ref{fig:MPG_fid_circ_2}, and \ref{fig:SPG_quarter_circ}.

The colored curves in the top panel of Figure \ref{fig:Pram} show that ram pressure in the fiducial SPG simulation drops below the median gas pressure after about 200~Myr.  That drop corresponds to a concurrent decline in jet power by an order of magnitude (see Figure \ref{fig:lum_mpg}). As a result, the weaker outbursts characteristic of the low-power feedback mode do not have a great enough momentum flux to drill much beyond 10~kpc. Note that raising the AGN efficiency parameter to $\epsilon_{\rm AGN} = 10^{-2.5}$ would also lower the jet momentum flux by more than an order of magnitude. In that case the resulting jets would have difficulty drilling to large radii, presumably eliminating the large-scale circulation pattern.

The colored curves in the middle panel of Figure \ref{fig:Pram} show that ram pressure in the fiducial MPG simulation eventually drops below the median gas pressure after about 400~Myr, also corresponding to a large drop in AGN power (see Figure \ref{fig:lum_mpg}). In this case, a power decline of roughly two orders of magnitude is needed for the jet momentum flux to drop below the ambient pressure. When that happens, weak outbursts no longer have enough momentum to drill much past $\sim 10$~kpc.

The bottom panel of Figure \ref{fig:Pram} tells a different story. It shows the quarter-angle SPG simulation, in which only $\theta_{\rm jet}$ differs from the fiducial SPG simulation. Reducing the jet opening angle by a factor of four increases the jet momentum flux by a factor of 16, which enables the jets to drill much more easily through the ambient atmosphere. The result is a large-scale circulation pattern that is both more spatially extended and longer lasting than in the fiducial SPG simulation.

\subsection{Regulation via Circulation}

With those features of momentum flux in mind, we now outline how large-scale circulation couples AGN feedback to the CGM in our simulations, while keeping a close eye on how the assumptions built into the simulations influence their behavior. 

\subsubsection{Natural Cooling Flow Rate}

The first thing to recognize is that each simulated atmosphere has a characteristic cooling rate determined by the ratio $K(r)/r$. Cooling depends on the entropy equation
\begin{equation}
    \frac {d \ln K} {dt} = - \frac {1} {t_{\rm cool}}
    \; \; ,
\end{equation}
where $t_{\rm cool}$ is the cooling time of the gas, and a cooling flow moving inward with a radial velocity $v_r < 0$ has an entropy gradient
\begin{equation}
    \alpha_K \equiv \frac {d \ln K} {d \ln r} = \frac {r} {v_r} \frac {d \ln K} {dt}
    \; \; .
\end{equation}  
Together, those equations yield 
\begin{eqnarray}
  \dot{M}_{\rm cool} & \: = \: & \frac {4 \pi r^3 \rho} 
                                 {\alpha_K t_{\rm cool}} \\
        & = & \frac {8 \pi \mu m_p} {3 \alpha_K}
              \left( \frac {n_e} {n_p} \right)
              \left[ (kT)^2 \Lambda \right]
              \left( \frac {K} {r} \right)^{-3}
    \; \; .
\end{eqnarray}
Plugging in values typical of the galactic atmospheres we are simulating then gives
\begin{equation}
  \dot{M}_{\rm cool} \approx  0.6 \, M_\odot \, {\rm yr}^{-1} \, 
        \left( \frac {K/r} {5 \, {\rm keV \, cm^2 \, kpc^{-1}}} \right)^{-3}
\end{equation}
for $kT \approx 1 \, {\rm keV}$ and $\Lambda \approx 3 \times 10^{-23} \, {\rm erg \, cm^3 \, s^{-1}}$. 

\subsubsection{Initial AGN Response}

The second thing to recognize is that the simulations have a natural AGN power output determined by $\dot{M}_{\rm cool}$ and $\epsilon_{\rm AGN}$. At the beginning of each simulation, central cooling triggers strong AGN feedback within 30~Myr because the initial cooling time of the gas at small radii is $\sim 10$--20~Myr. A steady cooling flow then supplies gas to the central engine at approximately the rate $\dot{M}_{\rm cool}$, resulting in a power output
\begin{equation}
 \epsilon_{\rm AGN} \dot{M}_{\rm acc} c^2 
   \: \sim \: 
           6 \times 10^{42} \, {\rm erg \, s^{-1}} \, 
           \left( \frac {\dot{M}_{\rm cool}} {1 \, M_\odot \, {\rm yr}^{-1}} \right)
\end{equation}
given $\epsilon_{\rm AGN} = 10^{-4}$.

In our SPG simulations (with $K/r \approx 3.2 \, {\rm keV \, cm^2 \, kpc^{-1}}$) the natural cooling rate is initially
\begin{equation}
    \dot{M}_{\rm cool} \approx ( 2.4 \, M_\odot \, {\rm yr}^{-1} ) 
    \; \; .
\end{equation}
Steady fuelling at that rate then results in a time-averaged power output 
\begin{equation}
    \dot{E}_{\rm AGN} \approx (1.4 \times 10^{43} \, {\rm erg \, s^{-1}}) 
    \; \; .
\end{equation}
The instantaneous power output fluctuates because it depends on the amount of cold gas within the central 0.5~kpc, and therefore upon the inhomogeneity of the inner cooling flow at any given moment. However, the smoothed AGN power changes slowly (see Figure \ref{fig:lum_spg}), because it is limited by the natural cooling flow rate of the hot gas.

The natural cooling rate of the MPG atmosphere (with $K/r \approx 1.2 \, {\rm keV \, cm^2 \, kpc^{-1}}$) is much greater and results in far more AGN power, at least at the outset.  For the initial MPG entropy profile, the natural cooling rate is 
\begin{equation}
    \dot{M}_{\rm cool} \approx ( 42 \, M_\odot \, {\rm yr}^{-1} ) 
    \; \; .
\end{equation}
And the resulting time-averaged power output is 
\begin{equation}
    \dot{E}_{\rm AGN} \approx (2.4 \times 10^{44} \, {\rm erg \, s^{-1}}) 
\end{equation}
(see Figure \ref{fig:lum_mpg}).

\subsubsection{Characteristic Momentum Flux}

The final thing determining the initial AGN momentum flux is the jet opening angle $\theta_{\rm jet}$.  In our fiducial simulations, that opening angle results in 
\begin{equation}
    \langle \rho v_r^2 \rangle_{\rm jet} \: \sim \: 
      5 \times 10^{-9} \, r_{\rm kpc}^{-2} \, {\rm erg \, cm^{-3}} 
\end{equation}
for the SPG case and
\begin{equation}
    \langle \rho v_r^2 \rangle_{\rm jet} \: \sim \: 
      1 \times 10^{-7} \, r_{\rm kpc}^{-2} \, {\rm erg \, cm^{-3}} 
\end{equation}
for the MPG case. Reducing the jet opening angle by a factor of 4 raises the characteristic momentum flux in the quarter angle SPG simulation to
\begin{equation}
    \langle \rho v_r^2 \rangle_{\rm jet} \: \sim \: 
      1 \times 10^{-7} \, r_{\rm kpc}^{-2} \, {\rm erg \, cm^{-3}} 
\end{equation}
In all of these expressions, $r_{\rm kpc}$ represents distance from the center of the galaxy in units of kiloparsecs.  These crude estimates compare reasonably well with the upper ram pressure curves in the panels of Figure \ref{fig:Pram}, except for the short-term fluctuations to unusually high AGN power not captured by the steady-state estimates.

\subsubsection{Circulation versus Cooling}

Circulation alone cannot raise the entropy of ambient galactic gas unless the radial mass flow associated with circulation ($\dot{M}_{\rm circ}$) exceeds the natural cooling rate of the ambient gas ($\dot{M}_{\rm cool}$).  Otherwise radiative cooling would reduce the entropy of circulating gas on a timescale $t_{\rm cool}$ shorter than the circulation timescale $r/v_{\rm circ}$ determined by the average radial speed $v_{\rm circ}$ of circulating gas.  That requirement places a lower limit $v_{\rm circ} > r / t_{\rm cool}$ on the typical speed of radial circulation. For an atmosphere with $kT \approx 1$~keV and $\Lambda \approx 3 \times 10^{-23} \, {\rm erg \, cm^3 \, s^{-1}}$, the relationship between entropy and cooling time is
\begin{equation}
    t_{\rm cool} \approx 5 \, {\rm Myr} \left( \frac {K} {\rm keV \, cm^2} \right)^{3/2}
\end{equation}
with $t_{\rm cool} \propto T^{-1/2} \Lambda^{-1}$. The circulation condition therefore translates to
\begin{equation}
    v_{\rm circ} \gtrsim 34 \, r_{\rm kpc}^{-1/2} \, {\rm km \, s^{-1}}
\end{equation}
for the SPG simulations and
\begin{equation}
    v_{\rm circ} \gtrsim 146 \, r_{\rm kpc}^{-1/2} \, {\rm km \, s^{-1}}
\end{equation}
for the MPG simulation.

\subsubsection{Lateral Momentum Transport}

Satisfying the circulation condition requires radial momentum to propagate laterally away from the jets and into the ambient medium. Analytical estimates of this non-linear process are challenging, and lateral momentum propagation in numerical simulations depends on viscosity assumptions built into the code. However, comparing the fiducial SPG simulation with the quarter-angle SPG simulation demonstrates that the simulation with narrower jets has more difficulty lifting low-entropy gas out of the central few kiloparsecs, presumably because the radial momentum introduced by the jets has not propagated as far from the jet axis. 

In Figure \ref{fig:entropy}, the entropy level of gas within 3~kpc is systematically lower in the quarter-angle SPG simulation during the first 500~Myr, indicating that lateral momentum transfer in the quarter-angle SPG simulation is not lifting out low-entropy gas on a timescale shorter than $t_{\rm cool}$. Meanwhile, ambient entropy at 5--20~kpc is rising, indicating that lateral momentum transfer is sufficiently rapid to drive effective circulation farther out. Furthermore, the orange and yellow lines depicting the entropy profiles from 400~Myr to 500~Myr in the quarter-angle SPG simulation suggest that the low-entropy gas within 3~kpc is gradually being depleted by accretion onto the central black hole. Depletion of the inner gas allows the higher-entropy gas originally transported within 10~kpc by circulation to gradually sink further inward, ultimately boosting the entropy level at 1~kpc to $\sim 10 \, {\rm keV \, cm^2}$.

\subsubsection{Locus of Heating-Cooling Equality}

Boosting entropy at 1~kpc to that level is significant because it allows stellar heating to exceed radiative cooling at $\sim 1$~kpc. \citet{voit2020} showed that the locus in the $K$--$r$ plane along which stellar heating locally equals radiative cooling approximately corresponds to
\begin{equation}
 K_{\rm eq}(r) \approx (5 \, {\rm keV \, cm^2} ) \, \sigma_{240}^{4/3} r_{\rm kpc}^{2/3}
    \; \; ,
\end{equation}
where $\sigma_{240}$ is the stellar velocity dispersion in units of $240 \, {\rm km \, s^{-1}}$. All of our simulations begin with entropy levels at 1~kpc below that locus.  The fiducial SPG exceeds it after 40~Myr and then starts to progress to a low-power state, but the quarter-angle SPG does not exceed it until $t \sim 500$~Myr.

\subsubsection{Raising Entropy at 1~kpc}

Once the ambient entropy level at 1~kpc rises to about $10 \, {\rm keV \, cm^2}$ the natural cooling rate at that radius falls to
\begin{equation}
    \dot{M}_{\rm cool} \approx 0.07 \, M_\odot \, {\rm yr}^{-1} 
    \; \; .
\end{equation}
The same natural cooling rate applies to an atmosphere with $5 \, {\rm keV \, cm^2}$ at 0.5~kpc because of how that rate depends on $K/r$.  AGN feedback power fueled by gas cooling at that rate is
\begin{equation}
    \dot{E}_{\rm AGN} \approx 4 \times 10^{41} \, {\rm erg \, s^{-1}}
    \; \; ,
\end{equation}
which is comparable to the time-averaged AGN power of the fiducial SPG simulation in its low-power state (see Figure \ref{fig:lum_spg}). Likewise, the fiducial MPG reaches $K \approx 10 \, {\rm keV \, cm^2}$ at 1~kpc by 500~Myr and settles into a temporary state with similar time-averaged power thereafter.

The upper panels of Figures \ref{fig:lum_spg} and \ref{fig:lum_mpg} show that stellar heating exceeds radiative cooling at 1~kpc during those low-power periods and also that cooling often exceeds heating within the central 0.5~kpc. Cold gas clouds can form in that central region and feed the central black hole.  However, the time-averaged rate at which they can form is limited by the natural cooling rate within 1~kpc, not the natural cooling-flow rate at larger radii. In other words, the entropy rise above $K_{\rm eq}$ at $\sim 1$~kpc has decoupled AGN fueling from the natural cooling-flow rate at larger radii. 

\subsubsection{Stellar Mass Loss}
In our simulation, the old stellar population in the central galaxy sheds gas at a rate $\sim 1 \, M_\odot \, {\rm yr}^{-1}$,
which greatly exceeds the natural cooling-flow rate associated with $K \sim 10 \, {\rm keV \, cm^2}$ at 1~kpc. The ejected gas is a net increase in gas mass in the simulation.
Most of the ejected stellar gas must therefore be pushed out of the galaxy during the low-power mode of AGN feedback. Stellar heating can accomplish that task on its own. But as discussed in \S \ref{sec:accumulation}, the formerly stellar gas swept out of the galaxy accumulates in the CGM at 30--100~kpc from the center, causing a slow rise in CGM pressure. In the fiducial MPG simulation, that slow rise in pressure forces the gas density to rise and entropy to drop at 1~kpc, causing radiative cooling to exceed stellar heating there at 1.1~Gyr (see Figure \ref{fig:lum_mpg}). AGN fueling is once again coupled to the natural cooling-flow rate at larger radii, initiating a second episode of strong AGN feedback and large-scale circulation.

\subsection{Achieving Greater Fidelity}

The observed entropy profiles of massive elliptical galaxies show that real AGN feedback can keep entropy levels at 1~kpc closer to 5~keV~cm$^2$, the level at which stellar heating approximately matches radiative cooling \citep{voit2015,voit2020}.  One of the reasons why our simulations have difficulty matching those observations is that the size scale of our AGN feedback mechanism, with its jets starting 1~kpc from the center, is comparable to the scale at which deviations from the observations are most pronounced in fiducial SPG simulations.  Notice, for example, the pronounced minima in $\tilde{L}_{\rm rad}$ at 1~kpc in the upper panels of Figure \ref{fig:lum_spg}.  A high-resolution simulation would allow the jet mechanism to be smaller in size, and perhaps would lead to more realistic results, but the computational cost would be considerably greater.  Also, the jets would need to drill outward from smaller radii without excessively flattening the inner entropy profile, through either dissipative heating or by driving excessive circulation.

Another possibility is that our simulations do not yet have the right combination of $\epsilon_{\rm AGN}$ and $\theta_{\rm jet}$.  Our comparison of the fiducial and quarter-angle SPG simulations has shown that making the jets too narrow can hamper their ability to raise the central entropy level and close the AGN feedback loop.  However, the low value of $\epsilon_{\rm AGN}$ used in our simulations results in jets that easily drill to large radii because of their large momentum flux. It is therefore possible that raising $\epsilon_{\rm AGN}$ while simultaneously reducing $\theta_{\rm jet}$ would produce jets that do not propagate too far before dumping their energy into the CGM while still being able to drill effectively through the inner 10~kpc of the galaxy's atmosphere. Inclusion of the weak magnetic fields certain to be present in the atmospheric gas is also likely to affect how jet momentum propagates away from the jet axis, with implications for AGN-driven CGM circulation that are difficult to predict without performing MHD simulations.

\subsection{Comparisons with Analogous Work}
\label{sec:comparison}
An early set of AGN feedback simulations with poorly coupled jets was presented by \citet{VernaleoReynolds_2006ApJ...645...83V}. Their work showed that narrow bipolar jets could drill out to distances well beyond 100~kpc without substantially reducing the cooling flow that fuels them.  In those simulations, the jets that failed to prevent strong cooling flows had an opening angle of $\theta_{\rm jet} = 15^\circ$ and efficiency factors of $\epsilon_{\rm AGN} = 10^{-5}$ and $\epsilon_{\rm AGN} = 10^{-4}$ (in our notation). Therefore, they are examples of outflows with large momentum flux per unit power, analogous to the ones in our quarter-angle SPG simulation.

\citet{VernaleoReynolds_2006ApJ...645...83V} performed additional simulations equivalent to cases with $\epsilon_{\rm AGN} = 10^{-2}$ and $\epsilon_{\rm AGN} = 10^{-1}$.  With these larger efficiency factors, bipolar jets were able to suppress strong cooling flows for longer periods of time, but not indefinitely. We therefore infer that a greater efficiency factor, which leads to jets with less momentum flux per unit power, results in feedback that does not drill as effectively through the ambient gas. The lower-momentum jets therefore dump their power into the surrounding atmosphere at smaller radii.

\citet{YangReynolds_2016ApJ...829...90Y} revisited these issues with a more sophisticated simulation, implementing cold-gas feedback and applying an efficiency factor equivalent to $\epsilon_{\rm AGN} = 10^{-3}$.  AGN feedback in that simulation succesfully self-regulated, with large-scale circulation playing an important role.  However, given that the accretion efficiency is an order of magnitude higher compared to our fiducial run, the resultant AGN jet is also lighter. Consequently, the atmospheric circulation found by \citet{YangReynolds_2016ApJ...829...90Y} is driven primarily by the buoyancy of heated gas rather than by direct momentum transfer, as in our simulations. Further, \citet{YangReynolds_2016ApJ...829...90Y} found that heating exceeded radiative cooling within the jet cone, while cooling exceeded heating outside of the jet cone. However, cooling did not become catastrophic because of circulation patterns resembling those we have found in our own simulations.

A more recent set of AGN feedback simulations by \citet{wang2019} is much more similar to the ones presented here, because \citet{wang2019} simulated galaxies nearly identical to our SPG and MPG, whereas \citet{VernaleoReynolds_2006ApJ...645...83V} and \citet{YangReynolds_2016ApJ...829...90Y} simulated the cores of massive galaxy clusters. Comparisons of the \citet{wang2019} simulations with observations show that they produce galaxies with long lasting self-regulated entropy profiles that more closely match real ones \citep{Frisbie_2020ApJ...899..159F}. That may be because \citet{wang2019} applied an efficiency factor equivalent to $\epsilon_{\rm AGN} = 10^{-2.3}$ and a precession angle of approximately 10 degrees.  Their jets therefore had a momentum flux per unit power approximately an order of magnitude higher than those we implemented and were also narrow. Those features enable drilling through the central few kiloparsecs but probably induce weaker circulation.

The bottom line of these comparisons is that the entropy profiles of simulated galactic atmospheres regulated by jet feedback depend on both the jet opening angle and the momentum flux per unit power of the jets. Both our simulations and those of \citet{wang2019} successfully self-regulate and reproduce some of the observed features of massive elliptical galaxies centered within halos of mass $\sim 10^{13.5} \, M_\odot$. However, certain details of the simulated entropy profiles are model-dependent and may affect the timescale on which self-regulation happens. Therefore, further exploration of the $(\epsilon_{\rm AGN}, \theta_{\rm jet})$ parameter space could potentially provide a basis for constraining jet momentum fluxes and opening angles with X-ray observations.  

\section{Conclusions}
\label{sec:conc}

This paper has analysed how simulated AGN feedback achieves self-regulation in the halos of mass $\sim 10^{13.5} \, M_\odot$ simulated by \citet{prasad2020}. Galaxies with two distinct initial configurations were simulated. One of them (the single phase galaxy/SPG) starts with a lower density, higher entropy atmosphere that has a deeper potential well at the center ($\sigma_v \approx 280 \, {\rm km \, s^{-1}}$). The other (the multiphase galaxy/MPG) has a higher density, lower entropy atmosphere with a shallower potential well at the center ($\sigma_v \approx 230 \, {\rm km \, s^{-1}}$). That setup allowed us to test the black hole feedback valve mechanism outlined in \citet{voit2020} and to explore the role of large-scale atmospheric circulation.

The following points summarize our findings:
\begin{itemize}

\item All of our simulations begin with a strong feedback outburst that raises the central gas entropy and lowers the central gas density (see Figures \ref{fig:init_spg} and \ref{fig:init_mpg}). In our fiducial SPG and MPG simulations, the initial outburst lowers the central gas density enough for local stellar heating to exceed local radiative cooling within the central few kiloparsecs, in alignment with the proposed black hole feedback valve mechanism. However, those simulations develop entropy excesses at 1--10~kpc compared to observations of the galaxies we aimed to model \citep[see][]{prasad2020}.

\item After the initial outburst allows stellar heating to exceed radiative cooling within the central galaxy, both fiducial simulations enter a low-power mode in which minor AGN outbursts help to maintain the galaxy's atmosphere in a quasi-steady state (see Figures \ref{fig:lum_spg} and \ref{fig:lum_mpg}). The general features of that low-power mode agree with the predictions of the black hole feedback valve mechanism.

\item In our fiducial simulations, AGN feedback flips itself from the high-power mode to the low-power mode by reconfiguring the galaxy's atmosphere out to distances exceeding 100~kpc (see Figures \ref{fig:dmde_SPG} and \ref{fig:dmde_MPG}). Most of the reconfiguration is accomplished through large-scale circulation, not heat input. Radial momentum introduced by the strong jets propagates into the ambient gas and lifts it outward, transporting lower-entropy gas from the central regions to greater altitudes. And as lifting removes lower-entropy gas from the center, high-entropy gas slides inward perpendicular to the jet axis and replaces it (see Figures \ref{fig:SPG_fid_circ}, \ref{fig:MPG_fid_circ_1}, and \ref{fig:MPG_fid_circ_2}). The result is a lowering of the central gas density that raises its cooling time and sharply reduces the central cooling-flow rate.

\item The importance of momentum transfer in our simulations results from a relatively low efficiency factor for conversion of accreted rest-mass energy into AGN power ($\epsilon_{\rm AGN} = 10^{-4}$). Our AGN feedback algorithm conserves gas mass, and so low-efficiency jets have a large radial momentum output per unit power. Consequently, the momentum flux of the jets 
allows them to drill through the CGM out to distances exceeding 100~kpc when in the high-power mode (see Figure \ref{fig:Pram}).

\item Reducing the opening angle of the jets ($\theta_{\rm jet}$), as in our quarter-angle SPG simulation, further increases the jet momentum flux, enabling much of the AGN feedback energy to drill to large radii without coupling as effectively to the central few kiloparsecs. As a result, the quarter-angle SPG simulation does not enter a low-power mode. Instead, the feedback algorithm continually consumes low-entropy gas at the center and ejects the consumed gas to large radii through the jets.

\item These numerical experiments demonstrate that choices of $\epsilon_{\rm AGN}$ and $\theta_{\rm jet}$ significantly affect how AGN feedback introduced via bipolar jets couples with the CGM of a massive elliptical galaxy. Outflows with excessively large momentum flux couple poorly with the CGM, while outflows with a low momentum flux cannot reach the CGM without decelerating and thermalizing the bulk of their energy. 

\item Our adopted values of $\epsilon_{\rm AGN}$ and $\theta_{\rm jet}$ allow self-regulation to naturally arise via the black hole feedback valve mechanism. However, the entropy profiles generated in those simulations deviate from observations by a factor $\sim 2$ at 1--10~kpc. Those entropy anomalies are produced by large scale circulation patterns that depend on both $\epsilon_{\rm AGN}$ and $\theta_{\rm jet}$, as well as the size scale of the jet-input mechanism. Further refinements of the mechanism may therefore result in self-regulating galactic atmospheres at this halo mass scale in better agreement with observations \citep[see also][]{wang2019}.

\item Observations capable of detecting circulation patterns in the CGM may be capable of constraining the momentum fluxes of AGN jets. For example, circulation patterns similar to the eddies shown in Figure \ref{fig:MPG_fid_circ_4} would signal prior episodes of large-scale uplift, followed by infall of uplifted low-entropy gas along the jet axis.

\end{itemize}

\begin{acknowledgments}
DP is supported by {\it Chandra} theory grant no. TM8-19006X (G. M. Voit as PI) and NSF grant no. AST-1517908 (B.W.O'Shea as PI). BWO acknowledges further support from NASA ATP grants NNX15AP39G and 80NSSC18K1105 and NSF grant AST-1908109. This work used the Extreme Science and Engineering Discovery Environment (XSEDE), which is supported by National Science Foundation grant number TG-AST090040 and TG-AST190022, as well as the resources of the Michigan State University High Performance Computing Center (operated by the Institute for Cyber-Enabled Research).   Computations described in this work were performed using the publicly-available Enzo \citep{Bryan2014,Enzo_2019} and YT \citep{YT} codes, which are the products of the collaborative effort of many independent  scientists  from  numerous  institutions  around  the world.
\end{acknowledgments}

\bibliography{reference}

\begin{thebibliography}{}
\expandafter\ifx\csname natexlab\endcsname\relax\def\natexlab#1{#1}\fi
\providecommand{\url}[1]{\href{#1}{#1}}

\bibitem[{{Babyk} {et~al.}(2018){Babyk}, {McNamara}, {Nulsen}, {Russell},
  {Vantyghem}, {Hogan}, \& {Pulido}}]{babyk2018}
{Babyk}, I.~V., {McNamara}, B.~R., {Nulsen}, P.~E.~J., {et~al.} 2018, \apj,
  862, 39

\bibitem[{{Balbus}(1986)}]{Balbus1986ApJ...303L..79B}
{Balbus}, S.~A. 1986, \apjl, 303, L79

\bibitem[{{Balbus}(1988)}]{Balbus1988ApJ...328..395B}
---. 1988, \apj, 328, 395

\bibitem[{{B{\^\i}rzan} {et~al.}(2004){B{\^\i}rzan}, {Rafferty}, {McNamara},
  {Wise}, \& {Nulsen}}]{birzan2004}
{B{\^\i}rzan}, L., {Rafferty}, D.~A., {McNamara}, B.~R., {Wise}, M.~W., \&
  {Nulsen}, P.~E.~J. 2004, \apj, 607, 800

\bibitem[{{Brummel-Smith} {et~al.}(2019){Brummel-Smith}, {Bryan}, {Butsky},
  {Corlies}, {Emerick}, {Forbes}, {Fujimoto}, {Goldbaum}, {Grete}, {Hummels},
  {Kim}, {Koh}, {Li}, {Li}, {Li}, {OShea}, {Peeples}, {Regan}, {Salem},
  {Schmidt}, {Simpson}, {Smith}, {Tumlinson}, {Turk}, {Wise}, {Abel},
  {Bordner}, {Cen}, {Collins}, {Crosby}, {Edelmann}, {Hahn}, {Harkness},
  {Harper-Clark}, {Kong}, {Kritsuk}, {Kuhlen}, {Larrue}, {Lee}, {Meece},
  {Norman}, {Oishi}, {Paschos}, {Peruta}, {Razoumov}, {Reynolds}, {Silvia},
  {Skillman}, {Skory}, {So}, {Tasker}, {Wagner}, {Wang}, {Xu}, \&
  {Zhao}}]{Enzo_2019}
{Brummel-Smith}, C., {Bryan}, G., {Butsky}, I., {et~al.} 2019, The Journal of
  Open Source Software, 4, 1636

\bibitem[{{Bryan} {et~al.}(2014){Bryan}, {Norman}, {O'Shea}, {Abel}, {Wise},
  {Turk}, {Reynolds}, {Collins}, {Wang}, {Skillman}, {Smith}, {Harkness},
  {Bordner}, {Kim}, {Kuhlen}, {Xu}, {Goldbaum}, {Hummels}, {Kritsuk}, {Tasker},
  {Skory}, {Simpson}, {Hahn}, {Oishi}, {So}, {Zhao}, {Cen}, {Li}, \& {Enzo
  Collaboration}}]{Bryan2014}
{Bryan}, G.~L., {Norman}, M.~L., {O'Shea}, B.~W., {et~al.} 2014, \apjs, 211, 19

\bibitem[{{Davies} {et~al.}(2019){Davies}, {Crain}, {McCarthy}, {Oppenheimer},
  {Schaye}, {Schaller}, \& {McAlpine}}]{davies2019}
{Davies}, J.~J., {Crain}, R.~A., {McCarthy}, I.~G., {et~al.} 2019, \mnras, 485,
  3783

\bibitem[{{Fabian}(1994)}]{fabian1994}
{Fabian}, A.~C. 1994, \araa, 32, 277

\bibitem[{{Field}(1965)}]{Field1965}
{Field}, G.~B. 1965, \apj, 142, 531

\bibitem[{{Frisbie} {et~al.}(2020){Frisbie}, {Donahue}, {Voit}, {Connor}, {Li},
  {Sun}, {Lakhchaura}, {Werner}, \& {Grossova}}]{Frisbie_2020ApJ...899..159F}
{Frisbie}, R. L.~S., {Donahue}, M., {Voit}, G.~M., {et~al.} 2020, \apj, 899,
  159

\bibitem[{{Gaspari} {et~al.}(2012){Gaspari}, {Brighenti}, \&
  {Temi}}]{Gaspari2012_ellipticals}
{Gaspari}, M., {Brighenti}, F., \& {Temi}, P. 2012, \mnras, 424, 190

\bibitem[{{Hillel} \& {Soker}(2017)}]{hillsok}
{Hillel}, S., \& {Soker}, N. 2017, \apj, 845, 91

\bibitem[{{Kormendy} \& {Ho}(2013)}]{kormendy2013}
{Kormendy}, J., \& {Ho}, L.~C. 2013, \araa, 51, 511

\bibitem[{{Lewis} {et~al.}(2000){Lewis}, {Babul}, {Katz}, {Quinn}, {Hernquist},
  \& {Weinberg}}]{lewis2000}
{Lewis}, G.~F., {Babul}, A., {Katz}, N., {et~al.} 2000, \apj, 536, 623

\bibitem[{{McCourt} {et~al.}(2012){McCourt}, {Sharma}, {Quataert}, \&
  {Parrish}}]{mccourt12}
{McCourt}, M., {Sharma}, P., {Quataert}, E., \& {Parrish}, I.~J. 2012, mnras,
  419, 3319

\bibitem[{{McNamara} \& {Nulsen}(2007)}]{McNamara2007}
{McNamara}, B.~R., \& {Nulsen}, P.~E.~J. 2007, \araa, 45, 117

\bibitem[{{Pizzolato} \& {Soker}(2005)}]{pizz2005}
{Pizzolato}, F., \& {Soker}, N. 2005, \apj, 632, 821

\bibitem[{{Prasad} {et~al.}(2020){Prasad}, {Voit}, {O'Shea}, \&
  {Glines}}]{prasad2020}
{Prasad}, D., {Voit}, G.~M., {O'Shea}, B.~W., \& {Glines}, F. 2020, \apj, 905,
  50

\bibitem[{{Rafferty} {et~al.}(2006){Rafferty}, {McNamara}, {Nulsen}, \&
  {Wise}}]{rafferty2006}
{Rafferty}, D.~A., {McNamara}, B.~R., {Nulsen}, P.~E.~J., \& {Wise}, M.~W.
  2006, \apj, 652, 216

\bibitem[{{Sharma} {et~al.}(2012){Sharma}, {McCourt}, {Quataert}, \&
  {Parrish}}]{sharma12}
{Sharma}, P., {McCourt}, M., {Quataert}, E., \& {Parrish}, I.~J. 2012, MNRAS,
  420, 3174

\bibitem[{{Turk} {et~al.}(2011){Turk}, {Smith}, {Oishi}, {Skory}, {Skillman},
  {Abel}, \& {Norman}}]{YT}
{Turk}, M.~J., {Smith}, B.~D., {Oishi}, J.~S., {et~al.} 2011, \apjs, 192, 9

\bibitem[{{Vernaleo} \&
  {Reynolds}(2006)}]{VernaleoReynolds_2006ApJ...645...83V}
{Vernaleo}, J.~C., \& {Reynolds}, C.~S. 2006, \apj, 645, 83

\bibitem[{{Voit} {et~al.}(2015{\natexlab{a}}){Voit}, {Bryan}, {O'Shea}, \&
  {Donahue}}]{voit15L}
{Voit}, G.~M., {Bryan}, G.~L., {O'Shea}, B.~W., \& {Donahue}, M.
  2015{\natexlab{a}}, ApJl, 808, L30

\bibitem[{{Voit} {et~al.}(2015{\natexlab{b}}){Voit}, {Donahue}, {Bryan}, \&
  {McDonald}}]{voit15N}
{Voit}, G.~M., {Donahue}, M., {Bryan}, G.~L., \& {McDonald}, M.
  2015{\natexlab{b}}, Nature, 519, 203

\bibitem[{{Voit} {et~al.}(2015{\natexlab{c}}){Voit}, {Donahue}, {O'Shea},
  {Bryan}, {Sun}, \& {Werner}}]{voit2015}
{Voit}, G.~M., {Donahue}, M., {O'Shea}, B.~W., {et~al.} 2015{\natexlab{c}},
  \apjl, 803, L21

\bibitem[{{Voit} {et~al.}(2017){Voit}, {Meece}, {Li}, {O'Shea}, {Bryan}, \&
  {Donahue}}]{voit2017}
{Voit}, G.~M., {Meece}, G., {Li}, Y., {et~al.} 2017, \apj, 845, 80

\bibitem[{{Voit} {et~al.}(2020){Voit}, {Bryan}, {Prasad}, {Frisbie}, {Li},
  {Donahue}, {O'Shea}, {Sun}, \& {Werner}}]{voit2020}
{Voit}, G.~M., {Bryan}, G.~L., {Prasad}, D., {et~al.} 2020, arXiv e-prints,
  arXiv:2006.09381

\bibitem[{{Wang} {et~al.}(2019){Wang}, {Li}, \& {Ruszkowski}}]{wang2019}
{Wang}, C., {Li}, Y., \& {Ruszkowski}, M. 2019, \mnras, 482, 3576

\bibitem[{{Werner} {et~al.}(2012){Werner}, {Allen}, \&
  {Simionescu}}]{werner2012}
{Werner}, N., {Allen}, S.~W., \& {Simionescu}, A. 2012, \mnras, 425, 2731

\bibitem[{{Werner} {et~al.}(2014){Werner}, {Oonk}, {Sun}, {Nulsen}, {Allen},
  {Canning}, {Simionescu}, {Hoffer}, {Connor}, {Donahue}, {Edge}, {Fabian},
  {von der Linden}, {Reynolds}, \& {Ruszkowski}}]{werner2014}
{Werner}, N., {Oonk}, J.~B.~R., {Sun}, M., {et~al.} 2014, \mnras, 439, 2291

\bibitem[{{Yang} \& {Reynolds}(2016)}]{YangReynolds_2016ApJ...829...90Y}
{Yang}, H. Y.~K., \& {Reynolds}, C.~S. 2016, \apj, 829, 90

\end{thebibliography}
\bibliographystyle{aasjournal}

\end{document}